\documentclass[epsfile,prc,aps,nofootinbib,showpacs]{revtex4}
\usepackage{graphicx}
\usepackage{epsfig}
\usepackage{amsmath}
\usepackage{amssymb}
\newcommand{\be}{\begin{equation}}
 \newcommand{\ee}{\end{equation}}
\newcommand{\bqa}{\begin{eqnarray}}
\newcommand{\eqa}{\end{eqnarray}}

\newcommand{\noi}{\noindent}
  
\begin{document}

\draft
\date{\today}

\title{Photon or meson formation in $J/\psi$ decays into  $p {\overline{p}}$}

\author{ J.-P. Dedonder  \thanks{e-mail "dedonder@univ-paris-diderot.fr"}}
\affiliation{Sorbonne Universit\'e, Campus Pierre et Marie Curie,\\
Sorbonne Paris Cit\'e, Universit\'e
 Paris Diderot, et IN2P3-CNRS, UMR 7585, Laboratoire de Physique Nucl\'eaire et de Hautes \'Energies,  
4 place Jussieu, 75252 Paris, France}

\author{ B. Loiseau \thanks{e-mail " "}}
\affiliation{Sorbonne Universit\'e, Campus Pierre et Marie Curie, \\Sorbonne Paris Cit\'e, Universit\'e
 Paris Diderot, et IN2P3-CNRS, UMR 7585, Laboratoire de Physique Nucl\'eaire et de Hautes \'Energies,  
4 place Jussieu, 75252 Paris, France}

\author{S. Wycech\thanks{e-mail "wycech@fuw.edu.pl"}}
\address{National Centre for  Nuclear Studies,
Warsaw, Poland}

\begin{abstract}
  The measurements of the $J/\psi \rightarrow \gamma p \overline{p}$ decays by the BES Collaboration indicate an enhancement at the $p \overline{p}$ threshold which, however, is not present in the J/$\psi$ decays into $\omega p \overline{p}$ and into $\pi p \overline{p}$.  
Here, two processes for describing the decays $J/\psi \rightarrow \mathcal{B} p \overline{p}$ where  $\mathcal{B} = \gamma, \omega$ are presented  in some detail and the cases $\mathcal{B} =\phi , \pi $  are briefly touched on. The first one, applied not only to the radiative decay to reproduce the threshold peak but also to the $\omega p \overline{p}$ decay channel to improve the description of the spectrum, postulates a direct emission of the boson before the baryon pair is formed. The second process assumes that the boson $\mathcal{B}$ is emitted from the baryon pair following the $J/\psi$ decay and includes for the decays into $\gamma p \overline{p}$ a final state nucleon-antinucleon interaction based on the Paris $N \overline{N}$ potential. The reproduction of the $p \overline{p}$ distribution in the $J/\psi \rightarrow \omega p \overline{p}$ decays needs a final state interaction involving a $N(2050)\ 3/2^-$ resonance. The photon- and meson-emission rates are reproduced in a semi-quantitative way. 
 
 \end{abstract}

\maketitle

\section{Introduction} \label{intro}

\vspace{1cm}

The $J/\psi$ decays with a  proton-antiproton ($p\overline{p}$) pair in the final state are interesting for at least two reasons:
\begin{itemize}
\item  They are related to the  searches for exotic states in the  nucleon-antinucleon ($N \overline{N}$)  systems. Such searches  have been pursued for a few decades, but significant results have been obtained only recently;

\item  They are closely related to the $p\overline{p}$ reactions, planned at
FAIR~\cite{fair}, aiming at the formation  of the $J/\psi$  in atomic nuclei.
\end{itemize}

The first topic is  discussed in this paper but the model developed here may be useful to describe the second one. Indication of  exotic states below the $N \overline{N}$ threshold may be given by scattering lengths for a given spin and isospin 
state. However, a clear separation of quantum states in scattering experiments is not easy. Equivalent measurements of the X ray transitions in the antiproton hydrogen atoms could also  select some partial waves if the fine structure of atomic levels is resolved. 
So far, only partial selections have been achieved~\cite{gotta2004}. On the basis of the existing data, the present authors have argued that even averaged fine structure atomic level widths in the lightest atoms indicate the existence of quasi-bound $N \overline{N}$ states~\cite{wycech2015}. Full resolution of the hyperfine structures should be the purpose of future experiments. 

To reach specific states one  can also use formation experiments. For instance, in the radiative $J/\psi$ decay,  
\be \label{1i}
 J/\psi \to \gamma p\overline{p},
\ee
an enhancement close to the $ p\overline{p} $ threshold has been observed by the BES~\footnote {Beijing Electron Sprectrometer} Collaboration~\cite{bai03,abl12}. We note that both the $J/\psi$ and the photon have $J^{PC}= 1^{--}$. There are three  final $p \overline{p}$ states allowed by  parity, $P$, and charge-conjugation, $C$, conservations in the $\gamma p \overline{p}$ channel:
 $^{1}S_0, \ ^{3}P_1$, and $ ^{3}P_0$. Table \ref{table0}    and Table \ref{tableCP} indicate  the allowed   $p \overline{p}$ states, denoted by  $^{2S+1}L_{J}$ or $^{2I+1,2S+1}L_{J}$, where $S,L,J$ denote the spin, angular momentum, total momentum of the pair, respectively and $I$  the isospin. Two isospin states,  $I=0, 1$,  enter the $p \overline{p}$ system. A first indication that the system is in an  $I=0$ state was obtained in a simple quark model in  Ref.~\cite{datta2003}. In Ref.~\cite{loi05} a unified picture and a limited  description of the radiative decays has been achieved in a semi-quantitative way. It suggests that the final $\gamma p \overline{p} $ state is dominated by the $^{11}S_0$ partial wave. In this partial wave the Paris potential  generates a $52$ MeV broad quasi-bound state at $4.8$ MeV below threshold~\cite{lac09}. The conclusion that a near threshold peak is formed in  the $^1S_0$ wave  has been reached by the  J\"{u}lich group although the Bonn-J\"ulich potential does not generate a bound state in this wave~\cite{jul06} 
 \footnote{For completeness we want to mention  that earlier in Ref.~\cite{sibirtsev2005} this group had claimed that within a Watson-Migdal approach they could reproduce the near threshold spectrum with an $I=1$ state. Also, later in Ref.~\cite{kang2015}, using for the  $N\overline{N}$ interaction a potential derived within chiral effective field theory fitted to results of a partial-wave analysis of  $p\overline{p}$ scattering data, the authors claim that the near-threshold spectrum observed in various decay reactions can be reproduced simultaneously and consistently by their treatment of the $p\overline{p}$ final-state interactions and that the interaction in the Isospin-1 $^1S_0$ channel, required to fit the decay $J/\psi \to \gamma p \overline{p}$, predicts an $N\overline{N}$ bound state.}
 and by Chen {\it et al.}~\cite{chen10} in the framework of an effective $N\overline{N}$ interaction model. Another study of the near threshold enhancement performed in Ref.~\cite{MSNPA996_54} finds a quasibound state to be the explanation.  The Bonn-J\"ulich group found recently a good description of the threshold behaviour in all mesic channels with a chirally motivated $N\overline{N}$ potential~\cite{kang2015}. The conclusion reached is similar to that obtained with the Paris potential, the near threshold enhancement indicates the presence of a quasi-bound state.  \\

To understand  better the nature of the $p\overline{p}$ states involved, one should look directly into the subthreshold energy region. This  may be achieved  in the antiproton-deuteron or the antiproton-helium reactions at  zero or  low energies. Another way to look below the threshold is the detection  of $N \overline{N}$ decay products. The specific decay mode
\be \label{3i}
 J/\psi \rightarrow  \gamma  \pi^+ \pi^- \eta'
\ee
has been studied by the BES Collaboration~\cite{abl05}.
 This reaction  is attributed by BES to an  intermediate $p \overline{p}$ configuration in the  $J^{PC}( p \overline{p} ) = 0^{-+}$ state which corresponds to spin singlet $S$-wave. The peak observed in the invariant  mass of the mesons 
  has been  interpreted as a new baryon state  and  named  $X(1835)$.

Under the assumption that all  mesons are produced in relative $S$-waves, reaction~(\ref{3i}), if attributed to an intermediate $p \overline{p}$ state, is even more restrictive than reaction (\ref{1i}). It allows only  one intermediate state, the $p\overline{p}$ $^{1}S_0$,  which coincides with the previous findings. The intermediate $p \overline{p}$  state  in reaction~(\ref{3i}) is possible but not warranted. In Ref.~\cite{ded09} a more consistent interpretation is obtained with the dominance of the $^{11}S_0$ state which  is a mixture of $p\overline{p}$ and $n\bar{n}$ pairs. It has been  argued  that the peak  is due to an interference of a quasi-bound, isospin 0,  $ N \overline{N}$ state with a background amplitude. This quasi-bound state was found in Loiseau and Wycech~\cite{loi05} to be  responsible for the threshold  enhancement in reaction~(\ref{1i}). A recent BES III experiment~\cite{abli2018} has studied the radiative decay $J/\psi \to \gamma \gamma \phi$ and observes a broad bump in the $M(\gamma \phi)$ invariant mass distribution. The shape of this bump is consistent with that observed in the absorptive $N \overline{N}$  amplitude obtained in Ref.~\cite{ded09}. A related strong enhancement of the 	absorption is observed in the light antiprotonic atoms.The comparison of atomic level widths in a series of atoms ($H,^2H, ^3H, ^3He,^4He$) allows to test the absorption  of antiprotons on more and more strongly bound protons up to subthreshold energies of $E_{p\overline{p}}$ down to $-\ 40 $ MeV~\cite{wycech2015}. The enhancement of absorption below the $p\overline{p}$ threshold is consistent with both results from Refs.~\cite{abli2018} and~\cite{wycech2015}. This, in our view, provides evidence that the $X(1835)$ meson is due to attraction in the $N\overline{N}$ system.

A similar decay mode
\be \label{4i}
 J/\psi \rightarrow \pi^0 p\overline{p}
\ee
displays no near threshold enhancement~\cite{bai03}. Recent  BES III experiments~\cite{abl07a,abl13a}  have extended these measurements to the reaction
\be \label{5i}
 J/\psi \rightarrow \omega p\overline{p}.
\ee
 No clear near threshold enhancement is found although Haidenbauer~{\it et al.}~\cite{jul08} claim the existence of a small signal above phase space very close to this threshold. Beyond, a depression at low  $ p\overline{p}$ energies  is seen in the data. These two reactions indicate a strong  $P$-wave dominance in reaction~(\ref{4i}) and  a sizable $P$ wave in reaction~(\ref{5i}). Both find a natural explanation in the model developed in 
 the present work. Recent experiments find no $ p\overline{p}$ threshold structure in the 
\be \label{6i} 
\psi' \rightarrow \gamma p\overline{p}
\ee 
decay~\cite{abl12,ale10}. This result is puzzling as final  $ p\bar{p}$ states in this process are the same as the final states in $J/\psi \rightarrow \gamma p\overline{p}$ decay. Within the model discussed here we find a qualitative explanation for this difference (see Section~\ref{radiative}).\\

\noi Different experimental branching fractions for the $J/\psi$ decay modes implying a $p\overline{p}$ pair based on Fermi Lab~\cite{PDG16} and BES experiments~\cite{abl07a,abl13a,PDG16,abl_PRD93} are shown in Table~\ref{tableCP}. One notable fact from this Table is that the radiative decay is comparable to the decay into strongly interacting mesons. We will see that this is due to a balance between the phase space (see Appendix \ref{phasespace}), the coupling constants and the fact of strong $N\overline{N}$ interactions and a direct emission process.\\
 
\begin{table}[ht]
\caption{ The states  of low energy  $p \bar{p}$   pairs allowed in the $ J/\psi \rightarrow \gamma  p\overline{p}$ and  $ J/\psi \rightarrow \pi^0  p\overline{p}$ decays. The first column gives  decay  modes and specifies the internal states of the $ p\overline{p} $ pair.  Both the $ J/\psi$ meson  and the photon have $J^{PC}  = 1^{--}$.  The second column gives $J^{PC}$ for the $p \overline{p}$ system, the last column gives the relative angular momentum of the photon or pion  vs. the $p\overline{p}$ pair. } 
\vspace{.3cm}
\begin{tabular}{lcc}
 \hline
 \hline
Decay mode                               &\  $J^{PC}( p \overline{p}) $ &\ Relative $l$      \\
\hline
$\gamma  p \overline{p} (^1S_0) $               &  $0^{-+}$               &  1              \\
$\gamma  p \overline{p} (^3P_0) $               &  $ 0^{++}$              &  0             \\
$\gamma  p \overline{p} (^3P_1)$                &  $ 1^{++}$              &  0           \\
\hline
 $\pi^0   p \overline{p} (^{31}P_1) $     &  $ 1^{+-} $             &  0            \\
 $\pi^0   p \overline{p} (^{33}S_1) $           &   $1^{--}   $         &   1          \\
\hline
 \hline
\end{tabular}
\label{table0}
\end{table}

\begin{table}[ht]
\caption{Experimental branching fractions for some decay modes of the $ J/\psi$ meson into channels implying $N \overline{N}$  pairs and the corresponding allowed states of the $N \overline{N}$ pair. All data from Ref.~\cite{PDG16} but for the $  p \overline{p} \phi$   channel recently measured in Ref.~\cite{abl_PRD93}}
\vspace{.3cm}
\begin{tabular}{lcc}
\hline
\hline Decay                               & \ Experimental  &\ $N \overline{N}$ allowed  \\
mode &\  branching fractions &  states   \\\hline
$ p \overline{p}  \pi^{0}  $           &  $1.19(0.08)\times 10^{-3}$  & $^{33}S_1 , ^{31}P_1$                 \\
$  p \overline{n} \pi^{-}  $           &  $2.12(0.09)\times10^{-3}$  & $^{33}S_1 , ^{31}P_1$                 \\
$ p \overline{p}  \gamma $            &  $3.8(1.0)\times10^{-4}$   &  $^{1}S_0,  ^{3}P_1, ^{3}P_0$         \\
$ p \overline{p} \omega  $             &  $9.8(1.0)\times10^{-4}$ &  $^{11}S_0,  ^{13}P_1, ^{13}P_0$     \\
$ p \overline{p} \phi $              &   $5.23(0.34)\times10^{-5}$   & $^{11}S_0 , ^{13}P_1, ^{13}P_0$     \\
$ p \overline{p}  $                   &   $2.120(0.029)\times10^{-3}$  & $^{13}S_1 $     \\
$ n \overline{n}  $                   &   $2.09(0.16)\times10^{-3}$ & $^{13}S_1 $     \\
\hline
\hline
\end{tabular}
\label{tableCP}
\end{table}

The purpose of the present  work  is to discuss and correlate  the physics of $N \overline{N}$ states produced  in the  $J/\psi$ decays. The main assumption is that the bosons (photon and mesons) are emitted after the 
 $N \overline{N}$ baryons have been produced. In this way one obtains  branching ratios   $ \Gamma(N\overline{N} \mathcal{B} )/\Gamma(N \overline{N})$ consistent with experimental data for the $\pi^0, \pi^-
 $ and $\phi$ mesons formation, listed in Table~\ref{tableCP}. One free parameter $R_0$ - the size of initial $N \overline{N}$ source enters this model and it comes out with a reasonable value of $0.28$~fm. On the other hand, to obtain the invariant $p \overline{p}$ mass spectra in the decays and in particular to generate the threshold peak it is necessary to include an additional mechanism  for the photon emission before the baryon formation phase. The peak of interest arises as a result of $ p\bar{p} $  final state interaction in the way described in Refs.~\cite{loi05,jul06}. The rate of this decay enters as another free parameter.\\
 
The content  of this paper goes as follows. Section \ref{Jpsippbar} recalls briefly the derivation of the width of the $J/\psi \to p \overline{p}$ decay mode.
Section \ref{internal}   develops a  model for radiative decay which assumes the photon to be emitted at an early stage of the process. This internal emission model explains the two maxima in the final  $p\bar{p}$ spectrum;  one is  due to  baryonium while the other represents a shape resonance in  the  $p\bar{p}$ interaction. It can be extended to the case of emission of any meson. Section \ref{baryonc} discusses the photon or meson $( \omega, \phi,\pi) $ emission from the final baryon currents, i.e., once the baryons are formed following the decay of the $J/\psi$. 
Section \ref{results}  collects the results. In the case of the $J/\psi \to \omega p \overline{p}$ decay, the description of the $p\overline{p}$ spectrum requires final state interactions with a $N^*(3/2^-)$ resonance while that of the $\omega p$ spectrum requires a contribution of the mechanism of $\omega$ emission before the baryon pair formation occurs.
 A brief summary together with some outlook are given in  Section \ref{conclude}.
 Finally, appendices tackle a number of technical questions.

\section{The  $J/\psi \rightarrow p\overline{p}$ amplitude and its width} \label{Jpsippbar}

Let  the initial  $J/\psi$ wave  function in momentum space $\psi_i$ be  normalized as
\be \label{n2}
 \psi_i (\mathbf{P})= \frac{1}{\sqrt{\mathcal{V}_0}} \ (2\pi)^3 \delta^{(3)}(\mathbf{P}),
\ee
where $ {\mathcal{V}_0}$ is the normalization volume. In the rest frame of the $J/\psi$, the amplitude $ A_{N\overline{N}}(\bf{q}_1, \bf{q}_2)$ that describes  the $J/\psi \to (N\overline{N})_{I=0}$ reaction is given by 
\be \label{n0}
A_{N\overline{N}}({\bf{q}_1, \bf{q}_2})=  \langle {{N(\bf{q}_1}) \ \overline{N}({\bf{q}_2} )\vert} 
\widehat{A}_{N\overline{N}} \vert \psi_i \rangle = 
 (2\pi)^3\  \delta^{(3)} \left({\bf{q}_1+\bf{q}_2} \right)\ 
\frac{1}{\sqrt{\mathcal{V}_0}}\  {\mathcal{F}_{J/\psi}(\mathbf{q}_r)}, 
   \ee
where $\mathcal{F}_{J/\psi}$ denotes the source function associated to the creation of the $N\overline{N}$ pair from the initial $J/\psi$ meson and  where $\bf{q}_1$ and  $\bf{q}_2$ denote the momenta of the nucleon and the antinucleon respectively. This source function is assumed to depend only on the relative $N\overline{N}$ momentum $\mathbf{q}_r$
\be \label{dir2}
\mathbf{q}_r = {\frac{\mathbf{q}_1-\mathbf{q}_2}{2}}.
 \ee
 
\noi We postulate furthermore the following smooth phenomenological form for the source function 
\be \label{dir3}
\mathcal{F}_{J/\psi}(\mathbf{q}_r) = \mathcal{F}_{J/\psi}(q_r) =  F_0 \exp(-q_r^2  R_0^2/2), 
\ee
where $R_0$ is the radius of the source for the formation of the $N\overline{N}$ pair and  $F_0$  is a normalization constant. \\

The probability for the  $J/\psi \to p\overline{p}$ decay channel can be written as
\be \label{n4}
 \Gamma (p\overline{p}) = \frac{1}{2}\ \int  \frac{d\mathbf{q}_1}{(2\pi)^3} \ \frac{d\mathbf{q}_2}{(2\pi)^3}
\frac{\delta( M_{J/\psi} - E(\mathbf{q}_1)-E(\mathbf{q}_2) )} {2E(\mathbf{q}_1) \ 2 E (\mathbf{q}_2)}
|A_{p\overline{p}}({\bf{q}_1, \bf{q}_2})|^2 ,
\ee
where we have taken into account the probability to find $p\overline{p}$ in the isospin $0$ state, $|<I=0|p\overline{p}>|^2 = 1/2$~
 \footnote{The isospin structure of the $N\overline{N}$ states, following the convention used in the Paris potential model,  is given by 
$ \vert I=0 \rangle = (\vert p\overline{p}\rangle - \vert n\overline{n} \rangle)/\sqrt{2}$ and $\vert I=1 \rangle = (\vert p\overline{p}\rangle + \vert n\overline{n} \rangle)/\sqrt{2} $, 
so that one  has
$$\langle 0\vert p\overline{p}\rangle = \langle 1\vert p\overline{p}\rangle = \langle 1\vert n\overline{n}\rangle= 1/\sqrt{2} \hspace{0.5cm}{\rm and}\hspace{0.5cm} 
 \langle 0\vert n\overline{n}\rangle = -  \langle 0\vert p\overline{p}\rangle.$$}. Using Eq.~(\ref{n0}) and the relation $(2\pi)^3 \delta^{(3)}({\mathbf{0}}) = \mathcal{V}_0$, one gets
\bqa \label{n5}
\Gamma (p\overline{p}) &=& \frac{1}{2} \ \frac{\delta^{(3)}(\textbf{0})}{\mathcal{V}_0}  \int d\textbf{q}
 \frac{\delta( M_{J/\psi}- 2E (q))} {[2 E(q)]^2}\  |\mathcal{F}_{J/\psi}({\bf{q}})|^2
 = \frac{1}{4 \pi^2} \int q^2 \ dq\ \frac{\delta( M_{J/\psi}- 2E (q))} {[2 E(q)]^2}\  |\mathcal{F}_{J/\psi}(q)|^2 \nonumber \\
 &=&   \frac{1}{16 \pi^2} \ \frac{q_m}{M_{J/\psi}}\ |\mathcal{F}_{J/\psi}(q_m)|^2, 
 \eqa
 where the delta function has provided $ q= q_m =  \frac{1}{2}\sqrt{M_{J/\psi}^2- 4\ m^2}$ where $M_{J/\psi}$ denotes the mass of the $J/\psi$ meson and $m$  the nucleon mass.
\noi This derivation is recalled here to ascertain that the same factors are used for the particle $\mathcal{B}$ formation reactions $J/\psi \rightarrow  p\bar{p} \mathcal{B} $. The corresponding decay rates will be referred to the prime  $J/\psi \rightarrow p\bar{p}$ rate. 

\section{Direct (internal)  emission  amplitudes } \label{internal}

\begin{figure}[ht]
\includegraphics[scale=0.8]{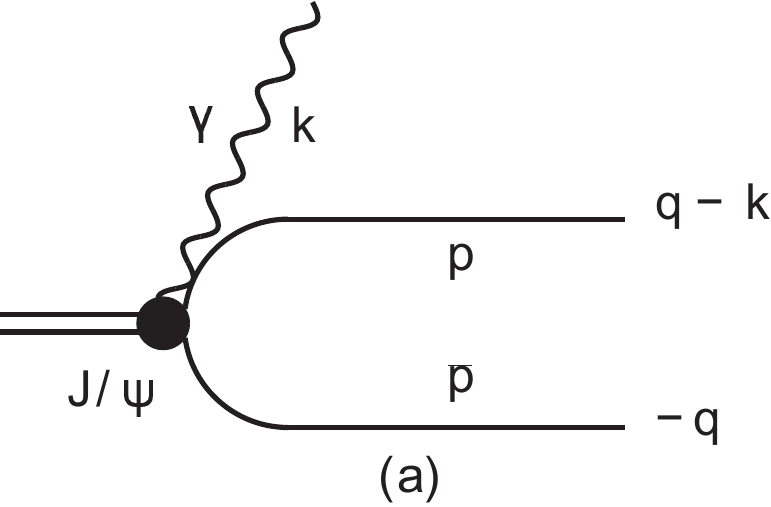}
\hspace{1cm} 
\includegraphics[scale=0.8]{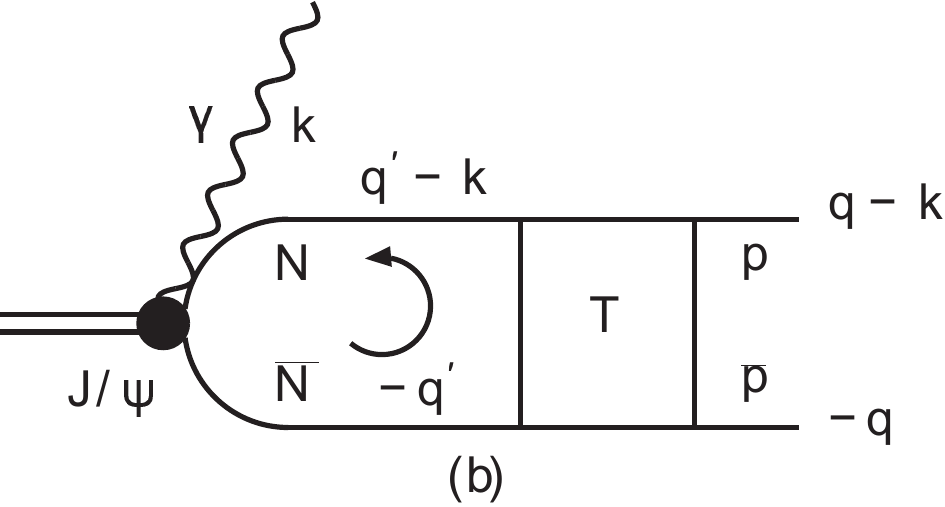}
\caption{Photon emission from the $J/\psi$ : the left panel (a) corresponds to the Born term while the right one (b)  includes final state corrections. The nucleon (antinucleon) line is denoted $N$ ($\overline{N}$) with respective momenta $\bf{q}'-\bf{k}$ ($-\bf{q}'$) while $p$ ($\overline{p}$) represents a proton (antiptroton) propagating with momentum $\bf{q}-\bf{k}$  ($-\bf{q}$). The wavy line is associated to the photon of momentum $\bf{k}$.}\label{figdirect}
\end{figure}

The essence of this approach is presented in figure~\ref{figdirect} where we illustrate the processes at stake in the case of the photon. The photon is emitted before the $p\overline{p}$ pair is formed. It has been shown in references~\cite{loi05} and~\cite{jul06} that this assumption allows to reproduce the near threshold enhancement in the $p\overline{p}$  invariant mass ($M_{p\overline{p}}$) distribution. This enhancement is due to the final state interaction of the two protons.  The interactions, Paris potential in~\cite{loi05} and Bonn/J\"{u}lich potential in~\cite{jul06}, are strongly attractive. In the Paris  potential case a quasi-bound state is generated while none appears in the Bonn potential case. An extension of these calculations to larger values of $ M_{p\overline{p}}$ is presented below.

In this approach, which will be referred to as the direct emission (DE) model, the direct internal emission process  arises either from the charmed $c\overline{c}$ quark pair or from the quark rearrangement stage of the process and its rate is hard to calculate. Here, this rate is fixed by an optimal description of the ratio ${\Gamma( p\overline{p}~\gamma)}/{\Gamma( p\overline{p})}$ and  of the magnitude of the  threshold peak. 
 The spectrum is generated by a Born operator, $\widehat{A}_{p\overline{p}\gamma}^{B,DE}$,  
and final state interactions (FSI) summed in the operator $\widehat{A}_{p\overline{p}\gamma}^{FSI,DE}$ and collected into the full internal emission operator $ \widehat{A}^{DE}_{p\overline{p}\gamma}$ which can be formally written as
\be
\label{dir1}
 \widehat{A}_{p\overline{p}\gamma}^{DE}=  \widehat{A}_{p\overline{p}\gamma}^{B,DE}  + \widehat{A}_{p\overline{p}\gamma}^{FSI,DE} = 
 \widehat{A}_{p\overline{p}\gamma}^{B,DE} \ \left [ 1 + G^+_{0, N\overline{N}\gamma} T_{[N\overline{N}]}(E_{N\overline{N}})\right ] = \left [1 + T_{[N\overline{N}]}(E_{N\overline{N}})G^+_{0,N\overline{N}\gamma} \right ] \ \widehat{A}_{p\overline{p}\gamma}^{B,DE}, 
 \ee
where $ G^+_{0,N\overline{N}\gamma} $ is the free $N \overline{N}$ propagator at the energy $E_{N\overline{N}}$ in the presence of the photon of momentum $\bf{k}$ and $T_{[N\overline{N}]}(E_{N\overline{N}})$ is  the $N \overline{N}$ scattering $T$-operator. This operator can act in both $I=0$ and $I=1$ state, which occurs for the $N\overline{N}$ pair in the diagram representing the final state interactions (right panel in Fig.~(\ref{figdirect})) and will be written when necessary $T_{[N\overline{N}]_I}(E_{N\overline{N}})$. The Born operator $\widehat{A}^{B,DE}_{N\overline{N}\gamma}$ is factorized into two contributions : the $N\overline{N}$ pair creation from the $J/\psi$ meson described through the operator $\widehat{A}_{N\overline{N}}$ and the direct photon emission from the $J/\psi$ meson  given by the operator  $\widehat{V}_{\gamma}^{DE}$
 \be \label{A0}
\widehat{A}_{N\overline{N}\gamma}^{B,DE}= \widehat{A}_{N\overline{N}}\ \widehat{V}_{\gamma}^{DE}. 
\ee

The direct photon emission operator  $\widehat{V}_{\gamma}^{DE}$ has to conserve the charge-conjugation-parity, $CP$, symmetry. In momentum space one has  three vectors available:   $\bf{k}$, ${\mbox {\boldmath $\xi$}}$ - the initial orientation of the $J/\psi$ spin -,  and the vector product  ${\mbox {\boldmath $\xi$}}\wedge \bf{k}$  to be combined with the polarization vector of the photon of helicity $\lambda$,  ${\mbox {\boldmath $\epsilon$}^*(\lambda)}$. The matrix element of the operator $\widehat{V}_{\gamma}^{DE}$ associated to a transition to reach a $^1S_0$ state should then be of the form 
\be \label{Vdir}
V^{DE}_{\gamma} {(\bf{k})} = g_{DE}\ {\mbox {\boldmath $\epsilon$}^*(\lambda)}\cdot
({\mbox {\boldmath $\xi$}}\wedge \bf{k}), 
\ee
where the constant $g_{DE}$ is a free parameter.\\ 

The initial $J/\psi$ meson  at rest is described by the momentum space wave function $\psi_i$ given in Eq.~(\ref{n2}) and the Born amplitude $ A_{p\overline{p}\gamma}^{B,DE}(\mathbf{q}_r,\mathbf{k})$ is given by the relations 
\bqa
\label{A_0Psi}
\langle p(\mathbf{q}_1)\ \overline{p}(\mathbf{q}_2)\ \gamma(\mathbf{k})\vert \widehat{A}_{p\overline{p}\gamma}^{B,DE}\vert \psi_i \rangle &=& 
(2\pi)^3 \ \delta^{(3)}( \mathbf{q}_1+\mathbf{q}_2+\mathbf{k}) \ A_{p\overline{p}\gamma}^{B,DE} (\mathbf{q}_r,\mathbf{k}), \nonumber \\
A_{p\overline{p}\gamma}^{B,DE} (\mathbf{q}_r,\mathbf{k})&=&  \frac{1}{\sqrt{\mathcal{V}_0}} \ 
 \mathcal{F}_{J/\psi}(\mathbf{q}_r) \ V_{\gamma}^{DE}(\mathbf{k}).
\eqa

\noi The semi-relativistic three particle  free $N\overline{N}\gamma$ propagator matrix elements read here
\be \label{G0NNbar}
\langle \mathbf{q}_1\ \mathbf{q}_2 \ \mathbf{k} \vert G^+_{0, N\overline{N}\gamma} \vert \mathbf{q'}_1\ \mathbf{q'}_2  \ \mathbf{k} \rangle = (2\pi)^6 \ \delta^{(3)}(\mathbf{q}_1 -\mathbf{q'}_1) \ 
\delta^{(3)}(\mathbf{q}_2 -\mathbf{q'}_2)\ G^+_{0, N\overline{N}\gamma}( \mathbf{q}_1, \mathbf{q}_2, \mathbf{k}), 
\ee
where 
\be \label{G0}
G^+_{0, N\overline{N}\gamma}( \mathbf{q}_1, \mathbf{q}_2, \mathbf{k})= \frac{1}{E_{N\overline{N}}+i \epsilon - \sqrt{\mathbf{q}_1^2 + m^2} - \sqrt{\mathbf{q}_2^2 + m^2}},
\ee 
 $E_{N\overline{N}}= M_{J/\psi} -k$ being the $N\overline{N}$ pair energy  and $k= \vert \mathbf{k} \vert $ the emitted photon energy. 
In the evaluation of the final state interaction contribution, the $N\overline{N}$ pair may be in either isospin  $I=0$ or $I=1$ state.
Thus we may write this contribution as 
\bqa \label{TGAPsi}
\langle p(\mathbf{q}_1)\ \overline{p}(\mathbf{q}_2)\ \gamma(\mathbf{k})\vert T_{[N\overline{N}]_I}(E_{N\overline{N}})\ G^+_{0, N\overline{N}\gamma}\  \widehat{A}_{[N\overline{N}]_I \gamma}^{B,DE} \vert \psi_i\rangle &=& \frac{1}{\sqrt{\mathcal{V}_0}} \int
\langle \mathbf{q}_1\ \mathbf{q}_2\vert T_{[N\overline{N}]_I}(E_{N\overline{N}}) \vert  \mathbf{q'}_1\ \mathbf{q'}_2\rangle \ (2\pi)^3\  \delta^{(3)}( \mathbf{q'}_1+\mathbf{q'}_2+\mathbf{k}) \nonumber \\
&\negthickspace& \negthickspace\negthickspace\negthickspace\negthickspace\negthickspace\negthickspace\negthickspace\negthickspace\negthickspace\negthickspace\negthickspace\negthickspace\negthickspace\negthickspace\negthickspace\negthickspace\negthickspace
\negthickspace\negthickspace\negthickspace\negthickspace\negthickspace\negthickspace\negthickspace\negthickspace\negthickspace\times \ \frac{ d\mathbf{q'}_1\ d\mathbf{q'}_2}{(2\pi)^6}  \
 \frac{1}{E_{N\overline{N}} +i \epsilon  - \sqrt{\mathbf{q}_1^{'2} + m^2} - \sqrt{\mathbf{q}_2^{'2} + m^2}} \ 
 \mathcal{F}_{J/\psi}(\vert \mathbf{q'}_r\vert) \ V_{\gamma}^{DE}(\mathbf{k}),
\eqa
where $\mathbf{q'}_r= (\mathbf{q'}_1-\mathbf{q'}_2)/2 = \mathbf{q}'-\mathbf{k}/2$ with $\mathbf{q}'_1= \mathbf{q}' - \mathbf{k}$ and $ \mathbf{q}'_2 = - \mathbf{q}'$. Since
\be \label{TNNbar}
\langle \mathbf{q}_1\ \mathbf{q}_2\vert T_{[N\overline{N}]_I}(E_{N\overline{N}})\vert  \mathbf{q'}_1\ \mathbf{q'}_2\rangle = (2\pi)^3\ \delta^{(3)}( \mathbf{q'}_1+\mathbf{q'}_2 - \mathbf{q}_1-\mathbf{q}_2) \ 
T_I(\mathbf{q}_r, \mathbf{q'}_r, E_{N\overline{N}}), \ee
we finally arrive at the loop integral that yields the contribution of final state interactions for the direct photon emission process
 \be 
  \langle p(\mathbf{q}_1)\ \overline{p}(\mathbf{q}_2)\ \gamma(\mathbf{k}) \vert T_{[N\overline{N}]_I}(E_{N\overline{N}})\ G^+_{0,N\overline{N}\gamma } \ 
  \widehat{A}_{p\overline{p}\gamma}^{B,DE} \vert \psi_i \rangle = (2\pi)^3 \ \delta^{(3)}( \mathbf{q}_1+\mathbf{q}_2+\mathbf{k}) \ A_{p\overline{p}\gamma}^{FSI,DE} (\mathbf{q}_r,\mathbf{k},E_{N\overline{N}}), 
  \ee
where
\bqa \label{TGAPsi2}
A_{p\overline{p}\gamma}^{FSI,DE} (\mathbf{q}_r,\mathbf{k},E_{N\overline{N}}) &=&  \frac{1}{\sqrt{\mathcal{V}_0}} \ 
 \ V_{\gamma}^{DE}(\mathbf{k}) \sum_{I=0,1}\int   \frac{d\mathbf{q'}_2}{(2\pi)^3} 
 \  T_I(\mathbf{q}_r, -\mathbf{q'}_2 - \frac{\mathbf{k}}{2}, E_{N\overline{N}})\ \nonumber \\  
 && \hspace{2cm} \times   
 \frac{1}{E_{N\overline{N}}+i \epsilon - \sqrt{({\mathbf{q'}_2+ \mathbf{k})^{2}+ m^2}}-\sqrt{\mathbf{q}_2^{'2} + m^2} } 
 \ \mathcal{F}_{J/\psi}( \vert \mathbf{q'}_2+ \frac{\mathbf{k}}{2}\vert),
 \eqa
$\mathbf{q}_r$ being defined in Eq.~(\ref{dir2}).
\noi Then the full amplitude for the direct photon emission reads
\bqa \label{dir4}
\langle p(\mathbf{q}_1),\overline{p}(\mathbf{q}_2),\gamma(\mathbf{k}) \vert \widehat{A}_{p\overline{p}\gamma}^{DE}\vert \psi_i \rangle &=&
(2\pi)^3 \ \delta^{(3)}( \mathbf{q}_1+\mathbf{q}_2+\mathbf{k}) \ A_{p\overline{p}\gamma}^{DE}(\mathbf{q}_r, \mathbf{k}, E_{N\overline{N}}) \nonumber \\
&=&
(2\pi)^3 \ \delta^{(3)}( \mathbf{q}_1+\mathbf{q}_2+\mathbf{k}) \ \left [A_{p\overline{p}\gamma}^{B,DE} (\mathbf{q}_r,\mathbf{k}) + A_{p\overline{p}\gamma}^{FSI,DE} (\mathbf{q}_r,\mathbf{k},E_{N\overline{N}})
\right ], 
\eqa
where $A_{p\overline{p}\gamma}^{B,DE} (\mathbf{q}_r,\mathbf{k})$, given by Eq. (\ref{A_0Psi}), corresponds to  the Born amplitude while the effect of final state interactions is given by the loop integral  $A_{p\overline{p}\gamma}^{FSI,DE} (\mathbf{q}_r,\mathbf{k},E_{N\overline{N}})$ of Eq.~(\ref{TGAPsi2}). These results can be similarly extended for the internal emission of a vector meson $\mathcal{B}$ where one has simply to replace the potential  $V_{\gamma}^{DE}(\mathbf{k})$ by an appropriate potential.\\

The isospin symmetry is violated by the ``internal photon" and as we wrote above in this section, the intermediate state of the baryon pair in Fig.~1(b)  is a superposition of $I=0$ and $I=1$ $N \overline{N}$ states or of $p\overline{p}$ and $n\overline{n}$ ones.
However, the  $  n\overline{n} \to p\overline{p}$  transition is weak as the $  n\overline{n} \to p\overline{p}$   cross section  is   smaller by a factor  of the order of $1/15 $  as compared to  the  $  p\overline{p} \to p\overline{p}$   cross section,  see Ref.~\cite{lac09} for comparison. Hence, in our calculation, the small  correction due the  $n\overline{n}$  interaction is neglected.  \\

To complete this phenomenological approach, we will assume in addition that, in this process, the source radius has a weak energy dependence on the $p\overline{p}$ invariant mass that reads
\be \label{Mppbar}
M_{p\overline{p}} = \sqrt{(M_{J/\psi}-k)^2 - (\mathbf{q}_1 + \mathbf{q}_2)^2}  = \sqrt{M_{J/\psi}\ (M_{J/\psi} - 2 \ k)}.
\ee
\noi We thus write, with masses expressed in units of fm$^{-1}$,
\be \label{dir6}
 R(M_{p\overline{p}})  = R_0 + \beta \sqrt{M_{p\overline{p}} - 2m} =  R_0 + \beta\ \sqrt{M_{J/\psi} \left (1-\frac{2 k}{M_{J/\psi}}\right )^{1/2} -2 m}.
\ee 
The values $ R_0=0.28$ fm and $ \beta  = 0.175$ fm$^{3/2} $ are found to  represent the data fairly well. 
This expression (\ref{dir6}) can also be reinterpreted as a modification of the functional form of the source function $ \mathcal{F}_{J/\psi}(q_r) $ [see Eq.~(\ref{dir3})].  We stay with this parametrization  as it indicates a physical effect indicated below.\\

\begin{figure}[ht]
\includegraphics[scale=0.65]{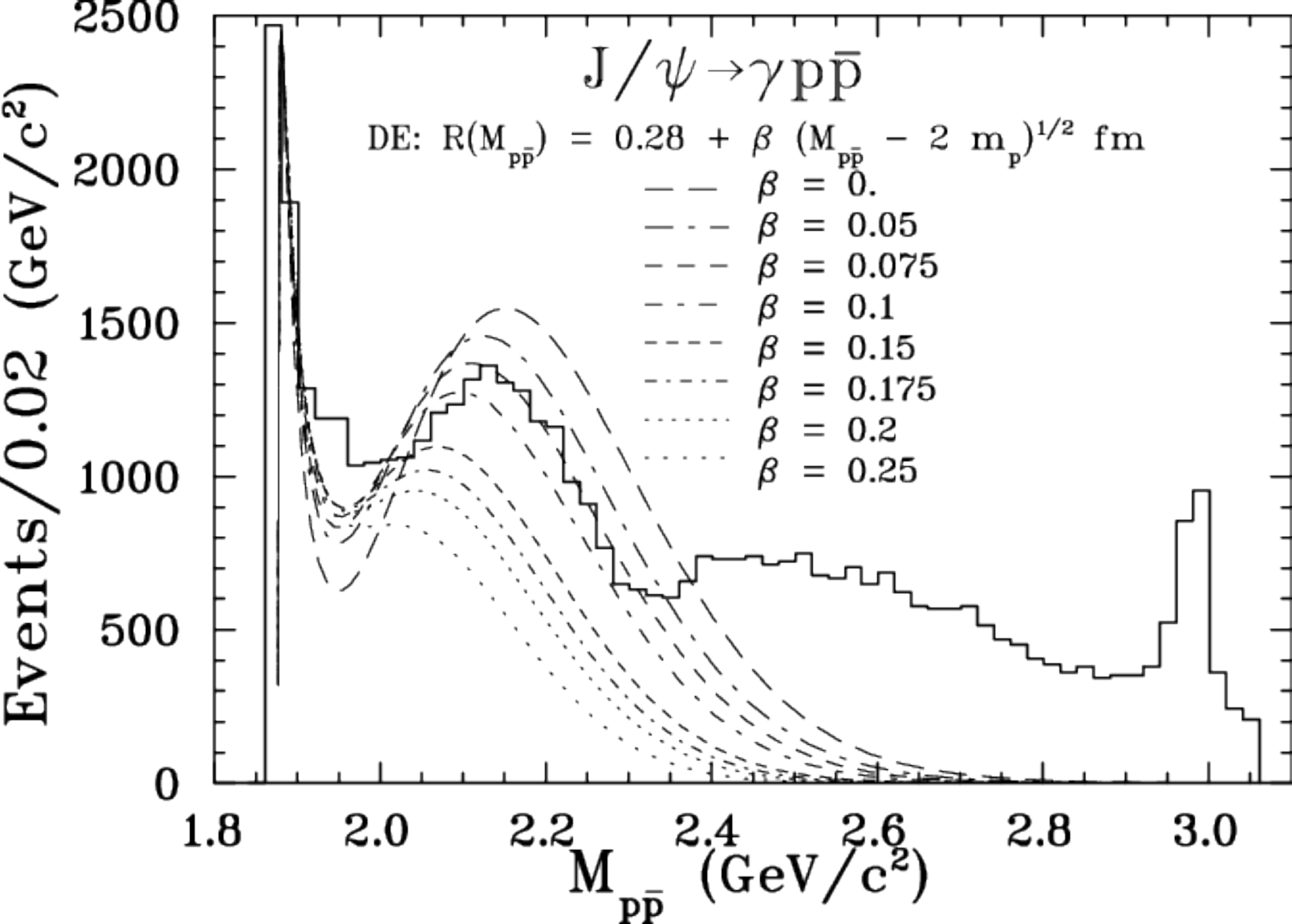}
\caption{  The invariant $ M_{p\overline{p}}$  mass distribution calculated within the direct emission DE model for different values of the parameter $\beta$. The data (histogram) is extracted  from  Fig. 1 in Ref.~\cite{abl12}.}
\label{MPPdirect}
\end{figure}

\begin{figure}[ht] 
\includegraphics [scale=0.35]{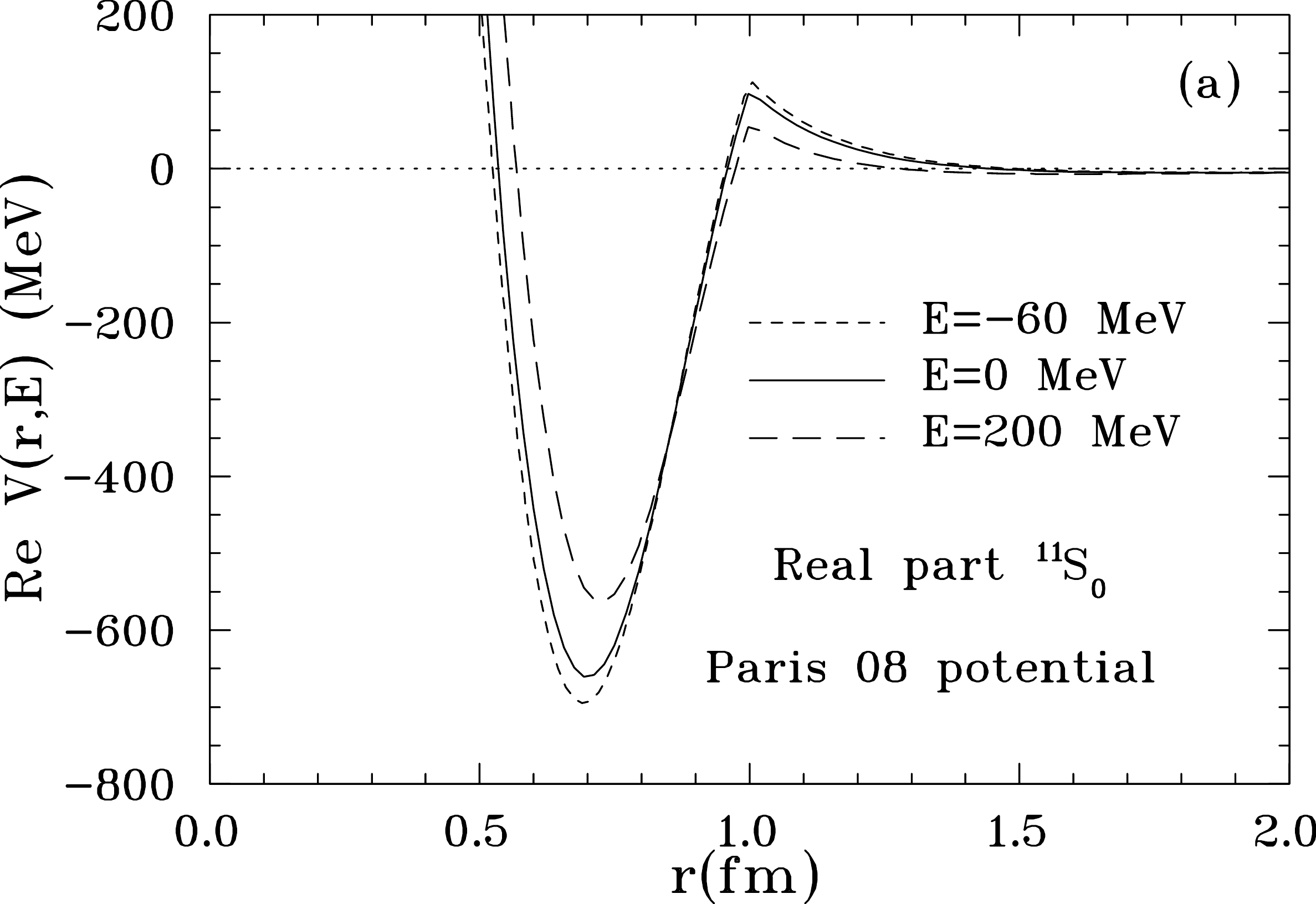}
\hspace{1cm}
\includegraphics [scale=0.35]{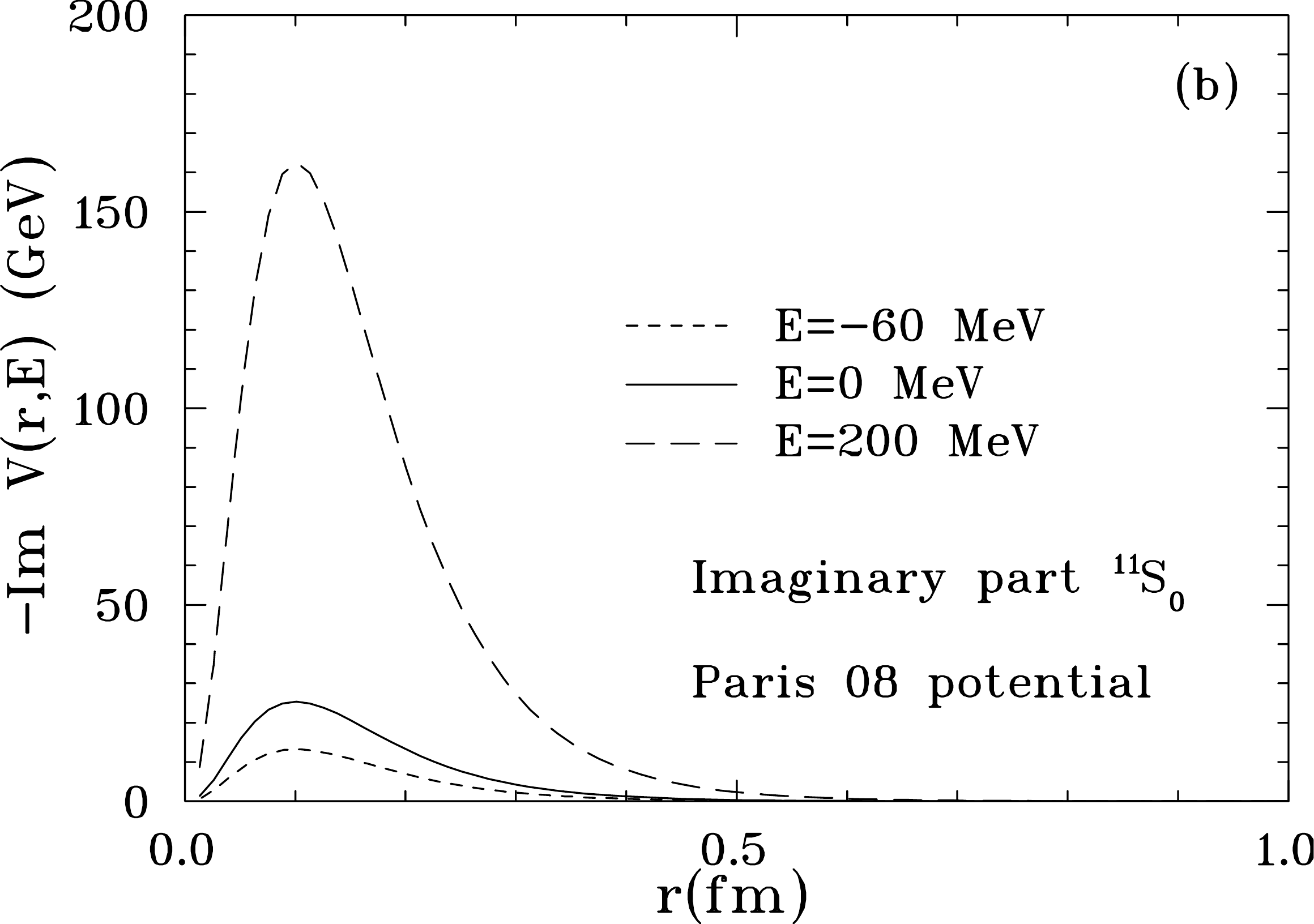}
\caption{The left panel (a) displays the real part  while the right one (b) displays the absorptive part of the Paris  $N\overline{N}$ potential in the isospin $0$ isosinglet $S$ wave. The deep well and the barrier are due to the interplay of theoretical one- and two-pion exchange forces supplemented with a short range phenomenological attraction. The well and barrier structure have the support of 4000 data but the detailed shape of the kink is an artifact of the phenomenological part and it cannot be determined very precisely~\cite{lac09}. The existence of the barrier is nevertheless indicated by the scattering data, in particular those of the $\bar n p$ total cross sections.} \label{Parpot}
\end{figure}

Calculating the related loop integral and averaging the probability over the phase space (see Appendices \ref{amplitudes} and \ref{numloop}), one obtains  the  $  M_{p\overline{p}}$ spectrum  plotted  in Fig.~\ref{MPPdirect} for different values of $\beta$. This spectrum has several interesting features summarized in Table~\ref{table3} and described below.
\begin{itemize}
\item   It displays two peaks.  The narrow peak that arises at the threshold is related to the near $p\overline{p}$ threshold ($E_{N\overline{N}}= -4.8$ MeV, $52$ MeV broad Paris-potential quasi-bound $^{11}S_0$ state~\cite{lac09}).
The other, broad,  peak is formed at  $ M_{p\overline{p}} \simeq 2130 $ MeV.  
It corresponds to a shape  resonance at which the wave length  equals to the size of the $p\overline{p}$  potential well in the Paris potential for the $^{11}S_0$ state \cite{lac09}. The isospin $0$ part of the potential well that generates such structures and the corresponding energy dependent absorptive part are shown in figure \ref{Parpot}.
\item  Figure \ref{MPPdirect} shows the expansion from the initial radius $ R_0 = 0.28 ~fm $ to some radius $ R_f$, i.e., when the probability of the photon emission falls to zero   and when  the  $ p\overline{p}$ pair is well formed. One sees from the curve on figure~\ref{MPPdirect} that the limiting radius $ R_f$ varies from 0.28 fm to  $ R_f\approx 0.61 $ fm with the invariant  $ M_{p\overline{p}}$  mass  varying from 2.90 to 2.60 Gev/c$^2$  when $\beta$ varies from 0.0 to 0.25 fm$^{3/2}$. 
\item The first minimum moves very slowly to slightly increasing invariant $ M(p\overline{p})$  mass but remains below the experimental value at about 1.97  GeV/$c^2$. 
\item The broad maximum in the spectrum moves to decreasing values of $M_{p\overline{p}}$ as $\beta$ increases, i.e. when $\beta$ goes from 0 to 0.25  fm$^{3/2}$, the maximum  moves from 2.15 to 2.01 GeV/$c^2$ when the experiment displays a maximum around 2.13 GeV/$c^2$. Furthermore the ratio of the height of the second maximum over the height of the first minimum  decreases and goes to 1 as $\beta$ reaches the value of 0.25 fm$^{3/2}$; for larger values of $\beta$ there is neither a minimum nor a maximum. 
 \end{itemize}

 \begin{table}[ht] \label{table3}
\caption{ Position and values of 1st minimum and second maximum of the direct emission (DE)  spectra, all normalized at first maximum value of 2450 events per $0.02$ GeV/$c^2$} 
\vspace{.3cm}
\begin{tabular}{lcccccccccc}  
\hline
\hline
\ \ DE  \ / $\beta$ in fm$^{3/2}$ & & 0&0.05&0.075&0.10&0.15&0.175 &0.20&0.25& exp\\ \hline
1st min &   pos  & 1.945         &  1.945              &  1.955& 1.955& 1.955 &1.9645&1.9645 & 1.974      &1.97       \\
& height & 628 &782               & 834            &  869 & 897.7& 896.4 & 883&  836   & 1040        \\ \hline
 2nd max & pos & 2.15  &2.12 & 2.11 & 2.10 & 2.067&2.058 & 2.04 & 2.01  & 2.13    \\
 & height &1547 & 1457 &1369 &1274& 1097& 1021 &953 & 849 & 1350 \\  \hline 
 ratio max/min & & 2.46& 1.86& 1.64 & 1.47& 1.22& 1.14 &1.08 & 1.02 & 1.3\\  \hline
  becomes negligible at &&2.90&2.85& 2.80 &2.75& 2.70& 2.67 &2.65 &2.60 & \\ 
\hline
\hline
\end{tabular}
\end{table}
\vspace{0.5cm}
\noi We will see further on that the contribution of the baryon current in this process shows a maximum at values of $M_{p\overline{p}}$ slightly above 2.15 GeV.\\

This internal emission model can be extended to the case of  a vector meson emission: the main change comes in the definition of the energy $E_{N\overline{N}}$, i.e.,  $E_{N\overline{N}}= M_{J/\psi} - \sqrt{m_{\mathcal{B}}^2 + k^2}$ where $m_{\mathcal{B}}$ is the mass of the emitted boson.\\

\section{Baryon current amplitudes } \label{baryonc}

This calculation is based on a model suggested  in Ref.~\cite{loi05} (similar ideas have been  developed quantitatively by Barnes {\it et al.}  in  Ref.~\cite{bar10}). The initial assumption is that the mesons are emitted after the $ N {\overline N}$ pair has been formed. In the decay process, the  initial heavy  $c\overline{c}$ quarks in the $J/\psi$ state of $ J^{PC} = 1^{--} $ have to disappear and form  another $q \overline{q} $ pair. The easiest way to do that is a three gluon intermediate state \cite{bol98}  which generates  a pair of the  same $J^{PC}$. Next this system generates  two extra $q \bar{q} $ pairs from the vacuum, e.g., by the $ ^3P_0$ mechanism. This leads to the  formation of a  $ ^3S_1$ state. The emission of $\gamma, \pi, \phi$ or $\omega$  is assumed to happen after the baryons have been formed. It turns out that this assumption yields a generally consistent description of the mesonic decays.
  Yet, in the case of the  $\gamma$ or $\omega$ bosons it represents only a sizable fraction of the decay rate and has to be completed by the contribution of the direct (internal) process just described in the preceding section for the photon case. The mechanism is visualized in Fig.~\ref{figFSIgraph} and to quantify it  one needs three basic ingredients:

\begin{figure}[ht]
\includegraphics[scale=0.65]{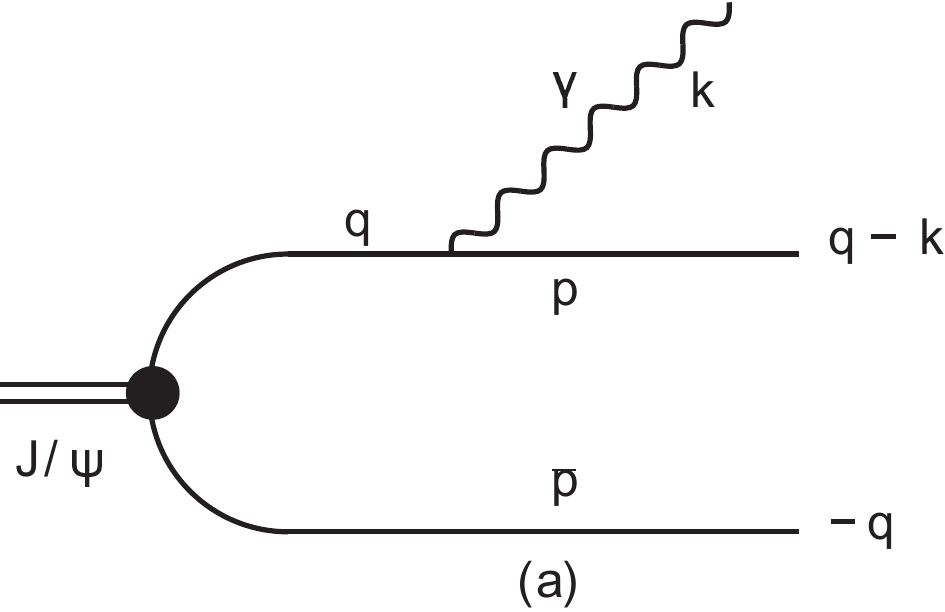}
\hspace{1cm}
\includegraphics[scale=0.65]{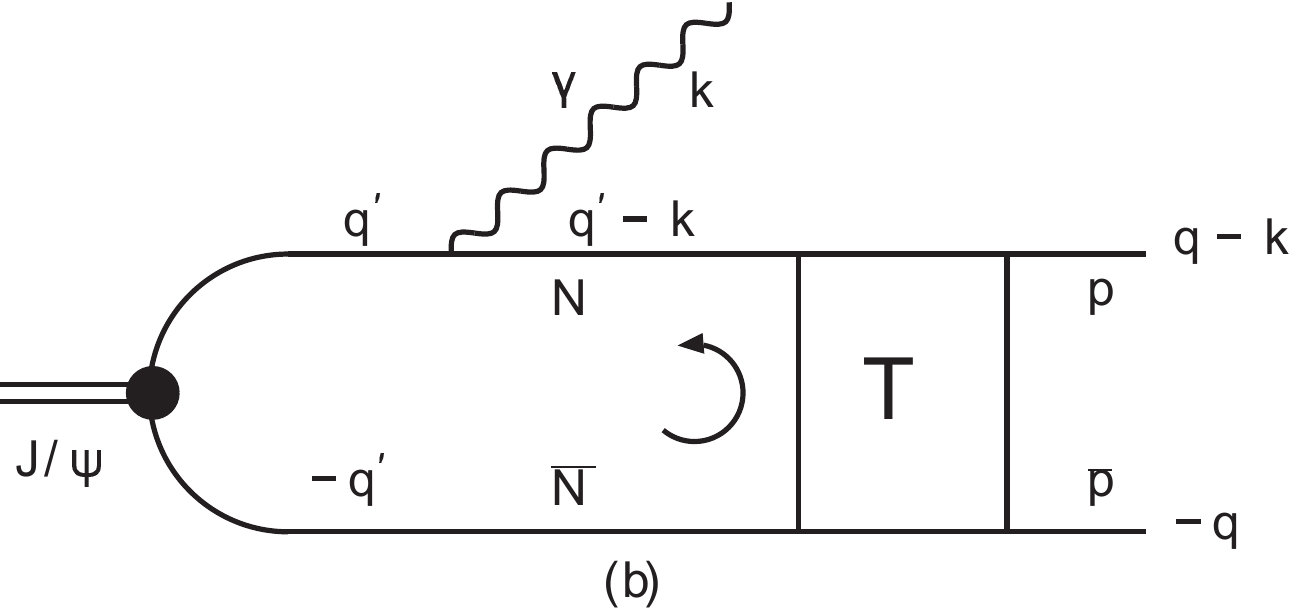}
\caption{Photon emission from intermediate baryons: the left graph (a) is associated to the Born term while the right one (b) includes final state corrections. The various labels have been defined in Fig.\ref{figdirect}. Similar diagrams for the emissions from anti-baryons are not drawn.  } \label{figFSIgraph} 
\end{figure}

\begin{itemize}
\item A wave function to describe the initial  $ N{\overline N}$ state. It is generated by the  
$c\overline{c}$ transition to the $^3S_1 $ $ N {\overline N}$  state of relative momentum $\mathbf{q}_r$,

\item A mechanism that  describes the emission of a  boson $\mathcal{B}$ from the initial $^3S_1$  to a final or intermediate   $ N {\overline N}$ system,

\item A  method to describe  $ N {\overline N}$ final-state interactions.
\end{itemize}

The boson $\mathcal{B}$ is emitted with momentum $\mathbf{k}$ from either the nucleon or the antinucleon of final momenta  $\mathbf{q}_1$ and $\mathbf{q}_2$, respectively [see Eq.~(\ref{dir2})]. The decay amplitude $A^{BC}_{p\overline{p}\mathcal{B}}(\mathbf{q}_1, \mathbf{q}_2,\mathbf{k})$ can then be expressed as 
\bqa \label{am0} 
A^{BC}_{p\overline{p}\mathcal{B}}(\mathbf{q}_1, \mathbf{q}_2,\mathbf{k})&=&  \langle p(\mathbf{q}_1)\  \overline{p}(\mathbf{q}_2)\ \mathcal{B}(\mathbf{k}) \vert \widehat{U}_{\mathcal{B}}\ G_{0,N\overline{N}}^+ \widehat{A}_{N\overline{N}}\vert \psi_i \rangle \nonumber \\
\label{am0bis} 
&=&   \langle p(\mathbf{q}_1)\  \overline{p}(\mathbf{q}_2)\ \mathcal{B}(\mathbf{k}) \vert \ \widehat{U}_{\mathcal{B}}\ \vert (N \overline{N})_{I=0}\rangle \ G_{0,N
\overline{N}}^+ \ \langle (N \overline{N})_{I=0} \vert \widehat{A}_{N \overline{N}} \vert \psi_i  \rangle, 
\eqa
where we assume the initial state to be an I = 0 state. We will refer to this as the baryon current (BC) model. The boson emission operator $\widehat{U}_{\mathcal{B}}$  includes final state interactions, $G_{0,N \overline{N}}^+$  is the  $ N \overline{N}$  Green's function before the emission of the boson and $\psi_i$  the wave function of the $J/\psi$ meson at rest. 

The amplitude in Eq.~(\ref{am0bis}) is built up from three factors: the last one, $\langle (N \overline{N})_{I=0} \vert \widehat{A}_{N \overline{N}} \vert \psi_i  \rangle$, corresponds to the creation of the $N\overline{N}$ pair in an isospin $0$ state, the middle one, $G_{0,N \overline{N}}^+ $ describes its propagation   while the first one, $ < p\overline{p}\mathcal{B} \vert \ \widehat{U}_{\mathcal{B}}\ \vert (N \overline{N})_{I=0}> $, describes the emission of the boson  of momentum $\mathbf{k}$, final-state interactions included and can be  written formally as
\bqa \label{amppbarV}
\langle p\overline{p}\mathcal{B} \vert \widehat{U}_{\mathcal{B}} \vert (N \overline{N})_{I=0}\rangle &=& 
\langle p\overline{p}\mathcal{B} \vert \widehat{U}_{\mathcal{B}}^0 \ \left [1+ G_{0,N \overline{N} \mathcal{B}}^+ \  T_{[N \overline{N}]}\right ]  \vert (N \overline{N})_{I=0} \rangle \nonumber \\
&=& \langle p\overline{p}\mathcal{B} \vert \left [ T_{[N \overline{N}]}\ G_{0,N \overline{N} \mathcal{B}}^+ +1\right ] \ \widehat{U}_\mathcal{B}^0\ \vert (N \overline{N})_{I=0} \rangle,  
\eqa
where $G_{0,N \overline{N} \mathcal{B}}^+$ denotes the free three-body Green's function (similar to Eq.~(\ref{G0}) in the photon case).\\ 

The intermediate $N\overline{N}$ pair being in an isospin $I=0$ state, the lowest order (Born) amplitude in the absence  of final state interaction (left panel of Fig. \ref{figFSIgraph}) is given, since in that case one has only an intermediate $p\overline{p}$ pair, formally by

\bqa  \label{am0ter}
 \langle p(\mathbf{q}_1)\  \overline{p}(\mathbf{q}_2)\ \mathcal{B}(\mathbf{k}) \vert \widehat{U}^0_{\mathcal{B}} \vert (p \overline{p})_{I=0}\rangle G_{0,p\overline{p}}^+ \ \langle (p \overline{p})_{I=0} \vert \widehat{A}_{ p\overline{p}} \vert \psi_i  \rangle 
&=& \int \frac{d\mathbf{q}'_1}{(2\pi)^3}\langle p(\mathbf{q}_1)\  \overline{p}(\mathbf{q}_2)\ \mathcal{B}(\mathbf{k}) \vert \widehat{U}^0_{\mathcal{B}}\ \vert [p(\mathbf{q}'_1)\overline{p}(-\mathbf{q}'_1)]_{I=0}\rangle\ \nonumber \\
 \hspace{8cm}&&  \times \ \frac{(2\pi)^3}{\sqrt{\mathcal{V}_0}} \ \delta^{(3)}( -\mathbf{q}'_1+\mathbf{q}_2)
 \ \frac{ \mathcal{F}_{J/\psi}({q}'_1) }{M_{J/\psi} + i\epsilon -2 \sqrt{q\mathbf{q}_1^{'2}+m^2}} \ \nonumber \\
&=&  (2\pi)^3 \ \delta^{(3)}( \mathbf{q}_1+\mathbf{q}_2+\mathbf{k}) \ \widetilde{A}^{B,BC}_{p\overline{p}\mathcal{B}}({\bf{q}_2,\bf{k}}), 
 \eqa
where 
\be  \label{Gtilde}
\widetilde{A}^{B,BC}_{p\overline{p} \mathcal{B} }({\bf{q}_2, \bf{k}}) = \widetilde{G}_{p\overline{p}}(q_2)\ U^0_{p\overline{p}\mathcal{B}}(\mathbf{q}_2,\mathbf{k})
\hspace{1cm}{\rm with}\hspace{1cm} 
\widetilde{G}_{p\overline{p}}(q_2)= \frac{1}{\sqrt{\mathcal{V}_0}} \ 
\frac{\mathcal{F}_{J/\psi}(q_2)}{M_{J/\psi}+ i\epsilon -2 \sqrt{\mathbf{q}_2^2+m^2}}. 
 \ee
 In these equations, the photon (meson) is emitted from the nucleon of momentum $\bf{q}_1$  and the antinucleon is the spectator with momentum $\bf{q}_2$ such that $\bf{q}_2 = - \bf{q}$,  $\bf{q}_1= -\bf{q}_2-\bf{k}= \bf{q} - \bf{k}$ and ${\bf{q}}_r = {\bf{q}-\bf{k}}/2$. \\

The FSI contribution contains the formal expression
\be
 \langle p  \overline{p} \mathcal{B} \vert T_{N \overline{N}}\ G_{0,N \overline{N} \mathcal{B}}^+ \ \widehat{U}^0_{\mathcal{B}}\ \vert (N \overline{N})_{I=0}\rangle =  \sum_I \ 
 \langle  p  \overline{p} \mathcal{B} \vert (N\overline{N})_I \rangle \ 
 \langle (N\overline{N})_I \vert T_{[N \overline{N}]_I}\ G_{0,N \overline{N} \mathcal{B}}^+ \vert (N\overline{N})_I \rangle \ 
 \rangle (N\overline{N})_I \vert \widehat{U}^0_{\mathcal{B}}\ \vert (N \overline{N})_{I=0}\rangle 
\ee  
where the sum over the isospin $I$ is restricted to $I=0,1$. We may then write explicitly 

\bqa \label{ampfsi}
A^{FSI,BC} _{p\overline{p}\mathcal{B}}(\mathbf{q}_1, \mathbf{q}_2,\mathbf{k})&=& 
   \langle p(\mathbf{q}_1)\  \overline{p}(\mathbf{q}_2)\ \mathcal{B}(\mathbf{k}) \vert T_{N \overline{N}}\ G_{0,N \overline{N} \mathcal{B}}^+ \ \widehat{U}^0_{\mathcal{B}}\ \vert (N \overline{N})_{I=0}\rangle \ G_{0,N
\overline{N}}^+ \ \langle (N \overline{N})_{I=0} \vert \widehat{A}_{N \overline{N}} \vert \psi_i \rangle \nonumber \\
&=& \sum_{I=0,1} \ \int \langle p(\mathbf{q}_1)\  \overline{p}(\mathbf{q}_2)\ \mathcal{B}(\mathbf{k}) \vert T_{[N \overline{N}]_I}\ G_{0,N \overline{N} \mathcal{B}}^+ \vert \
 [N(\mathbf{q'}_1)\  \overline{N}(\mathbf{q'}_2)]_I \ \mathcal{B}(\mathbf{k'}) \rangle \
 \frac{d\mathbf{q'}_1}{(2\pi)^3}\ \frac{d\mathbf{q'}_2}{(2\pi)^3} \ \frac{d\mathbf{k'}}{(2\pi)^3} \nonumber \\
 & & \hspace{1cm} \times\ 
 \langle [N(\mathbf{q'}_1)\  \overline{N}(\mathbf{q'}_2)]_I \ \mathcal{B}(\mathbf{k'}) \vert \
  \widehat{U}^0_{\mathcal{B}}\ \vert (N \overline{N})_{I=0}\rangle \ G_{0,N \overline{N}}^+ \ \langle (N \overline{N})_{I=0} \vert \widehat{A}_{N \overline{N}} \vert \psi_i \rangle \nonumber \\
 &=& \sum_{I=0,1} \int \langle p(\mathbf{q}_1)\  \overline{p}(\mathbf{q}_2)\ \mathcal{B}(\mathbf{k}) \vert T_{[N \overline{N}]_I}\ \vert  [N(\mathbf{q'}_1)\  \overline{N}(\mathbf{q'}_2)]_I\rangle \
 G_{0,N \overline{N}\mathcal{B}}^+ ( \mathbf{q'}_1,\mathbf{q'}_2,\mathbf{k}) \nonumber \\
 & &\hspace{1cm} \times \  \frac{d\mathbf{q'}_1}{(2\pi)^3}\ \frac{d\mathbf{q'}_2}{(2\pi)^3} \
 (2\pi)^3 \ \delta^{(3)}( \mathbf{q'}_1+\mathbf{q'}_2+\mathbf{k}) \ A^{B,BC}_{N\overline{N}\mathcal{B}}({\bf{q'}_2},{\bf{k}}).
 \eqa
Then, from Eqs.~(\ref{TNNbar}), (\ref{am0ter}) and (\ref{Gtilde}), we obtain
\bqa  \label{ampfsi2}
A^{FSI,BC} _{p\overline{p}\mathcal{B}}(\mathbf{q}_1, \mathbf{q}_2,\mathbf{k}) &=& 
  (2\pi)^3 \ \delta^{(3)}(\mathbf{q}_1+\mathbf{q}_2+\mathbf{k})\ \widetilde{A}_{p\overline{p}\mathcal{B}}^{FSI,BC}({\mathbf{q}_r,\mathbf{k}}),  \nonumber \\
\widetilde{ A}_{p\overline{p}\mathcal{B}}^{FSI,BC}({\mathbf{q}_r,\mathbf{k}}) &=&  \sum_I \int  \frac{d\mathbf{q'}_2}{(2\pi)^3} \ T_I(\mathbf{q}_r,-\mathbf{q'}_2- \frac{\mathbf{k}}{2},E_{N\overline{N}}) \
 G_{0,N \overline{N} \mathcal{B}}^+ (-\mathbf{q'}_2-\mathbf{k}, \mathbf{q'}_2,\mathbf{k}) \ A^{B,BC}_{N\overline{N}\mathcal{B}}(\mathbf{q'}_2,\mathbf{k}),   
 \eqa
where

\be \label{GNNbarV}
G^+_{0,N\overline{N}\mathcal{B}}(-\mathbf{q'}_2-\mathbf{k}, \mathbf{q'}_2,\mathbf{k}) =
\frac{1}{M_{J/\psi} + i \epsilon - E_\mathcal{B}(k) -  \sqrt{(\mathbf{q}^{'2}_2 + \mathbf{k})^2+m^2}- \sqrt{\mathbf{q}^{'2}_2 +m^2}},     
\ee 
with  $E_\mathcal{B}(k)= \sqrt{k^2+m_\mathcal{B}^2}$. The FSI amplitude (\ref{ampfsi2}) requires integrations over the corresponding loop momenta 
 and its detailed form will be discussed in the appendices. 
Finally, the amplitude in the BC+FSI model is given by  
\be \label{Abctot}
A^{BC}_{p\overline{p}\mathcal{B}}(\mathbf{q}_1, \mathbf{q}_2,\mathbf{k})= 
 (2\pi)^3 \ \delta^{(3)}( \mathbf{q}_1+\mathbf{q}_2+\mathbf{k}) \left [\widetilde{A}^{B,BC}_{p\overline{p}\mathcal{B}}({\mathbf{q}_r,\mathbf{k}}) + \widetilde{A}^{FSI,BC}_{p\overline{p}\mathcal{B}}({\mathbf{q}_r,\mathbf{k}})  \right ], 
\ee
with $\widetilde{A}^{B,BC}_{p\overline{p}\mathcal{B}}({\mathbf{q}_r,\mathbf{k}}) $ given by Eq.~(\ref{Gtilde}) and $\widetilde{A}^{FSI,BC}_{p\overline{p}\mathcal{B}}({\mathbf{q}_r,\mathbf{k}}) $ by Eq.~(\ref{ampfsi2}). One has to add a similar contribution for the emission from the antinucleon. This specific final state interaction correction will be evaluated with the half off-shell scattering matrix~\cite{ded09} arising from the Paris potential~\cite{lac09}. It will  be applied in what follows to the case of the photon and of the $\omega$ meson. The explicit expression of the amplitude for the photon emission is calculated in Appendix~\ref{photoncase}. The lowest order amplitude is  enhanced by  the $N \overline{N}$ final state interactions and the effect is significant  in the spin singlet  $S$ wave.  As indicated by the summation over the isospin $ I$ states, it involves also radiation of magnetic photons from the intermediate $ N\overline{N}$ pairs   and generates a delicate interference pattern. Since the intermediate states involve $I=0$ this amplitude is expected to determine, or contribute significantly, to the shape of the threshold peak in the invariant $p\overline{p}$ mass distribution.\\

  \subsection {The initial $ N \overline{N}$  state   } \label{NNbar}

\noi In the two models describing  the $p \overline{p}$ threshold peak~\cite{loi05,jul06}, 
 it was assumed that in the course of radiative process the  $p \overline{p}$ final state is  formed in the spin singlet  $^1S_0$  state~\cite{ded09}. The near threshold enhancement arises as a result of the $I=0$, $p \overline{p}$ final state interaction. However, in order to understand  the full  energy spectrum and mesonic emission rates one needs a better description of the formation mechanism. We go one step further, in addition to the state indicated above,
the initial  $ N \overline{N}$ state is assumed to inherit the spin and isospin quantum numbers of
the $J/ \psi$, $S=1,I=0$, hence it is a $^3S_1$ state. \\

\noi Before the emission  of the photon, the process is given by the matrix element: 
\be \label{i3}  
\langle [N(\mathbf{q}_1)  \overline{N}(\mathbf{q}_2)]_{I=0} \vert G_{0,N \overline{N}}^+ \ \widehat{A}_{N \overline{N}} \vert \psi_i \rangle  = {(2\pi)^3}\ \delta^{(3)}(\mathbf{q}_1+\mathbf{q}_2)\ \widetilde{G}_{N\overline{N}}(q_2), 
\ee
where $\widetilde{G}_{N\overline{N}}(q_2)$ is given by Eq.~(\ref{Gtilde}) with, in the $J/\psi$ rest frame, $\mathbf{q}_1 = - \mathbf{q}_2 = \mathbf{q}_r= \mathbf{q}$. 
This free Green's function becomes singular when the momentum approaches its on-shell value. This singularity leads, in the case of electric photon emissions, to the  well known infrared catastrophe. The effect of infrared enhancement should  be  seen in the experimental data as a peak at the end of the spectrum, that is, in the soft photon limit. 
However, it is also clear~\cite{bai03,abl12} that contributions from the infrared photons ($k<50$ MeV/c) have been effectively cut out from the data. We refer the reader to the discussion at the end of Section~\ref{radiative}.\\

\subsection {The  emission vertices } \label{emisvert}

The electromagnetic current associated to the photon emission from the nucleon is given by
\be \label{Jem}
<q'|J_{\nu}|q> = e\ \overline{u}(q') [  \gamma_{\nu}+ \frac{\kappa}{2i m} \sigma_{\mu \nu} (q' -q)^{\mu}] u(q) =
 e\ \overline{u}(q') [  \gamma_{\nu}+ i \  \frac{\kappa}{2 m} \sigma_{\mu \nu} k^{\mu}] u(q),  
 \ee
where 
$$ \sigma_{\mu\nu} = \frac{i}{2} \ [\gamma_\mu,\gamma_{\nu}],$$ 
$e$ is the unit of charge and $\kappa$ is the anomalous magnetic moment of the nucleon ($\kappa_p=1.793$ for the proton and $\kappa_n=-1.913$ for the neutron). The final nucleon four-momentum $q'$ is related to the initial four-momentum $q$ by $q'= q - k$ where $k$ is the emitted boson four-momentum. The corresponding current for the photon emission from the antinucleon will be given by the substitution $e\rightarrow -e$ and $\mathbf{q} \rightarrow -\mathbf{q}$.

More generally, the emission of a vector particle  by a nucleon is described  by the operator 
\be \label{e1}
 L = g_V \gamma_{\nu} \epsilon ^{\nu*}(\lambda) +i\  \frac{g_T}{2m} \sigma_{\mu \nu} k^{\mu} \epsilon ^{\nu*}(\lambda), 
 \ee
where the four-vector ${\epsilon}^*(\lambda)$ denotes the polarization vector of the emitted particle,
 $\lambda$ its helicity while $g_V $ and $g_T= \kappa \ g_V$ are the vector
 and tensor coupling constants, respectively.\\

The  final photon or vector meson may be produced in a magnetic or an electric transition. The relevant formation amplitudes are obtained from the transition matrix elements of the operator  (\ref{e1}) reducing bispinors $u$ to spinors $\chi_S$.  We have
\be \label{e2} 
 \overline{u}(q') L u(q)  = \chi_{S'}^\dagger \widehat{U}_{\mathcal{B}}^0 \chi_S,   \ee
 where $S$ and $S'$ denote the initial and final nucleon spin, and one obtains  the vertex coupling in the two dimensional spin space
\be \label{defv}
\widehat{U}_{\mathcal{B}}^0 = g_V \ \widehat{A}_V + i\  \frac{g_T}{2 m} \ \widehat{A}_T.
\ee
\noi The explicit expressions for the vector $\widehat{A}_V$ and  the tensor $\widehat{A}_T$ parts are derived  in Appendix \ref{explicit} (see Eqs.~(\ref{AV2}) and~(\ref{AT}), respectively). \\

\subsubsection{The photon case}
  
The full  photon potential operator, ${U}_{\gamma}^0({\bf{k},\bf{q}})$, combining the photon emission from either baryon is given by Eqs.~(\ref{VNNbarg}) and (\ref{fgh}). The magnetic terms in ${U}_{\gamma}^0({\bf{k},\bf{q}})$, proportional to  $ {\mbox {\boldmath $\sigma$}}\cdot[\textbf{k}\wedge {\mbox {\boldmath $\epsilon$}}^*(\lambda)]$, change the $^3S_1 $ state of the initial $N\overline{N}$ system to the final $^1S_0 $ state. Electric terms in ${U}_{\gamma}^0({\bf{k},\bf{q}})$ proportional to  $ \textbf{q}\cdot {\mbox{\boldmath $\epsilon$}}^*(\lambda)$, change the $^3S_1 $ state  to  final $^3P $ states. Relativistic corrections generate additional  terms, most of these cancel in the $p \overline{p} $ system, some spin operators  bilinear in $\mathbf{q}$ lead to D waves but contribute corrections only on $1\%$ level and are not included in the calculations. As discussed in Appendix~\ref{explicit}  the two basic couplings add coherently for the proton and the antiproton. Thus the summary   coupling of photons to the $p \overline{p} $ system becomes 
\be
\label{fotonrel}
U^0_{\gamma} ({\bf{q},\bf{k}})={V}_{E,\gamma} +  {V}_{M,\gamma} 
\ee
with 
\bqa
V_{E, \gamma} &=&  \frac {e}{2m}\ [C_E \  {\textbf{q}\cdot{\mbox {\boldmath $\epsilon$}}^*(\lambda)}],  \label{VEgam} \\
V_{M, \gamma} &=&  i \frac {e}{2m}\ [\mathbf{C}_M\cdot(\textbf{k}\wedge {\mbox {\boldmath $\epsilon$}}^*(\lambda))],  \label{VMgam}
\eqa

which are still operators in the spin-isospin space and where [see Eqs.~(\ref{VMphoton}) and (\ref{VEphoton})]
\bqa
\label{fotonrel2}
 C_E & =&-\left \{ \zeta +\overline{\zeta}+   \frac{1}{\zeta} + \frac{1}{\overline{\zeta}} \right \} + 
\kappa\  \frac{k }{2m} \ \left \{ \zeta +\overline{\zeta} - \frac{1}{\zeta} - \frac{1}{\overline{\zeta}} \right \} +
\kappa\ \frac{k^2}{2m\Omega} \ \left \{   \frac{1}{\zeta} + \frac{1}{\overline{\zeta}}  \right \}, 
\nonumber \\
\mathbf{C}_M &=& r_- \ ({\mbox{\boldmath$\sigma$}}_1 - \ {\mbox{\boldmath$\sigma$}}_2) 
+ r_+ \ ({\mbox{\boldmath$\sigma$}}_1 + \ {\mbox{\boldmath$\sigma$}}_2),  
\eqa
with
\bqa \label{fotonrel3}
 r_- &=& \frac{1}{2}\ \left \{ \left (\frac{1}{\zeta}+ \frac{1}{\overline{\zeta}} \right )\ \left (1 + \kappa\ \frac{k}{2m} \right )\  +\kappa\ \frac{\Omega}{2m} ( \zeta + \overline{\zeta})\right \},  \nonumber \\
r_+ &=& \frac{1}{2}\ \left \{ \left (\frac{1}{\zeta}- \frac{1}{\overline{\zeta}} \right )\ \left (1 + \kappa\ \frac{k}{2m} \right )\  +\kappa\ \frac{\Omega}{2m} ( \zeta - \overline{\zeta})\right \}.  
\eqa 
The energies $\Omega$, $\Omega'$ and the coefficients  $\zeta$, $\overline{\zeta}$ are defined in Appendix \ref{explicit} [Eqs.~(\ref{om}), (\ref{zeta}) and (\ref{zetabar})]. The approximation leading to Eqs.~(\ref{VEgam}-\ref{fotonrel3}) may be acceptable close to the central region of the $M_{p\overline{p}}$  distribution (see Fig.~\ref{figMPPtot}). It is too crude in the threshold region where $k/m\approx1$ and at the other extremity where $q/m\approx1$. Nevertheless, some coefficients in 
${U}_{\gamma}^0({\bf{k},\bf{q}})$ [Eqs.~(\ref{VNNbarg}) and (\ref{fgh})]
display remarkable stability. This, in particular, concerns the terms $\zeta+\overline{\zeta}$, $\zeta + 1/\zeta$, $\overline{\zeta} + 1/\zeta$,  $ \overline{\zeta} + 1/\overline{\zeta}$ which are approximately $2$ within $1\%$ over all the phase space. On the other hand, there are a number of terms involving more complicated combinations of the spin and  momenta which are less stable, but small due to other reasons. The terms 
$$ \frac{ie}{2m}\    \frac{\kappa}{2m\Omega} \left [ \left ( \frac{{\mbox{\boldmath$\sigma$}}_1}{\zeta} -  \frac{{\mbox{\boldmath$\sigma$}}_2}{\bar{\zeta}}\right ) \cdot (\bf{k}\wedge\bf{q})\right ]\ \left [\bf{q}\cdot{\mbox{\boldmath$\epsilon$}}^*(\lambda)\right ],$$
$$  \frac{ie}{2m}\  \frac{\kappa}{2m\Omega} \ {\bf{q}\cdot\bf{k}}\ \left  \{\frac{{\mbox{\boldmath$\sigma$}}_1}{\zeta} - \frac{{\mbox{\boldmath$\sigma$}}_2} {\bar{\zeta}} \right \} \cdot [
{\bf{q}\wedge{\mbox{\boldmath$\epsilon$}}^*(\lambda)}],$$
and $$ \frac{ie}{2m}\ \frac{\kappa}{2m\Omega} \
{\bf{q}\cdot[{\bf{k}\wedge{\mbox{\boldmath$\epsilon$}}^*(\lambda)}]} \ 
 \left ( \frac{{\mbox{\boldmath$\sigma$}}_1}{\zeta} -  \frac{{\mbox{\boldmath$\sigma$}}_2}{\bar{\zeta}}\right )
 \cdot{\bf{q}},$$
involve a spin flip transition. According to $CP$ conservation (see Table~\ref{table0}) these terms lead predominantly to final spin singlet $S$-wave state. The resulting contribution would give an average $\langle q_iq_j\rangle= q^2/3$ and would mainly contribute at large $k$, i.e.,  in the threshold region where corrections will be of the order of $q^2/12 m^2$, i.e., about $ 2\%$. 
The term proportional to ${\bf{q}\wedge{\mbox{\boldmath$\epsilon$}}^*(\lambda)}$ in Eq.~(\ref{VNNbarg}) reads
\be
\left [g({\bf{q},\bf{k}}) \ {\mbox{\boldmath$\sigma$}}_1 +  \bar{g}({\bf{q},\bf{k}}) \ {\mbox{\boldmath$\sigma$}}_2 \right ] 
=\left [ \tilde{r}_-  \ ({\mbox{\boldmath$\sigma$}}_1 -   {\mbox{\boldmath$\sigma$}}_2) + \tilde{r}_+ \ ({\mbox{\boldmath$\sigma$}}_1 +   \ {\mbox{\boldmath$\sigma$}}_2) \right ] 
= 2\ ( \tilde{r}_-  \ {\mbox{\boldmath$\sigma$}}_- + \tilde{r}_+ \ {\mbox{\boldmath$\sigma$}}_+ ),
 \ee
 with
 \bqa \label{rtilde}
  \tilde{r}_-&=& \frac{1}{2} \left \{ \left (\zeta - \overline{\zeta}\right ) \left ( 1- \kappa \ \frac{k_0}{2m} \right )
-  \left (\frac{1}{\zeta} -\frac{1}{\overline{\zeta}}\right )  \left ( 1+ \kappa \ \frac{k_0}{2m} \right ) \left (1+ \frac{k_0}{\Omega}\right )
\right \}, \nonumber\\
 \tilde{r}_+  &=&  \frac{1}{2}  \left \{ \left (\zeta + \overline{\zeta}\right ) \left ( 1- \kappa \ \frac{k_0}{2m} \right )
-  \left (\frac{1}{\zeta} +\frac{1}{\overline{\zeta}}\right )  \left ( 1+ \kappa \ \frac{k_0}{2m} \right ) \left (1+ \frac{k_0}{\Omega}\right )
\right \}. 
\eqa
It  involves a dominant ${\mbox{\boldmath$\sigma$}}_+\cdot [{\bf{q}\wedge{\mbox{\boldmath$\epsilon$}}^*(\lambda)}]$ combination which generates spin triplet $P$-wave states. It could be contributing as much as $20\%$ of the dominant electric term. However, it is only important close to the threshold region where 
 \be
  \tilde{r}_+\approx -\frac{k_0}{\Omega} \left [
1 + \kappa \frac{\Omega}{m} \ \left( 1 + \frac{k_0}{2 \Omega} \right )  \right ], 
\ee
 but where the $P$-wave contributions are strongly suppressed by the phase space.  
  
In practical calculations it is sufficient to neglect small corrections of the  order of $k^2/4m^2$ which contribute about $ 3\%$ to the electric  rate, and about one percent to the total rate. On the other hand, a sizable, i.e., of the order of $10\%$,  relativistic correction is due to the $\kappa/2m$ term affecting  the anomalous magnetic  moment in equation (\ref{fotonrel3}).
Note that in the limit $k/ 2m << 1$, to order, $k^2/4m^2$,  Eq.~(\ref{fotonrel}) reduces to the simple expression  
\be \label{Vgamppbarapp}
{U}_{\gamma}^0 ({\bf{q},\bf{k}}) \approx \frac {e}{2m}\ \left [-4 \  {\textbf{q}\cdot{\mbox {\boldmath $\epsilon$}}^*(\lambda)} +i \left (1 + \frac{\kappa \Omega}{2m} \right ) \ ({\mbox{\boldmath$\sigma$}}_1-{\mbox{\boldmath$\sigma$}}_2)
 \cdot(\textbf{k}\wedge {\mbox {\boldmath $\epsilon$}}^*(\lambda)) \right ]. \ee

\subsubsection{The vector meson case} 

The emitted $ \omega$ meson has a negative $G$ parity and couplings to the proton or the antiproton differ in sign (this also applies to the case of the $\pi$ meson emission). Again the emission by a baryon or an antibaryon is predominantly  coherent as this sign is compensated by the momentum and/or  spin involved in the vertices. In the case of  $\omega$ meson the tensor coupling is known to be consistent with zero  and the main contribution comes from the vector coupling~\cite{dum83}. From Eqs.~(\ref{AVom}) 
 one infers that the first $\epsilon^*_0(\lambda)={\bf{k}\cdot{\mbox{\boldmath$\epsilon$}}^* (\lambda)}/k_0$  term [Eq.~(\ref{kepsmes})] almost  disappears, by the G-parity rule, when the emissions from nucleon and antinucleon are added. Retaining only the dominant pieces in Eq.~(\ref{AVomapp}) that is neglecting terms of the order of  $k^2/ 2 \Omega E_q$ or $k^2/4m^2$, one obtains, with $k_0$ defined in Eq.~(\ref{k0om}) and $E_q$ in Eq.~(\ref{Eq}),

\be \label{AVOm2}
{U}_{\omega}^0 ({\bf{q},\bf{k}}) \approx - \frac{g_{V\omega}}{2m}\ \left [4 \ {\bf{q}  \cdot{\mbox{\boldmath$\epsilon$}}^*(\lambda)} + 2 \ \frac{\bf{k}\cdot\bf{q}}{E_q}\ {\frac{\bf{k}\cdot{\mbox{\boldmath$\epsilon^*$}}(\lambda)}{k_0} } - i ({\mbox{\boldmath$\sigma_1$}}- {\mbox{\boldmath$\sigma_2$}})\cdot{\{ \bf{k}\wedge{\mbox{\boldmath$\epsilon$}}^*(\lambda)\} } \right ],
\ee
which implies that one  neglects the following contributions 
 \be  \label{AVOm2bis} 
- i \ \frac{ g_{V\omega}}{2m \Omega}\ \left [   \frac{\bf{k}\cdot\bf{q}}{E_q}\ \left \{ ({\mbox{\boldmath$\sigma_1$}}- {\mbox{\boldmath$\sigma_2$}})\cdot (\bf{q}\wedge{\mbox{\boldmath$\epsilon$}}^*(\lambda))\right \} +
 \ \frac{\bf{k}\cdot{\mbox{\boldmath$\epsilon$}}^*(\lambda)}{k_0}\ \left \{ ({\mbox{\boldmath$\sigma_1$}} + {\mbox{\boldmath$\sigma_2$}})
 \cdot(\bf{k}\wedge\bf{q})\right \}  \right ]
\ee
in addition to a term that disappears since it contains the expression 
$$ 1 - 2\  \frac{E_q}{\Omega} + \frac{q^2}{\Omega^2} =0.$$

\noi The first term in Eq.~(\ref{AVOm2bis})
$$i \ \frac{ g_{V\omega}}{2m \Omega}\  \frac{\bf{k}\cdot\bf{q}}{E_q}\ \left [ ({\mbox{\boldmath$\sigma_1$}}- {\mbox{\boldmath$\sigma_2$}})\cdot \{ \bf{q}\wedge{\mbox{\boldmath$\epsilon$}}^*(\lambda)\} \right ]$$
has to be compared to the basic magnetic contribution 
$$i \ \frac{ g_{V\omega}}{2m}\ \left [ ({\mbox{\boldmath$\sigma_1$}}- {\mbox{\boldmath$\sigma_2$}})\cdot \{ \bf{k}\wedge{\mbox{\boldmath$\epsilon$}}^*(\lambda)\} \right ],$$
since they both lead to $S$-wave magnetic transitions. For average momenta of the order of $500$ MeV/c the neglected term is of the order of $0.1$ of the dominant one. Taking into account the interference contribution in the probability, reduces further this contribution, justifying its neglect. \\

The second term  in Eq.~(\ref{AVOm2bis})
$$ i \ \frac{ g_{V\omega}}{2m \Omega}\  \frac{\bf{k}\cdot{\mbox{\boldmath$\epsilon$}}^*(\lambda)}{k_0}\ \left \{ ({\mbox{\boldmath$\sigma_1$}} + {\mbox{\boldmath$\sigma_2$}}) \cdot(\bf{k}\wedge\bf{q})\right \},$$
which gives rise to $^3 P$ waves has to be compared to the basic electric contribution 
$$- 4 \ \frac{g_{V\omega}}{2m}\ {\bf{q} \cdot{\mbox{\boldmath$\epsilon$}}^*(\lambda)},$$
and, in the absence of any interference contribution, leads to a very small contribution, of the order of $10^{-3}$ of the basic decay rate.

 Apart from the magnitude of the coupling constant there is one important difference with respect to the photon. The magnetic coupling is weak in comparison to the electric one. The reverse was true in the $\gamma$ case due to the large proton magnetic moment. This is the  basic reason making the transition to the final $^1S_0 $ state small (about 1/10 of the total). \\

\subsubsection{The pion case}
For $\pi$ mesons we use  the standard $ \gamma_5$  coupling
\be \label{e4}
{U}_{\pi}^0 ({\bf{q},\bf{k}}) \approx g_{\pi} \  {\mbox {\boldmath $\sigma$}} \cdot \left [ \frac{\bf{q}-\bf{k}}{E_{\bf{q}-\bf{k}}+m} -\frac{\bf{q}}{E_{\bf{q}}+m} \right ]  {{\mbox {\boldmath$\tau$}}\cdot{\mbox {\boldmath $\varphi$}}_{\pi}} \simeq - \frac{ g_{\pi}}{2m} \ ({{\mbox {\boldmath$\sigma$}}\cdot{\bf{k}}}) \ \ ({{\mbox {\boldmath$\tau$}}
\cdot{\mbox {\boldmath $\varphi$}}_{\pi}}).
\ee
In this  case,  the emission  requires a spin flip and a change of  nucleon angular momentum 
leading to the final $ p\overline{p}$   in the $^{31}P_1 $ state. This mechanism eliminates the
possibility of $^1S_0 $ states and does not produce any  threshold
enhancement as indicated by the BES experiments \cite{bai03,abl12,abl07a,abl13a}.
  
\subsection {The $ N \overline{N} $ final state interactions  } \label{NNbarf}

The emission of a magnetic photon from the nucleon in the  reaction
\be \label{l1}
 J/\psi \rightarrow \gamma p\overline{p}
\ee
generates, within our model,  the final spin $0$ state in the $ p
\bar{p} $ system. The corresponding operator in spin space is denoted, in the small $k/2m$ limit and up to the relativistic correction $\kappa \Omega/2m$, by [see Eq.~(\ref{Vgamppbarapp})]
\be \label{VMapp} 
{{V}_{M,\gamma}({\bf{q},\bf{k}})} \approx \frac{ie}{2m}~ ({\mbox{\boldmath $\sigma$}}_1-{\mbox{\boldmath $\sigma$}}_2)
\cdot({\textbf{k}}\wedge {\mbox {\boldmath$\epsilon$}}^*(\lambda)).
\ee
It shows no dependence on the nucleon momentum  before emission $\bf{q}$. 
The Born amplitude associated to this approximation of the magnetic contribution reads then in spin space, using Eqs.~(\ref{am0ter}) and (\ref{Gtilde}) with $\bf{q}_2 = - \bf{q}$, 
\be \label{ABMppbar}
{A}^{M,(B,BC)}_{p\overline{p}\gamma} ({\bf{q}, \bf{k}})= (2\pi)^3 \ \delta^{(3)}({\bf{q}_1- \bf{q} + \bf{k}})\ {{V}_{M,\gamma}(\bf{q},\bf{k})} \  \widetilde{G}_{p\overline{p}}(q) .
\ee

We consider the $^3S_1\to$ $^1S_0 $ transition in the $p\overline{p}$ system. As read from Eq.~(\ref{VMapp}) and  discussed in appendix \ref{explicit} the radiation from both baryons is described by  ${\mbox {\boldmath $\sigma$}}_1 -{\mbox
{\boldmath $\sigma$}}_2$.  The related transition matrix element may be expressed in terms  of  the direction of spin in the triplet state, $ {\mbox {\boldmath $\xi$}}$. Then, the relation 
\begin{equation}
\label{l3a} \langle 0\ 0| \frac{1}{2}({\mbox {\boldmath $\sigma$}}_1
-{\mbox {\boldmath $\sigma$}}_2)~| 1\  {\mbox {\boldmath
$\xi$}}\rangle \cdot (\textbf{k}\wedge {\mbox {\boldmath
$\epsilon$}}^*(\lambda) ) = {\mbox {\boldmath
$\xi$}}\cdot(\textbf{k}\wedge {\mbox {\boldmath
$\epsilon$}}^*(\lambda) ),
\end{equation}
leads to a  formula which may be used in the case of the limit defined by Eq.~(\ref{VMapp}) for a transition from the $^3S_1 \to ^1S_0$ state. Indeed, from Eqs.~(\ref{fotonrel}), (\ref{Vgamppbarapp}) and~(\ref{ABMppbar}), we obtain
\bqa \label{l4} 
\langle ^1S_0 \vert{ A}^{M,(B,BC)}_{p\overline{p}\gamma} ({\bf{q}_r, \bf{k}}) \vert ^3S_1 \rangle &=& (2\pi)^3 \ \delta^{(3)}({\bf{q}_1- \bf{q} + \bf{k}})\ \widetilde{G}_{p\overline{p}}(q_r)  < ^1S_0 | {V}_{M,\gamma}| ^3S_1  > \nonumber \\
&=&  (2\pi)^3 \ \delta^{(3)}({\bf{q}_1- \bf{q} + \bf{k}})\ \widetilde{A}^{M,(B,BC)}_{p\overline{p}\gamma} (\bf{q}_r,\bf{k}), 
\eqa
with
\be \label{I4bis}
{ \widetilde{A}^{M,(B,BC)}_{p\overline{p}\gamma} (\bf{q}_r,\bf{k})}
 = \frac{ie}{m}~ {\mbox{\boldmath $\xi$}}\cdot(\textbf{k}\wedge {\mbox {\boldmath$\epsilon$}}^*(\lambda)) 
 \  \widetilde{G}_{p\overline{p}}(q_r). 
\ee

The electric contribution to the amplitude is 
\be \label{l6} 
{A}^{E,(B,BC)}_{p\overline{p}\gamma} (\mathbf{q},\mathbf{k}) = (2\pi)^3 \ \delta^{(3)}({\bf{q}_1- \bf{q} + \bf{k}}) \ {{V}_{E,\gamma}(\bf{q})} \ \widetilde{G}_{p\overline{p}}(q).
 \ee
 with the electric potential given in Eq.~(\ref{VEgam}). In the same limit ($k/2m << 1$  and up to the relativistic correction $\kappa k^2/ \Omega m$), the approximate electric potential reads
 \be \label{VEppbar}
 {V}_{E,\gamma} ({\bf{q}, \bf{k}}  ) \approx - \frac{ e}{2m}~ 4\  \textbf{q}\cdot{\mbox{\boldmath $\epsilon^*$}}(\lambda).
 \ee
It leads to transitions from the $^1S_0$ state to $^3P$ states. However, final state interactions in the P wave state are weak,  at least in the Paris potential model and they are therefore neglected. \\

On the other hand, in the magnetic transitions the scalar amplitude $ {A^{M,(0,BC)}_{p\overline{p}\gamma} (\bf{q},\bf{k})}$ has to be corrected and these corrections turn out to be very important. They are described  by the loop integral (see Eq.~(\ref{ampfsi2}))
\be \label{s3}
\widetilde{A}^{M,(FSI,BC)}_{p\overline{p}\gamma}(\mathbf{q}_r,\mathbf{k}) = \sum_{I=0,1}
\int~ \frac{d\mathbf{q'}}{(2\pi)^3} \ T_I(\mathbf{q}_r,\mathbf{q'}-\mathbf{k}/2, E_{ N\overline{N}})\ 
G^+_{0, N\overline{N}\gamma}( \textbf{q}',\textbf{k }) \  { \widetilde{A}^{M,(B,BC)}_{p\overline{p}\gamma} (\bf{q}',\bf{k})}, 
 \ee
calculated with the recent Paris potential~\cite{lac09}, in a way described in Ref.~\cite{ded09}. 
On shell, this  $T$ matrix element is normalized to the scattering length. The numerical evaluation of the loop integral in the presence of two singular propagators has to be done  with considerable care. The procedure is described in Appendix \ref{numloop}. The full amplitude for magnetic transitions becomes
\be \label{s6}
A^{M,BC}_{p\overline{p}\gamma}(\mathbf{q}_1, \mathbf{q}_2,\mathbf{k}) = 
(2\pi)^3 \ \delta^{(3)}({\bf{q}_1+ \bf{q}_2 + \bf{k}})\ \left [\widetilde{A}^{M,(B,BC)}_{p\overline{p}\gamma} ({\bf{q}_r,\bf{k}})
+\widetilde{A}^{M,(FSI,BC)}_{p\overline{p}\gamma}(\mathbf{q}_r,\mathbf{k}) \right ],
\ee
with $\bf{q}_2$ and $\bf{q}_r$ given in Eq.~(\ref{dir2}).\\

For the results presented in the following section, the amplitude (\ref{s6}) has been evaluated with the potential $V_{M,\gamma}({\bf{q}, \bf{k}})$ given in Eq.~(\ref{VMgam}) including the $\kappa/2m$ relativistic correction but dropping the $r_+ \ ({\mbox{\boldmath$\sigma$}}_1 + \ {\mbox{\boldmath$\sigma$}}_2) $ since this term generates spin triplet $P$ waves only important close to the threshold region where they are strongly suppressed by the phase space.

\section {The results } \label{results}

\subsection {The radiative decays  } \label{radiative}

\begin{figure}[ht]
\includegraphics [scale = 0.65]{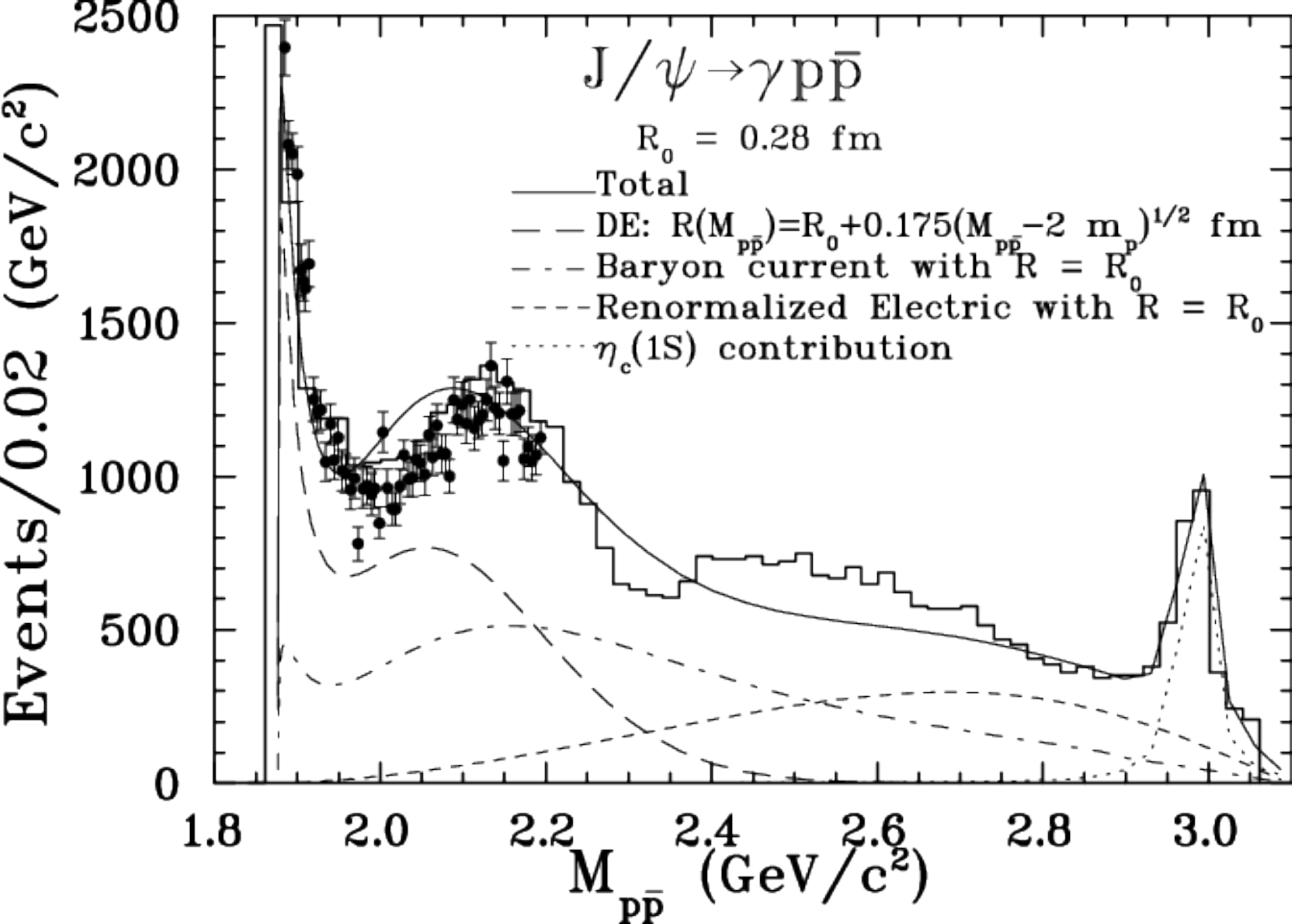}
\caption{ The $M_{p\overline{p}} $ spectra  obtained with the DE and Baryon Emission BC+FSI models. The peak at 3 GeV attributed to the sequential  $J/\psi \rightarrow \eta_c \gamma $ decays is discussed in the text. Histogram as in Fig.~\ref{MPPdirect}, data extracted from Ref.~\cite{abl12}.  }
\label{MPPphoton} 
\end{figure}

\begin{figure}[ht]
\includegraphics [scale = 0.65]{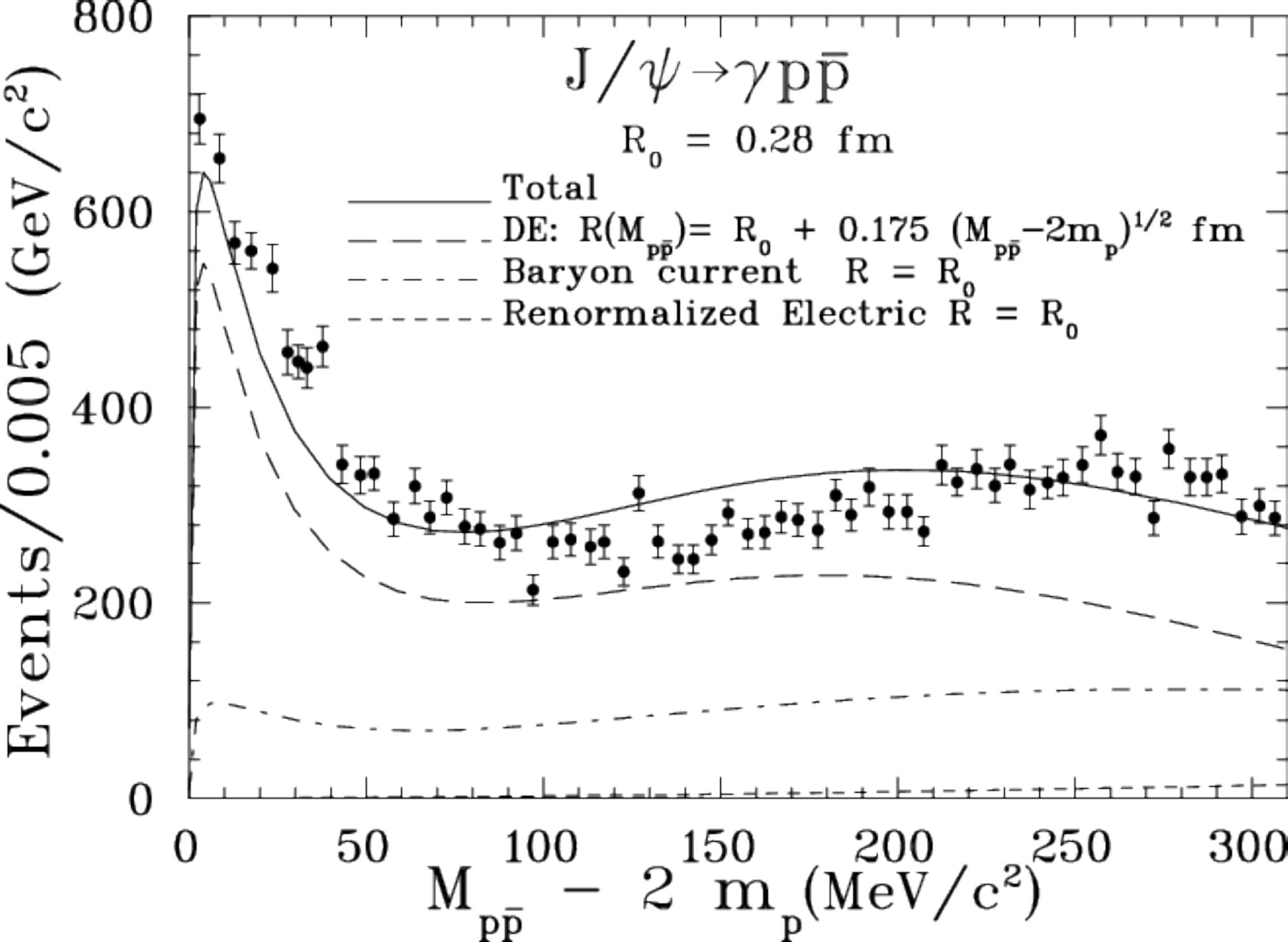} 
\caption{As in Fig.~\ref{MPPphoton} but for the near threshold region. 
Note the small $P$-wave renormalized electric contribution. }\label{figMPPtot} 
\end{figure}

The $M_{p\overline{p}}$ data  in the region of high photon energy  are
dominated by a peak  as can be seen in Figs.~\ref{MPPphoton} and ~\ref{figMPPtot}.  The explanation is related to strong nucleon-antinucleon attraction  essentially in $N\overline{N}$ the isospin-spin singlet  $^{11}S_0$  state but to a certain extent also in the  isotriplet spin singlet  the $^{31}S_0$.   Now, with the radiation  due to baryonic currents this peak is strongly suppressed due to interference  of the intermediate $p \overline{p}$ , $n \overline{n}$ channels and the cancellations  of the  magnetic moments involved.

The various contributions to the $M_{p\overline{p}}$ spectrum plotted in Fig.~\ref{MPPphoton} together with the experimental data have been obtained with the following procedure:

- a) $\vert F_0\vert $   is the overall normalization that is fixed by the  $J/\psi \to p\overline{p}$  decay rate, 

- b) magnetic and electric amplitudes are calculated independently  for the DE and BC  emission modes,

- c) the emission rates are added and the normalizations of the  DE and BC  rates are fixed to reproduce the experimental ratios  shown in Table IV and the invariant mass distribution. 

 We are  not able at present to evaluate the phase difference between the amplitudes associated to the DE and BC  mechanisms. However, the interference effects are most likely fairly weak for two reasons:

- a)  in the low $M_{p\overline{p}}$ region, characterized by magnetic photons the  contribution of the DE mechanism dominates largely the contribution of the BC mechanism,

- b) in the high $M_{p\overline{p}}$ region a similar, although less striking, dominance is attributed to the electric photons whereas there is practically no contribution any more of the DE mechanism.

The experimental data, displayed in Fig.~\ref{MPPphoton}, indicates possible interference effects in the region between $2.3$ and $2.6$ GeV. However, in view of the quality of the data, we hesitated to include an additional parameter to the already many parameters introduced in our description. Thus we have neglected the possible interferences and have simply added the contributions of the DE and BC  mechanisms on the probability level. \\

The  emission from final baryons (BC) - see table \ref{tphoton} - generates about half of the experimental rate and misses the full strength of the threshold enhancement. Photons may be emitted also by exchange currents related to charged mesons exchanged between baryons. Such processes are  known and well described in the $ NN$ systems, but to our best knowledge, have not been discussed in the $N \overline{N} $ systems. Calculations have been performed. We found effects of about $10 \%$ which do not change the overall picture in any significant way.  
The BC model  has to be supplemented by  the other internal emission DE mechanism discussed in Section \ref{internal}. The relative strength of the later is a free parameter which is set to try to reproduce 
the two  peaks in  $M_{p\overline{p}}$  spectrum (see Figs~\ref{MPPphoton} and~\ref{figMPPtot}).  In this way the total branching ratio becomes consistent with the data. The direct emission is characterized by different $N\overline{N}$ final state interactions, in particular there is no cancelation of $p$ and $n$ magnetic moments during the emission process. Hence the interaction in the $^{11}S_0$ wave is stronger and the two resonances  at  threshold and at $M_{p\overline{p}}\approx 2170$ MeV  are more distinct. As discussed previously in Section \ref{internal}, the first is due to quasi-bound state, the second is a shape resonance. Both are generated by the Paris potential model ~\cite{lac09}. \\

Now, a comparison of  radiative decays  $ J/\psi \rightarrow \gamma p\overline{p} $ and $ \psi(2S)\rightarrow \gamma   p\overline{p} $ could be discussed qualitatively.  The baryon current emission does not depend on the internal structure of the $ J/\psi \rightarrow \gamma p\overline{p} $ and $ \psi(2S)\rightarrow \gamma   p\overline{p} $ mesons. What is shown in the present work  is that the near threshold  peak is suppressed by the difference in the proton and neutron magnetic moments. On the other hand, the probability of internal photon radiation does depend on the internal structure. As a consequence, the relative weight of the two modes depends on the internal wave functions which  is nodeless  in the $ J/\psi $ case and has a node in the relative $c\overline{c}$ coordinate in the $ \psi(2S)$ case. We are not able to calculate this wave function. Our qualitative argument is that the internal emission from the $\psi(2S)$ meson (that yields a peak) has to be small. Apparently this node suppresses the magnetic radiative transitions   via  the  DE mode and no peak is generated. We see some, although not fully convincing argument for this suppression mechanism in the experimental  $\gamma p \overline{p}$ branching ratios  equal     $3.8(1.0)\times 10^{-4}$    for $J/\psi$ and much smaller     $3.9(0.5)\times 10^{-5}$ for $\psi(2S)$~(Ref.~\cite{PDG16}). \\

\begin{table}[ht]
 \caption{ Ratio   $ \mathcal{R} = \Gamma(p \bar{p} \gamma ) 
/\Gamma(p\bar{p})$ in $\% $ units with consecutive steps of improvement. 
The experimental ratio $ \mathcal{R} = 18 (5)$ is evaluated from the experimental branching fractions displayed in Table~\ref{tableCP}.
The first column indicates the involved mechanism: BC for the first line and BC+DE for the second.
The other columns  give the different contributions in these mechanisms (see text).
 The  additional DE radiation arises from
the quark phase and is fitted to the magnitude of the near threshold peak.}  

\vspace{0.5cm} 
\begin{tabular}{ccccr}
\hline \hline
 ${\rm Mechanism}$   &$ {\rm Electric} $ & $ {\rm Magnetic}$ (BC)   &$ {\rm Magnetic}$ (BC+FSI) & $ {\rm Total}$\\
\hline
$ BC $      &  $4.38$    & $1.65$&   $1.97$  & $ 6.35 $    \\
$  BC + DE $                &  $-$    & $5$&   $6.51$  & $ 12.86 $ \\ 
 \hline \hline
\end{tabular} \label{tphoton}
\end{table}

$\bullet$ \emph{The end of  $ p \bar{p}$  spectrum}

The origin of the experimental peak at the end of spectrum is the sequential decay
\be \label{end1} 
J/\psi\rightarrow \eta_c~\gamma \hspace{1cm}{\rm and}\hspace{1cm}
\eta_c~\rightarrow p\overline{p}
\ee
which generates  a   peak at the invariant mass $M_{p\overline{p}} =
M(\eta_c) = 2983.4 $ MeV. The decay rates are known experimentally and
$J/\psi\rightarrow \eta_c~\gamma $ is $ 1.7 \ (0.4)\times 10^{-2}$ of the total
while  $\eta_c ~\rightarrow p\overline{p}$ is $ 1.50\ (0.16) \times 10^{-3}$ of
its decay rate \cite{PDG16}. 
The dotted line in Fig.~\ref{MPPphoton} results from the modulus squared of the following relativistic Breit-Wigner amplitude for the description of the $\eta_c$
\be \label{Aetac}
A_{\eta_c}= -f_{\eta_c} \ \frac{m_{\eta_c} \Gamma_{el}}{M_{p\overline{p}}^2 - m_{\eta_c} ^2 + i \ m_{\eta_c} \Gamma_{tot}}, 
\ee
with $\Gamma_{tot} = 31.8 \ (0.8)$ MeV and  $\Gamma_{el} /\Gamma_{tot} = 1.50\  (0.16) \times 10^{-3}$. In the energy range of this $\eta_c$ contribution, the interference of the $A_{\eta _c}$ amplitude with the very small  magnetic $S$-wave is neglected and for the curve shown in Fig.~\ref{MPPphoton} the free parameter $f_{\eta_c}$ is fixed at the value $23.2 \ \times 10^3$ events/ $0.02$ (GeV/c$^2$). 
Together the expected area under the
end peak would amount to $  2\times 10^{-5}$ of the total decay rate, 
i.e.,  about $5\%$ of the $p\overline{p}\gamma$ decay rate. The first
experimental result of reference \cite{bai03} indicated   a $1\%$
effect but more recent measurements yield comparable results~\cite{abl12}.\\

In addition to the $\eta_c$ peak, another peak arises within the BC model. It is related to the infrared enhancement in the
intermediate state $p\overline{p}$ propagator. The real infrared catastrophe does not occur since the initial $J/\psi$ has a finite width. This effect produces an enhancement  in the region  $M_{p\overline{p}}> 2820 $ MeV and a narrow bump at the end  $ 3090 < M_{p\overline{p}} < 3100$ MeV.  The area under this enhancement  amounts to $3\%$  of the rate  $ \mathcal{R}_{\gamma}$ calculated with the baryon emission model. That is  about  $ 0.7\%$ of the experimental rate.
  The experimental check  is not easy as  the errors in the photon energy determination   - $\sigma_E$ -
 are large in this region  and these two effects overlap. The BES detector offers \cite{bai03}
\be \label{end2} 
\frac{\sigma_E}{E} = \frac{ 21\%}{ \sqrt{E/GeV}}, 
\ee
and  in the region of interest  $\sigma_E\approx  E \approx ~100 $ MeV and thus the position of the peak is not well determined. With a better resolution, the magnitude and shape of the infrared bump would be a  check  for decay  models. \\

As indicated in Section \ref{NNbar} in the comment below Eq.~(\ref{i3}), the BES data~\cite{abl12}, where a $k < 50$ Mev/c cut is applied, does not display the contribution of the infrared photon contribution. Thus, in the present work, the $M_{p\overline{p}}$ infrared pole  is eliminated by introducing a smooth non-relativistic Breit-Wigner function. In other terms, a phenomenological  final state interaction correction is applied to the $P$-wave electric amplitude $A^{E,(B,BC)}_{p\overline{p}\gamma}(\mathbf{q}, \mathbf{k})$  given by Eq.~(\ref{l6}). Hence, the short-dash line in Fig.~\ref{MPPphoton}
is the result of the renormalized electric photon amplitude where the $M_{p\overline{p}} = M_{J/\psi}$ pole has been eliminated
\be \label{GAME}
\widetilde{A}^{E,(B,BC)}_{p\overline{p}\gamma}(\mathbf{q}, \mathbf{k})= f_r\ A^{E,(B,BC)}_{p\overline{p}\gamma}(\mathbf{q}, \mathbf{k}), 
\ee
\be \label{GAME1}
f_r = {\vert}   N_E\  \frac{M_{J/\psi}- M_{p\overline{p}}}{M_{J/\psi}- M_{p\overline{p}} - i \Gamma_E} {\vert}.
\ee
 The free parameters are fixed at the respective values $\Gamma_E= 500$ MeV for the width and $N_E= 3.5$ for the normalization. 
 
 Despite the Breit-Wigner form we do not suggest that there is a new resonant mechanism involved. This form is used only for a parametrization that serves two purposes:\\
 
 1) removing the infrared enhancement since it is  removed in the experimental data and \\
 
 2) enhancing the electric photon emission approximately by a factor $3$ to reach consistency with the data.\\
 
 Although a new resonance is a possibility we are inclined to view $f_r$ in Eq.~(\ref{GAME1}) as a result of another DE mechanism. This possibility is discussed in the next subsection devoted to the $\omega$ emission and where, indeed, the broad bump in the spectrum is due to the DE decay mode and not to an intermediate resonance. A quantitative analysis may be performed in the  $\omega$ emission case as more data exist. It is not feasible in the radiative decay mode and we limit the analysis to the phenomenological formula (\ref{GAME}).

\subsection {The  $\omega, \phi, $  and $ \pi$   emission rates} \label{emisrates}

\subsubsection{The $\omega$ emission} 

Its rate seems to be easier to understand than those for the radiative decays  due to the weak tensorial  coupling which
favors  strongly the electric type transitions. The corresponding branching fraction 
\begin{equation}
\label{om1}  \mathcal{R}_\omega =  \frac{\Gamma(
p\overline{p}~\omega)}{\Gamma( p\overline{p})}
\end{equation}
obtained with the basic final state emission (BC) model is  shown in table \ref{tablem}. The electric type transitions (E)
lead to  $P$-wave $p \bar{p}$ states with small  final state
interactions. The  magnetic type transitions  (M) generate $S$ wave
states strongly affected by the final $N\overline{N} $ interactions. In comparison to the photon case  these
interactions are stronger as isospin conservation requires
baryons to be in isospin $0$ state. On the other hand, due to
large meson mass, the final $p $ and $\bar{p}$ are less strongly oriented close to threshold
and the  tensor $NN \omega$  coupling  is weak and consistent with zero.
Hence, effects of these final state interactions are not well
visible in the emission rate, see figure \ref{figEE}.  It is the electric  transition (labelled EE in Fig.~\ref{figEE} that dominates.
Now in distinction to the photon case the longitudinal component ${\mbox{\boldmath $\epsilon$}}^*(\lambda=0)$ exists and the corresponding
strength of the dominant electric transition is given in Eq.~(\ref{SVEom2}). It yields approximately $ (g_{V\omega}q/{2m})^2 $. 
One finds that the longitudinal component gives  a large contribution to the low part of the $ M_{p\overline{p}} $  spectrum which is not supported by the data.

\begin{table}[ht]
 \caption{ Calculated ratios  $\Gamma (p \overline{p}\ meson) / \Gamma(p\overline{p})$
of channel widths allowed in the $ J/\psi $ decays. The errors correspond to uncertainties of the  $p\overline{p}$ meson coupling $g$. Values of  $\mathcal{R}_{exp}$ calculated from the experimental branching fractions listed in Table~\ref{tableCP}. }
\vspace{0.5cm}
\begin{tabular}{lccccr}
\hline
\hline
meson                & $\mathcal{R}_{exp}$   &\ $ \mathcal{R}[BC]$ &$\ \mathcal{R}[BC,FSI]$ &\ $ \mathcal{R}[BC,FSI,DE]$ &\ $g^2/(4\pi)$\\
\hline
$p\bar{p}\pi^0   $   &  $0.575\ (0.05)$       &  $0.43$   & & & $ 13.8$  \ \cite{rij06}        \\
$p\bar{n} \pi^-  $   &  $0.966\ (0.06)$       &  $0.85$   & & & $ 13.8$     \cite{rij06}       \\
$p\bar{p}\omega  $   &  $0.507\ (0.07)$       &  $0.87\ (0.16)$ & 0.67\ (0.13) &  & $ 8.1\ (1.5)$ \ \cite{gre80}   \\
$p\bar{p}\omega  $   &  $0.507\ (0.07)$       &  $0.40   $ & 0.33 & 0.39   & $ 4.16$ \ \cite{shk13}         \\
$p\bar{p} \phi   $   & $0.0247 \ (0.0016)$     &  $0.023$  & & &  $5.5$ this work\\
 \hline
\hline
\end{tabular} \label{tablem}
\end{table}

The decay rates are given in table \ref{tablem} and these results are obtained with the coupling
constant~$g^2_V(NN\omega)/4\pi = 8.1\ (1.5), ~g^2_T(NN\omega)/4\pi =
0.16\ (0.46) $ obtained with dispersion relations \cite{gre80},  more recent values from $NN$ interactions are $
g^2_V(NN\omega)/4\pi = 9.73, ~g^2_T(NN\omega)/4\pi = 0.005 $
\cite{rij06}. In both cases  the tensorial coupling is  negligible and it was neglected.
However, the most significant  parameter is the source radius and  the rates of $\omega$  (and $\pi$)
meson emissions put  very strong limits on ${R}_0$.
 The final choice is  obtained from the best fit of
  $\mathcal{R}_\omega$ and $\mathcal{R}_\pi$ is $ R_0=0.28(1)$.
The $\omega$ coupling constants are uncertain, those indicated above are extracted from
NN scattering data.  On the other hand coupling constants derived from semi-phenomenological meson
formation data  are smaller. A value  $ g^2_V(NN\omega)/4\pi = 1.19$ has been obtained in Ref.~\cite{pen11} 
with a very small tensorial coupling although   Ref.~\cite{shk13} reaches a value $ g^2_V(NN\omega)/4\pi = 4.16$. In our calculations, we have chosen $ g^2_V(NN\omega)/4\pi  = 3.8 $ and  $g_T(NN\omega) =0$.

The omega emission case  differs strongly  from the photon emission as apparently the BC  mode dominates. However,  as already shown in the BES Collaboration paper~\cite{abl13a},  the emission of $\omega$ meson requires the involvement of excited nucleon states $N^*$  and the final state involves three interacting particles. The multiple scattering method presented here requires the leading corrections due to  $N^*$ to reach  some  25 \%  of the leading order.  Roughly the next  (missing) order is expected to reach about 10 \% . Such corrections are unlikely to be  kept under control
as the quantum numbers of  $N^*$  resonances are very uncertain. Thus one will have to resort to more appropriate methods than that of the simple DWBA used in the present work to achieve more reliable results.  Also, one has  to  keep in mind that the basic  term involves a rather uncertain $N N \omega $ coupling constant which makes the calculations even less (if not) reliable if this coupling turns out to be much smaller.

\subsubsection{The  $M_{p\overline{p}}$ and $M_{\omega p}$ spectra}

\begin{figure}[ht]

\includegraphics [scale=0.36]{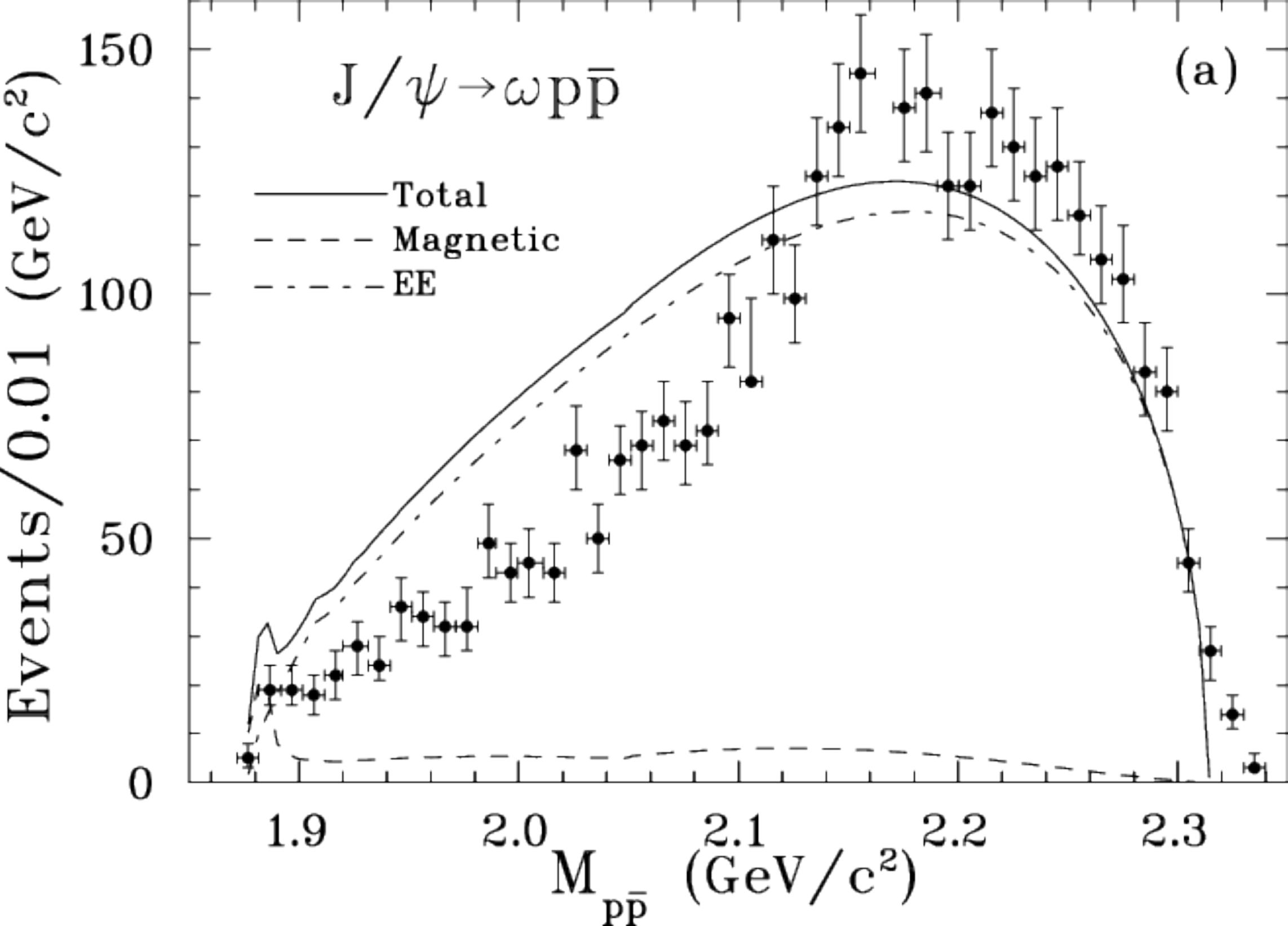}
\hspace{0.5 cm}
\includegraphics [scale=0.36]{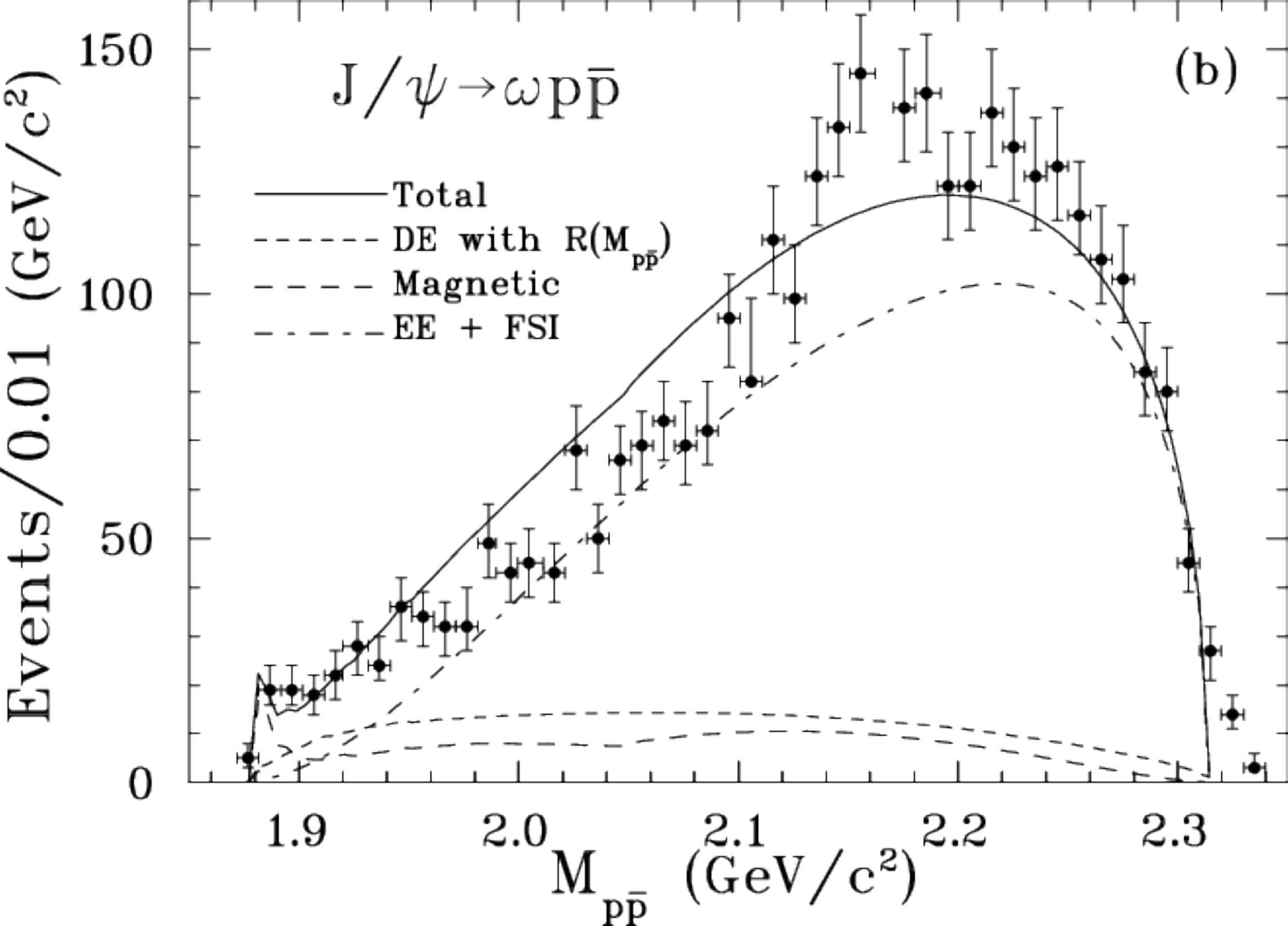}
\caption{Left panel (a):
the $M_{p\overline{p}}$ spectrum   obtained with the BC model. Right panel (b):
the $M_{p\overline{p}}$ spectrum   obtained with the BC+FSI model. The same  arbitrary normalization is used to fit the experimental  shape for both graphs. The electric contribution is labelled EE. No FSI contribution in the DE model but the weak energy dependance of the source radius [Eq.~(\ref{dir6})] has been kept. Here $R_0=0.28$ fm, $g_{V\omega}^2/(4\pi) = 3.8$ and $g_{T\omega}= 0$. Data extracted from Figs.~2  in \cite{abl13a}. }
\label{figEE}
\end{figure}

The BC model of meson emission from final baryons yields a  fair estimate of the decay rate, Table \ref{tablem},  unfortunately subject to large uncertainty in  the $\omega NN $ coupling constant.
On the other hand, the   spectra of  the invariant masses     $M_{p\overline{p}}$   and $M_{\omega p}$ pose a more difficult problem. The  $M_{p\overline{p}}$  distribution,  plotted in the left panel of Fig. \ref{figEE}, requires strong reduction in the lower mass region which may be generated only by a destructive interference. Such a possibility is offered by final state interactions involving a $ N(3/2^-)$ resonant state  expected to mediate  the $\omega$-$p$ interaction in the $2$ GeV energy region [see 
Fig.~\ref{figN3/2}(b)]. Now, the bulk of available phase space  covers region  between $ M_{\omega p} = 2052$ MeV at $p\overline{p} $ threshold and $ M_{\omega p} = 1805$ MeV at the end of  $M_{p\overline{p}} $ spectrum. Hence, interference of the intermediate $N(3/2^-)$ and the basic decay mode may be constructive in part and destructive in another part of the phase space.

The related mechanism is presented only schematically here. The Rarita-Schwinger particle propagation is given by
 \be \label{RS}
G^{\mu \nu} =   \frac{\gamma p + m_{3/2} }{m_{3/2}^2-p^2 +i m_{3/2}\Gamma} P^{\mu \nu}, 
\ee
where $P^{\mu \nu}$  projects on spin $3/2$  states. We follow Ref.~\cite{nimai89} which underlines some controversies in the formulation. These are of small concern as in the situation discussed here one finds this particle to be non relativistic  and~\cite{chu06}
\be \label{slowN}
 P^{\mu \nu} \rightarrow P^{ij}= 2/3 ~[ \delta^{ij}+i/2 \epsilon_{ijk}\sigma^k ].
  \ee
\vspace{0.5cm}

\vspace{2cm}

\begin{figure}[ht]
\includegraphics [scale=0.44]{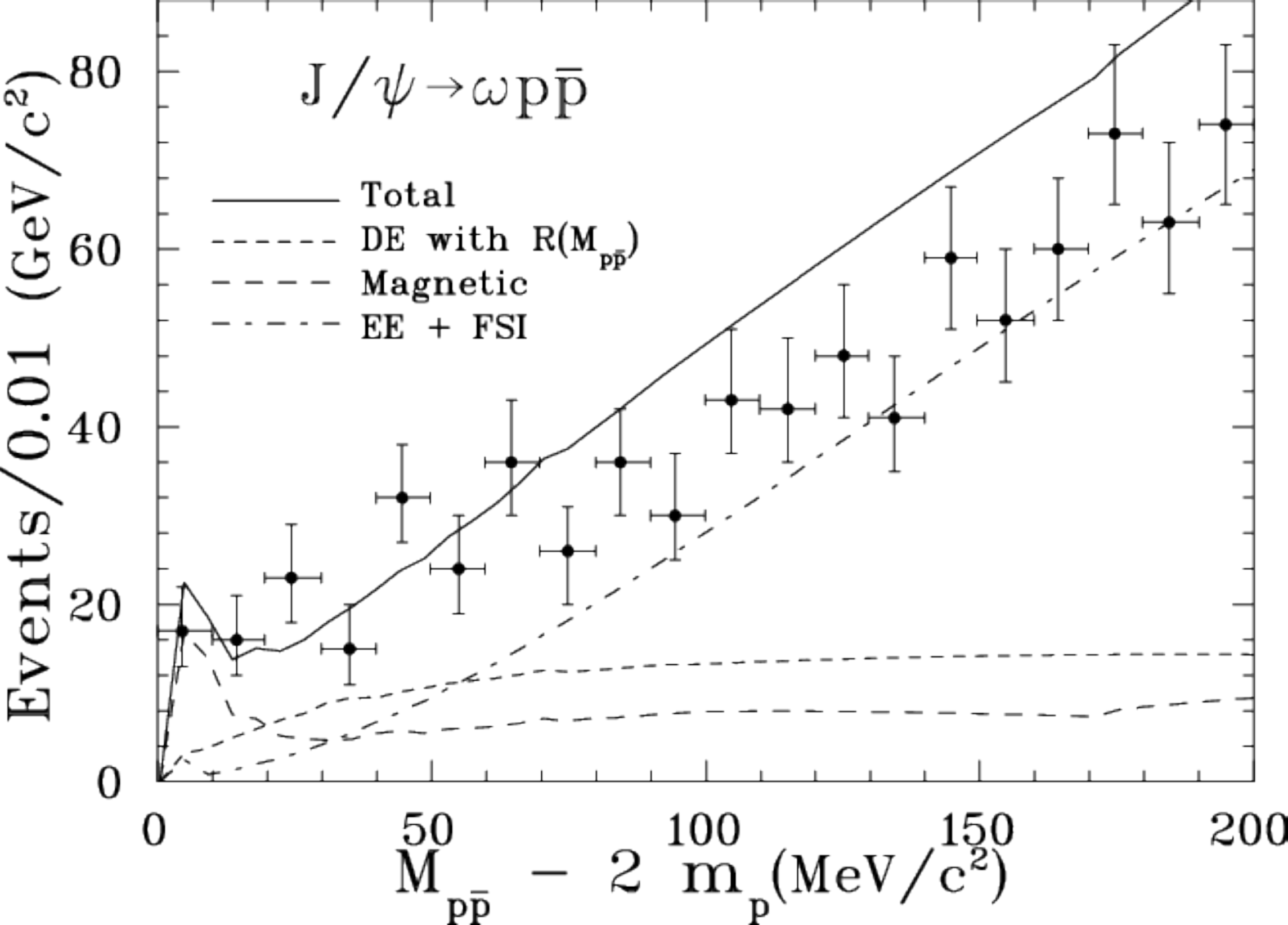}
\caption{The $M_{p\overline{p}}$ spectrum as in Fig.~\ref{figEE} (b) but for the near threshold region.} 
\label{figEEbis}
\end{figure}

\begin{figure}[ht]
\includegraphics[scale=0.65]{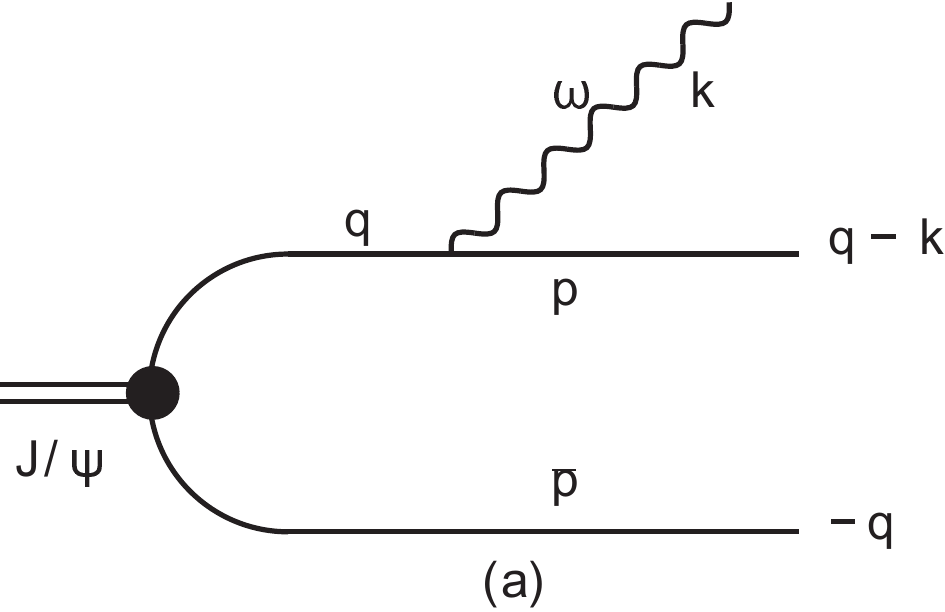}
\hspace{1cm}
\includegraphics[scale=0.65]{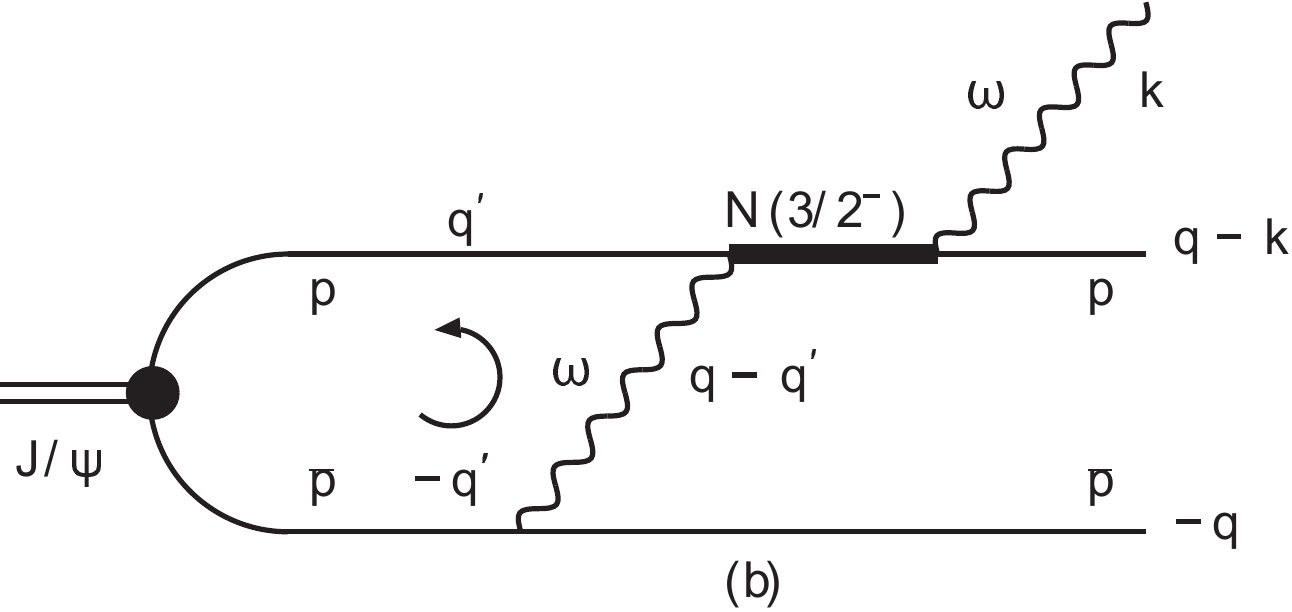}
\caption{Emission of the $\omega$ meson from intermediate baryons. The left graph (a) corresponds to the Born term while the right one (b) includes final state corrections involving the $N(3/2^-)$ nucleon resonance.} \label{figN3/2}
\end{figure}

\begin{figure}[ht]
\includegraphics [scale=0.45]{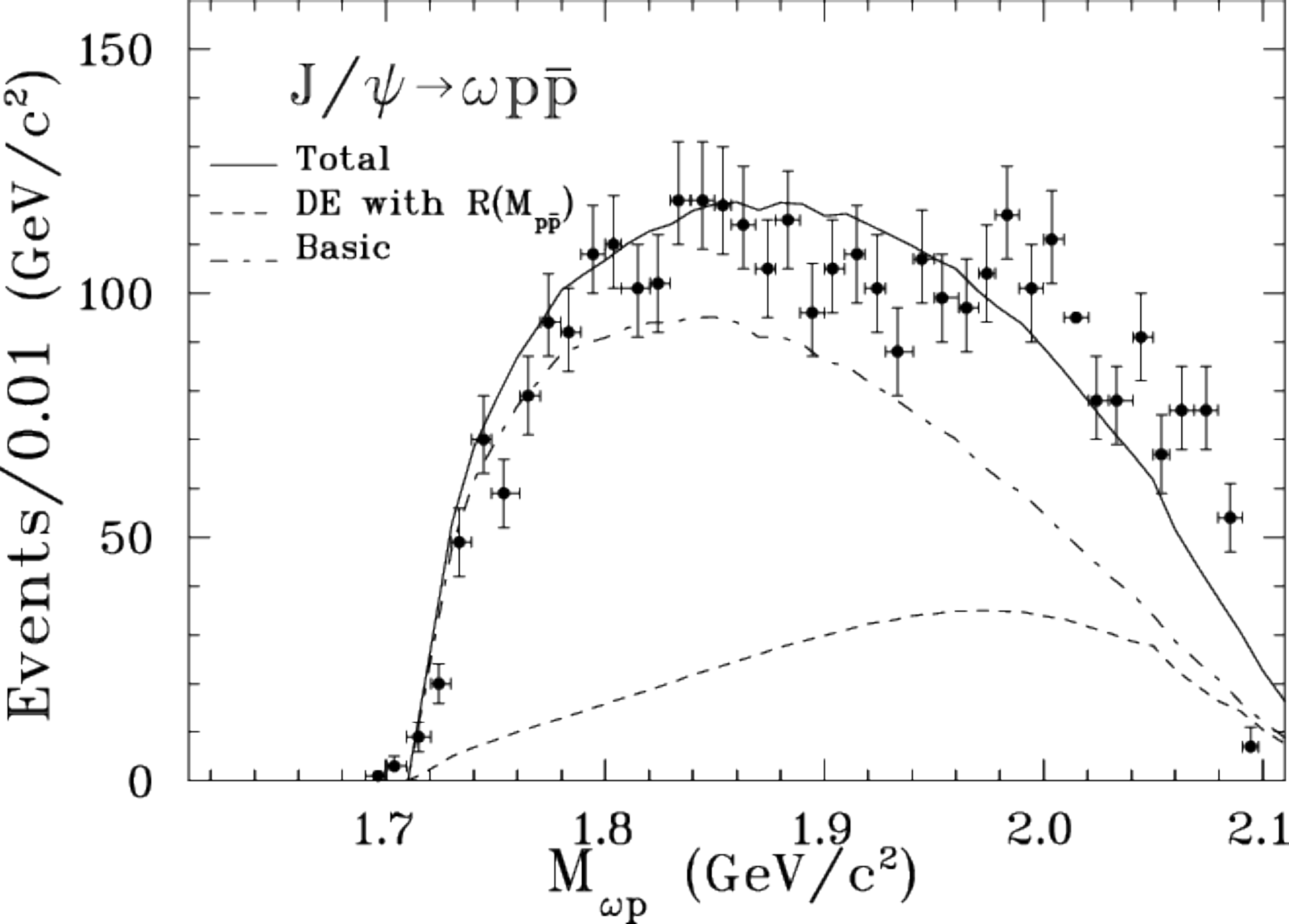}
\caption{ The $M_{\omega p} $ spectrum obtained with the BC+FSI+DE model (see Fig.~\ref{figEE} caption). The dash-dot curve  (Basic model) refers to the BC+FSI calculation.   }
\label{EEomp} \end{figure}

This formula sets  the main simplification of this final state interaction. In addition, we drop the second term which leads to spin flip leading to no interference with the basic Born amplitude.  The calculation involves a standard loop integral which follows the procedure of Appendix \ref{numloop}.  Three uncertain parameters are implied: we use~$m_{3/2}=2050$~MeV, $\Gamma =300 $~MeV for the $ N(3/2^-)$  position and width (nonessential),  and $3\%$ for the strength of  the $ N(3/2^-)$  coupling to the $\omega-N$ channel. The effect is shown in 
Fig.~\ref{figEE}(b) and in Fig.\ref{EEomp}. The resonance parameters ($m_{3/2}=2050$~MeV, $\Gamma =300$~MeV) are close to those of the $N(1875)$ and of the more uncertain $N(2120)$ $3/2^-$ resonances~(see Ref.~\cite{PDG16}).   \\

On the other hand the $\omega$ spectra are not reproduced
and the $M_{\omega p}$ and  $M_{\omega \overline{p}}$   distributions  miss  a bump in the data at large masses.
Inspection of  figure \ref{EEomp} indicates  a broad structure missing around  2 GeV.  Such resonance has been already introduced into our description of the final state interaction. Nevertheless,  its effect is not seen in the distribution  of    $M_{\omega \overline{p}}$ \cite{abl13a}.
The formalism developed so far indicates a strong correlation of $M_{\omega p}$   and $M_{p \overline{p}}$. The phase space region close to the $p \overline{p}$ threshold  overlaps with the region of $M_{\omega p} \sim 2.05 $ GeV. Thus enhancing the high energy tail of  $M_{\omega p}$ reduces the low energy end of $M_{p \overline{p}}$. Within the BC+FSI approach it is not possible to reproduce both distributions and another mechanism has to be found.
Another option tried  was  $ J/\psi \rightarrow \overline{N}(1/2^-) N^*(3/2^-)  \rightarrow \overline{N} N \omega $, but for the reason given above it was not able to explain simultaneously  the $M_{p \overline{p}}$ and  the $M_{\omega p} $  spectra.\\

 As the introduction of resonances brings  no  success we resort to another  formation mechanism   which
 was found useful in the study of radiative decays.   A fraction of $ \omega$  mesons is assumed to to be emitted internally i.e.  before  the baryons  are formed in a   $P$   wave state. This emission process is depicted in 
 Fig.~\ref{figdirect}(a), 
 the photon line being replaced by an $\omega$ meson line. Instead of the matrix element given by Eq.~(\ref{Vdir}) related to $S$ states,  that  associated to  $P $ states should  be of the form
\be \label{Vdirom}
V^{DE}_{\omega}(\lambda) = f\ {\mbox {\boldmath $\epsilon$}^*(\lambda)}\cdot
{\mbox {\boldmath $\xi$}}, 
\ee
where  $f$ is a free coupling constant.  In the center-of-mass (c.m.) system of $J/\psi$  the bilinear form of $V^{DE}_{\omega}$ averaged over directions of ${\mbox {\boldmath $\xi$}}$ leads to sum over polarizations
  \be \label{om3}
\sum_{\lambda=-1}^1 |<V^{DE}_{\omega}(\lambda)>|^2  = f^2 \sum_{\lambda} {\epsilon}_i^*  \epsilon^i = f^2 (3 - \textbf{k}^2/m_{\omega}^2).
\ee
Now the essential point  is that this coupling does not  depend  on $q^2$  in contrast to the helicity sum of $\vert V_E^{\omega}\vert^2(\lambda)$ in Eq.~(\ref{SVEom2}).   Jointly with the
assumed expansion of the system during the decay expressed in Eq.~(\ref{dir6}) one is able to avoid the unrequired correlation
of low energy $M_{p\overline{p}} $ and high energy $M_{\omega p}$. Here, final state interactions are not introduced in the direct emission DE model but the same weak energy dependance for the source radius [Eq.~(\ref{dir6})]  as for the photon case has been kept, i. e., $R_0= 0.28$ fm and $\beta = 0.175$ fm$^{3/2}$.  The strength of coupling  to $V^{DE}_{\omega}$ is obtained by the best fit to both $M_{p \overline{p}}$ and  $M_{\omega p}$ spectra.
The results are given in Figs.~\ref{figEE} and~\ref{EEomp}.  It is apparent that the bump in the last figure is not due to a resonance but to a different decay mode. This internal emission mode contributes about $22\%$ of the decay rate. 

\subsubsection{The  $\pi$ and $\phi$ emission rates} 

They are given in Table \ref{tablem}. The neutral pion is emitted coherently from the intermediate
$p\bar{p}$ system.  The negatively charged pion may be emitted from the
antiproton  only. However, in the intermediate $I=0$ state one has
also $n\bar{n}$ component and the $\pi^-$ may be emitted by the neutron.
These processes are coherent.  Therefore  the ratio $\Gamma(p\bar{n} \pi^-) /
\Gamma(p\bar{p}\pi^0) = 1.78 \ (0.22) $ within error limits equals 2 minus the square of the relevant pion-nucleon coupling constants.
 This indicates that pions are emitted predominantly in the baryonic phase of
decay  and that final state interactions are not essential, the $p\bar{p}$ and
$p\bar{n}$ interactions being different. 
The BES data indicate effects of $N^*$ resonances in the invariant mass distribution which, 
depending on the way of description, amount to some $25 \%$ of the total rate. 
 
The $\phi$ experimental branching ratio is small as the allowed phase space is small. Table
\ref{tablem} shows that it may be obtained  with the value  $g_V^2/4 \pi = 5.5, ~g_T=0. $ which compares well
 with   $ g_V^2/4 \pi= 5.1 , g_T^2/4 \pi = 0.2 $ obtained in Ref.~\cite{hoe76}. The experimental spectra obtained by the BES III Collaboration~\cite{abl_PRD93}  for the $J/\psi \to p \overline{p} \phi$ are qualitatively very close to those obtained in the $J/\psi \to p \overline{p} \omega$ case (see Figs. 8 and 10). So the basic BC mode is likely to require corrections on the same 25 \% level as in the $\omega$ case, norwithstanding that the missing knowlege of the $\phi$ coupling to $N^*$ or $ \Delta^*$ resonances and uncertainties  in the $\phi N N$ coupling constants do not allow a more precise discussion. The present accurate experimental value~ \cite{abl_PRD93} for the ratio
$\mathcal{R}(p\overline{p}\phi)/\mathcal{R}(p\overline{p})$ favors clearly the necessity of a more elaborate approach than just relying on the Born term of the baryon current considered in this study. But this would be the subject of a research project by itself.

\section {Summary and outlook  } \label{conclude}
\subsection{Summary}
In the present work, the $J/\psi \rightarrow \mathcal{B}p\overline{p}$ decays where $ \mathcal{B}= \gamma, \omega,\phi,\pi$ have been studied. Two processes have been introduced to describe the BES Collaboration data on the photon \cite{bai03,abl12} or $\omega$ meson~\cite{abl07a,abl13a} formation in  $J/\psi$ decays into $p\overline{p}$. 
For the radiative decays both processes include final state nucleon-antinucleon interactions with $S$-wave half-off shell functions~\cite{ded09} based on the Paris $N\overline{N}$ potential~\cite{lac09}. The $J/\psi$ source is described in momentum space by a phenomenological Gaussian function with radius $R_0$ (see Eq.~{\ref{dir3}). The value $R_0=0.28$ fm is found to be the best choice to reproduce the different particle $\mathcal{B}$ decay rates as compared to that of the  $J/\psi \rightarrow p\overline{p}$ decay.
Before presenting some outlook, the description of the two mechanisms and of the free parameters for the photon and meson emissions are summarized below.

\begin{enumerate}
\item \textit{Direct emission process}.
Here, the emission of photons or $\omega$ mesons occurs before the final baryons are formed. 
In the radiative decay, the final state interactions generate two distinct final resonant states in the $p\overline{p}$ system. One close to the threshold (very sharp peak in the $p\overline{p}$ spectra) is explained as an effect of baryonium - a broad $52$ MeV wide quasi-bound state  at $4.8$ MeV below threshold generated in the $^{11}S_0$ wave  by the Paris potential. Another - a resonant state at 2170 MeV - is formed as a shape resonance in the same partial wave.
The Born contribution of the direct emission process allows to achieve a good description of the full $\omega$ spectrum at large $M_{p\overline{p}}$  and $M_{\omega p}$ invariant masses.
For the $\gamma$ or $\omega$ meson it is found necessary to introduce, for the source radius, a weak energy dependence on $M_{p\overline{p}}$ [see Eq.~(\ref{dir6})], in both case the same dependence is used. 

\item \textit{Emission from baryonic current}.
The second mechanism assumes the emission of photons (or mesons) by the baryonic currents in the final state of the $J/\psi \rightarrow p\overline{p}$ decay. This emission occurs thus after the initial decay of the $J/\psi$ into an $N\overline{N}$ pair.
In the radiative decay channel this process is not sufficient to provide a fair reproduction of final resonant states. This is the reason why this model has to be completed by the direct emission model just described above. 
For the $\omega$ meson production case, the Born term of this process is the dominant mode as it is in the $\pi$ or $\phi$ formation case. However, the $\omega$ invariant mass distribution $M_{p\overline{p}}$ requires a strong reduction in the lower mass region. This is obtained by introducing a specific final state interaction involving a $N^*(3/2)$ or $\overline{N}^*(3/2)$ resonance created by an $\omega$-$p$ ($\omega$-$\overline{p}$) interaction through an $\omega$ meson exchange between the $\overline{p}$($p$) and $p$($\overline{p}$) pairs [see Fig.~\ref{figN3/2}(b)].

\item \textit{Free parameters for the radiative emission}.
For the photon emission case, to reproduce the $M_{p\overline{p}}$ spectra (see Figs.~\ref{MPPphoton},~\ref{figMPPtot}) and the relative decay rate (see Table~\ref{tphoton}) {\it seven free parameters} are used: the initial radius of the source function, $R_0=0.28$ fm, the slope parameter associated to the energy dependence of this radius $\beta=0.175$~fm$^{3/2}$~, the normalization of the DE model contribution [Eq.~(\ref{dir4})], that of the baryon current contribution [Eq.~(\ref{Abctot})], the coefficients $\Gamma_E=500$ MeV and $N_E= 3.5$ entering the renormalized electric photon amplitude [Eq.~(\ref{GAME})] and the normalization  $f_{\eta_c}$ for the $\eta_c$ Breit-Wigner parametrization in Eq.~(\ref{Aetac}).

\item \textit{Free parameters for the meson emissions}.
In the case of the $\omega$ meson emission, to fit the invariant masses $M_{p\overline{p}}$ and $M_{\omega p}$ distributions (see Figs.~\ref{figEE} and~\ref{EEomp}) and the relative decay rate (see Table~\ref{tablem}), {\it five free parameters} are introduced: the normalization of the BC+FSI model [see Eqs.~(\ref{Abctot}, \ref{AVOm2}) and~(\ref{SVEom2}, \ref{SVMom2})], the two $N(3/2^-)$ parameters in Eq.~(\ref{RS}) viz.,  $m_{3/2}= 2050$ MeV, $\Gamma= 300$ MeV plus the 0.3\% strength of its coupling to $N\omega$ and the direct $\omega$ emission coupling constant $f$ [see Eq.~(\ref{om3})].
Looking at the ratio of the decay rates $\mathcal{R}(p\overline{p}\omega)/\mathcal{R}(p\overline{p})$ given in Table~\ref{tablem}, a $g^2_{V\omega}/4\pi$ coupling between $4.16$ and $8.1$ would bring this ratio closer to the experimental value. This table also shows that the Born amplitude [Eq.~(\ref{am0ter})] allows to reproduce well the relative decay rates for the cases of the $\pi$ or $\phi$ emission for known values of $g^2(p\overline{p}\pi)$ or $g^2(p\overline{p}\phi)$.

\item \textit{Uncertainties, shortcomings}.
 The basic mechanism  for  pion emission from the  baryonic currents yields  decay rates smaller than the experimental ones  (see Table~\ref{tablem}). 
A proper description  of the proton-pion invariant mass distribution requires at least 
three pion-nucleon resonant states  and a good  control  over relative phases~\cite{ablPRD80}.  
 The limitation to single dominant  final partial wave is not sufficient to describe the rather precise data.

For radiative decays, the separation of  two formation mechanisms  is only approximate due  to two effects:
  
(a) unknown relative phase of both amplitudes   may affect  the region $M_{p\overline{p}}  \approx 2.45 \pm 0.15$ GeV where  these mechanisms give comparable rates,  
 
(b)  there might be some  presumably weak effect  of the isospin symmetry violation in the course of the internal photon 
emission. Both effects are difficult to calculate.\\

\end{enumerate}

\subsection{Outlook}

The internal structure of the $J/\psi$ and $\psi(2S)$ being different, the direct emission model is less likely  to give a reasonable description of the radiative decay of the $\psi(2S)$ state. This might explain, in a qualitative way,  why no resonant states are visible in this process.

The phenomenological part of the $N\overline{N}$ Paris potential~\cite{lac09} has been determined to reproduce the $N\overline{N}$ data up to $E_{N\overline{N}} \approx 200$ MeV, i.e., $M_{p\overline{p}}\approx 2.1$ GeV. It is interesting to observe that, nevertheless, it produces reasonable results beyond the region tested in scattering experiments. The present approach could also be applied with other ${N\overline{N}}$ scattering matrices, for instance that of Ref.~\cite{dai17}. Furthermore, with more accurate experimental results, effects of weakly populated final ${N\overline{N}}$ states might enter. In the present study, they do not seem to give sizable contributions. 

Complications were found for the $\omega$ emission channel: the $M_{p\overline{p}}$, $M_{\omega p}$ and $M_{\omega \overline{p}}$ spectra~\cite{abl13a} might indicate contributions of two $N^*$ states. The description of these is complementary to that of the mesic and radiative excitations of the nucleon. With increased precision of the BES Collaboration measurements, the extraction of the resonance parameters and nucleon-meson coupling constants should be more accurate. 

Spectra of the $J/\psi \to p\overline{p} \pi^0$ decays (see, e. g., Ref.~\cite{ablPRD80}) albeit not discussed here, indicate at least effects of established $N(1535)$ and $N(1650)$ states. The description of these decays seems to be a pressing question which may yield more information that those arising from the uncertain $\omega$ case. 

The infrared catastrophe is approached by the baryon current model. It would thus be interesting to improve the energy resolution at the end of the spectrum to validate or disprove the photon (light meson) emission process from the final baryons.

Finally, the present work should allow to approach the related $\overline{p} p\to J/\psi + \ meson$ reaction on nuclei which sooner or later will be studied experimentally~\cite{fair}.

\section*{Acknowledgments}
\noi Support from the French-Polish COPIN/IN2P3-PAN Collaboration Agreement 05-115 is gratefully acknowledged. 


\appendix

\section {Phase space} \label{phasespace}

Let $p= (p_0, \bf{p})$, 
$$p^2 = p_0^2 - {\bf p}^2 \hspace{2cm}{\rm and}\hspace{2cm}  p_0=  {E(\mathbf{p})}= E(\vert\mathbf{p} \vert) = E(p)=  \sqrt{{\mathbf p}^2+m^2}.$$ 
Then the restricted two body phase space for  the $ J/\psi \rightarrow  p\overline{p} $ decay at rest reads, with $q_1$ and $q_2$ denoting the four-momenta of the nucleon and the antinucleon with masses $m_1=m_2=m$, 
\bqa
\label{a0}
 L_2 &=& (2\pi)^4  \int \prod_{i=1,2} {\frac{d^4q_i}{(2\pi)^3}} \theta(q^0_i)\delta(q_i^2 -m_i^2) \ \delta^{(3)}(
\textbf{q}_1+ \textbf{q}_2)\ \delta(M_{J/\psi}- {E(\bf{q}_1})- {E(\bf{q}_2)})  \nonumber \\
&=& \frac{1}{4\pi}\ \int \frac{q^2\ dq}{E^2(q)}\  \delta(M_{J/\psi}- 2\ E(q)) = \frac{1}{8\pi}\ \frac{\sqrt{M_{J/\psi}^2 -4 m^2}}{M_{J/\psi}}.
\eqa
which is numerically equal to 0.03164.\\
For a $J/\psi$  at rest decaying into a $ p\overline{p} \mathcal{B}$ channel with respective four-momenta  $q_1, q_2$ and $k$, the three-body phase space reads 
\begin{eqnarray}
\label{a1}
 L_3 &=& {(2\pi)^4} \int \prod_{i=1,2} \frac{d^4q_i}{(2\pi)^3} \theta(q^0_i)\delta(q_i^2-m_i^2) \ \frac{d^4k}{(2\pi)^3} \  \theta(k^0_i)\delta(k^2-m_ \mathcal{B}^2) \
  \delta^{(3)}( {\bf q_1} + {\bf q_2}+ {\bf k}) \ \delta(M_{J/\psi} - E(\mathbf{q}_1)-E(\mathbf{q}_2)-E_ \mathcal{B}(\mathbf{k}))  \nonumber \\
&=& \frac{1}{(2\pi)^5}\ 
\int \frac{d{\bf q_1}}{2 E(\mathbf{q}_1)} \  \int \frac{d{\bf q_2}}{2 E(\mathbf{q}_2)} \ \int \frac{d{\bf k}}{2E_ \mathcal{B}(\mathbf{k})}   \ \delta( {\bf q_1} + {\bf q_2}+ {\bf k})\ \delta(M_{J/\psi} - E(\mathbf{q}_1)-E(\mathbf{q}_2)-E_ \mathcal{B}(\mathbf{k}))
\nonumber \\
&=&   \frac{1}{(2\pi)^5}\ 
\int \frac{d{\bf q}}{2\ E(\mathbf{q})} \ \int \frac{d{\bf k}}{2\ E_ \mathcal{B}(\mathbf{k})\ 2\ E(-\mathbf{q}-\mathbf{k})} \ 
 \ \delta(f(x))
\end{eqnarray}
with 
\be \label{coskq}
x =  \frac{ {\bf k}\cdot  {\bf q}}{k\ q} \ee and  
\be \label{fx}
f(x) = M_{J/\psi} - E(q)-\sqrt{k^2 + q^2 + 2 \ kq\ x + m^2}-E_ \mathcal{B}(k).
\ee
\noindent Thus,  energy conservation implies that $f(x)$ cancels for 
\be \label{x0} 
x = x_0 = \frac{M_{p\overline{p}}^2 - 2 \ E(q) \ [M_{J/\psi} - E_ \mathcal{B}(k)]}{2 \ k \ q} 
 \ee
and  we have introduced the invariant nucleon-antinucleon invariant mass squared 
\be \label{mppbar2}
M_{p\overline{p}}^2= s = (M_{J/\psi} - E_ \mathcal{B}(k))^2 - k^2.
\ee
The invariant $N\overline{N}$ mass spans the interval $[2m,M_{J/\psi} - m_ \mathcal{B}]$. Then, we have

\bqa   \label{lph3}
 L_3 &=& \frac{1}{(2\pi)^3}  \int_0^\infty \frac{k^2\ dk}{2 E_\mathcal{B}(k)}
\int_0^\infty  \frac{q^2\ dq}{2 E(q)} \int_{-1}^1
 dx  \frac{\delta \left(f(x)\right)}{\sqrt{k^2 + q^2 + 2 \ kq\ x + m^2}}\nonumber \\
& =& \frac{1}{(2\pi)^3} \int_0^\infty \frac{k\ dk}{2 E_ \mathcal{B}(k)} \int_0^\infty  \frac{q\ dq}{2 E(q)} \int_{-1}^1
dx \  \delta(x-x_0).
\eqa 
From  Eq.~(\ref{x0}) where one has $-1\leq x_0 \leq 1$, one gets $1-x_0^2 \geq 0$ so that
$$ 4 \ k^2 \ q^2 (1-x_0^2) = - 4 \ M_{p\overline{p}}^2 \ E^2 (q)+ 4 \ [M_{J/\psi} - E_ \mathcal{B}(k)] \  E(q) \  M_{p\overline{p}}^2 - (M_{p\overline{p}}^4 + 4 \ m^2 \ k^2)$$ 
 must be positive. Hence $E(q)$ will lie between the two positive roots, $E_+$ and $E_-$, of the trinomial
 \be  \label{Epm}
 E_{\pm}= \frac{(M_{J/\psi} - E_ \mathcal{B}(k)}{2} \pm \frac{k}{2 \ M_{p\overline{p}}} \ \sqrt{M_{p\overline{p}}^2- 4 m^2},
 \ee
where $E_-> m$ and, from Eq.~(\ref{mppbar2}), $k$ is a function of the invariant mass $M_{p\overline{p}}$
\be \label{kmppbar}
k = k(M_{p\overline{p}}) = \frac{\sqrt{[(M_{J/\psi} + m_ \mathcal{B})^2 - M_{p\overline{p}}^2 ] \ [(M_{J/\psi} - m_ \mathcal{B})^2 - M_{p\overline{p}}^2]}}{2 M_{J/\psi}}
= \frac{\sqrt{\lambda(m_ \mathcal{B}^2, M_{J/\psi}^2, M_{p\overline{p}}^2) }}{2 M_{J/\psi}}, 
\ee
 where we have introduced the standard triangle (K\"allen) function $\lambda(x,y,z)$~\footnote{One has $\lambda(x,y,z) = x^2+y^2+z^2 -2xy-2yz-2zx = (x+y-z)^2-4xy= (-x+y+z)^2- 4yz  =(x-y+z)^2 - 4zx. $}.

 \noindent We may now transform the remaining integrations using
 $ q\ dq / E(q) = dE $ and from Eq.~(\ref{mppbar2}) changing  the variable $k$ to $M_{p\overline{p}}$
to obtain
 \be \label{l3bis}
 L_3 =\frac{1}{(2\pi)^3} \int_{2m}^{M_{J/\psi} - m_ \mathcal{B}} 
 \frac{M_{p\overline{p}} \ dM_{p\overline{p}}}{4 M_{J/\psi}} \ (E_+ - E_-)   =
  \frac{1}{(2\pi)^3} \  \frac{1}{4 M_{J/\psi}} \ \int_{2m}^{M_{J/\psi} - m_ \mathcal{B}}  k(M_{p\overline{p}}) \ \sqrt{M_{p\overline{p}}^2- 4 m^2} \ dM_{p\overline{p}}
   \end{equation}
and arrive at  
\bqa \label{l3final}
 L_3 &=&\frac{1}{(2\pi)^3}\  \frac{1}{8 \ M_{J/\psi}^2} \ \int_{2m}^{M_{J/\psi} - m_ \mathcal{B}} \ \sqrt{\lambda(m_ \mathcal{B}^2, M_{J/\psi}^2, M_{p\overline{p}}^2) \  [M_{p\overline{p}}^2- 4 m^2]} \ dM_{p\overline{p}} \nonumber \\
&=&  \frac{1}{(2\pi)^3} \  \frac{1}{16 \ M_{J/\psi}^2} \ \int_{4m^2}^{(M_{J/\psi} - m_ \mathcal{B})^2} \frac{du}{u} \ \sqrt{\lambda(m_ \mathcal{B}^2, M_{J/\psi}^2, u) \  \lambda(m^2, m^2, u)}.
 \eqa

In the  case  where the vector particle is a photon, the integral~(\ref{l3final}) can be calculated exactly (see, e.g.,  
Ref.~\cite{GR}) and gives 
\be \label{l3photon2}
 L_{3, \gamma} = \frac{1}{32 \pi^3}\ \Big \{ \frac{M_{J/\psi}^2 + 2\ m^2}{8 \ M_{J/\psi}} \ \sqrt{M_{J/\psi}^2- 4 m^2} - \frac{m^2\ (M_{J/\psi}^2 - m^2)}{M_{J/\psi}^2} \ \ln \Big [\frac{M_{J/\psi} +  \sqrt{M_{J/\psi}^2 -4m^2}}{2\ m}\Big ]  \Big \},
\ee
which is numerically equal to $261.718$ (MeV)$^2$.\\

\section{The width for $ J/\psi\rightarrow  p\bar{p}\mathcal{B}$ decay process} \label{amplitudes}

Let us evaluate the decay amplitude in the  Born approximation, the boson being radiated either from the proton  or from the antiproton. If we let the proton radiate, left panel on figure \ref{figFSIgraph} indicates that  $\mathbf{q}_1 = \mathbf{q} -\mathbf{k} 
$ denotes the momentum of the proton after the photon emission while $\mathbf{q}_2 = - \mathbf{q}$ 
is the spectator antiproton final momentum,  $\mathbf{k}$ the boson momentum. Then, the potential 
 $U^0_{p\overline{p}\mathcal{B}}(\mathbf{q}_1,\mathbf{k}) $ can be expressed 
as [Eq.~(\ref{defv})]
\be \label{U0g}
U^0_{p\overline{p}\mathcal{B}}(\mathbf{q}_1,\mathbf{k}) 
= A_{V,\mathcal{B}} (\mathbf{q}_1, \mathbf{k}) +  \frac{i \kappa}{2\  m}\ A_{T,\mathcal{B}} (\mathbf{q}_1, \mathbf{k})
\ee
with $\kappa$ the anomalous magnetic moment. From this expression one then evaluates the associated amplitude [Eq.~(\ref{Gtilde})]  $A^{B,BC}_{p\overline{p}\mathcal{B}}({\bf{q}_1,\bf{k}})$ which is still an operator in the spin-isospin space. From Eqs.~(\ref{am0ter}) and~(\ref{Gtilde}), we obtain the probability for the decay with a boson emission  

\be \label{g8}
\Gamma ( p\overline{p}\mathcal{B}) = \frac{1}{(2 \pi)^5} ~\int \frac{ d\textbf{q}_1\ d\textbf{q}_2 \ d\textbf{k}}{2E(q_1) 2 E (q_2) 2 E_{\mathcal{B}}(k)} \delta( M_{J/ \psi} - E(q_1)-E(q_2)-k) \ 
\delta^{(3)}(\textbf{q}_1 + \textbf{q}_2 +  \textbf{k}) 
 | A^{B,BC}_{p\bar{p}\mathcal{B}}(\mathbf{q}_2,\mathbf{k}) |^2.
 \ee

For a boson emission of mass $m_{\mathcal{B}}$ with an energy $E_\mathcal{B}(k)= \sqrt{m_\mathcal{B}^2+k^2}$, we have 
\bqa \label{gppbarV}
\Gamma ( p\overline{p} \mathcal{B} ) &=& \frac{1}{(2\pi)^5 } \int \frac{d\textbf{q} \ d\textbf{k}}{2E(q) \ 2 E (\vert{\mathbf{q}+\mathbf{k}}\vert)\ 2E_\mathcal{B}(k)} \   \delta(  M_{J/\psi} - E(q)-E(\vert{\mathbf{q}+\mathbf{k}}\vert)-E_\mathcal{B}(k)) \vert A^{B,BC}_{p\bar{p}\mathcal{B}}(\mathbf{q},\mathbf{k}) \vert^2 \nonumber \\
&=& \frac{1}{(2\pi)^4 } \int \frac{d\mathbf{q}} {2E(q)} \int \frac{k^2dk}{2E_\mathcal{B}(k)}\int dx\  \  \frac{ \delta( f(x)))}{2E (\vert{\mathbf{q}+\mathbf{k}}\vert)}\  \vert A^{B,BC}_{p\bar{p}\mathcal{B}}(\mathbf{q},k,x) \vert^2, 
\eqa
where $x$ has been defined in Eq.~(\ref{coskq}) 
and $f(x)$ is given in Eq.~(\ref{fx}). The $x$ integration, based on the energy conservation relation and assuming, $ \vert A^{B,BC}_{p\bar{p}\mathcal{B}}(\mathbf{q},k,x)\vert $ to be independent of $x$ and depend mainly on $q$, i.e.,  $ \vert A^{B,BC}_{p\bar{p}\mathcal{B}}(\mathbf{q}_r,\mathbf{k})\vert \approx \vert h(q) \vert $ gives 
\be \label{a3b}
\int ~\delta(f(x))\ \vert A^{B,BC}_{p\bar{p}\mathcal{B}}(\mathbf{q}_r,\mathbf{k})\vert ^2~\frac{dx}{E(\textbf{-q-k})}\approx \frac {\vert h(q)\vert^2}{qk},
\ee
where $x$ is fixed now at the value $x_0$ given by Eq.~(\ref{x0}) with the condition $ -1 \leq x_0\leq 1$.
Therefore, we have 
\be \label{gppbarV2}
\Gamma ( p\overline{p}\mathcal{B}) =  \frac{1}{32 \pi^3\  M_{J/\psi}} \int M_{p\overline{p}} \ dM_{p\overline{p}}\int  \frac{q dq}{E(q)}\ |h|^2,
\ee

The experimental spectrum of the $ p\overline{p}$  invariant mass is rather complicated and its description is a check for the theory in question.  Since we may write the width for the emission of a vector particle as 
\be \label{sp1}
\Gamma ( p\overline{p}\mathcal{B})  =  \int dM_{p\overline{p}}~S(M_{p\overline{p}}), 
\ee
where $S(M_{p\overline{p}})$ denotes the spectral function, we have
\be \label{Smppbar}
S(M_{p\overline{p}}) =   \frac{M_{p\overline{p}} }{32 \pi^3\   M_{J/\psi}}  \int \frac{q dq}{E(q)}\ |h(q)|^2 = 
 \frac{M_{p\overline{p}}} {32 \pi^3\   M_{J/\psi}}  \int_{E_-}^{E_+} dE\ |h(q)|^2, 
\ee
 where the integration limits $E_-$ and $E_+$ are given in Eq.~(\ref{Epm}) while the invariant mass $M_{p\overline{p}}$ is defined by Eq.~(\ref{mppbar2}) and the emitted particle momentum $k(M_{p\overline{p}})$ is read from Eq.~(\ref{kmppbar}). Numerical calculations at the endpoints require care and the following approximate expression is helpful to check the accuracy

\be \label{Smppbar2}
S(M_{p\overline{p}}) \approx \frac{M_{p\overline{p}}} {32 \pi^3\   M_{J/\psi}} \ (E_+ - E_-) \ |h|^2,
\ee
assuming further that $|h|^2$ depends only weakly on $E$. From Eq.~(\ref{Epm}), the difference $E_+ - E_-$ is simply 

$$E_+ - E_- =k(M_{p\overline{p}})  \ \frac{\sqrt{M_{p\overline{p}}^2 - 4 m^2}}{M_{p\overline{p}}}.$$
So, finally, the spectral function reads
\be \label{Smppbarfinal}
S(M_{p\overline{p}}) =\frac{\sqrt{\lambda(M_\mathcal{B}^2,M_{J/\psi}^2,M_{p\overline{p}}^2)\ [{M_{p\overline{p}}^2 - 
4m^2}]}}{64 \pi^3\   M_{J/\psi}^2} \ |h|^2
\ee
and this formula  is useful to understand the end points. Close to the $ p\overline{p}$ threshold,  $M_{p\overline{p}}= 2m$,  one finds $ S \sim \sqrt{k_M-k} $ where $ k_M = (M_{J/\psi}^2-4m^2)/2M_{J/\psi} = 979.14$ MeV is the  maximal value of  $k$  reached at the threshold. This limit leads to $ k = 2q$ and the partners in the $p\overline{p}$ pair run parallel in the direction opposite to the photon direction. Such a configuration enhances final state interactions.    
 This  dependence in $k_M-k $ determines the position of the threshold peak in $S$. At the other end of the spectrum, $M_{p\overline{p}}= M_{J/\psi} - m_\mathcal{B}$,  one has $ S \sim  k = 0$.

 \section{Explicit expressions for the electromagnetic operators}  \label{explicit}

Let $q$ be the initial nucleon four momentum and $q'= q-k$ that of the nucleon after emission of the boson 
with four-momentum $k$. The following Lorentz condition applies
 \be \label{keps0}
 {\bf{k} \cdot{\mbox{\boldmath$\epsilon$}}^* (\lambda)} - k_0 \epsilon_0^*(\lambda) = 0, 
 \ee
 which in the case of the photon leads to 
 \be \label{kepsphot} {\bf{k}\cdot{\mbox{\boldmath$\epsilon^*$}} (\lambda) }  = 0, 
 \ee
 and for a massive vector particle to
  \be \label{kepsmes}
  \epsilon_0^*(\lambda) = {\bf{k} \cdot{\mbox{\boldmath$\epsilon^*$}} (\lambda)} /k_0.
  \ee

\noindent The vector part of the current reads 
\be \label{AV0}
 \mathcal{A}_V(q',q,\epsilon^*(\lambda)) = \bar{u} ({q'}) \ \gamma_{\mu} \epsilon^{\mu*}(\lambda)\ u({q}) =  \chi_{S'}^{\dagger} \  A_V({\bf{q},\bf{k}}, \epsilon^*(\lambda)) \ \chi_S, 
\ee
where $\chi_S$ and $\chi_{S'}$ denote the standard two-dimensional spin vectors. The four-dimensional spinors read 
\begin{eqnarray} \label{uubar}
u({q}) &=& \sqrt{\frac{\Omega}{2 \ m}} \ \left( \begin{array}{c} \chi_S \\ {\displaystyle\frac{{\mbox{\boldmath$\sigma$}}
\cdot\bf{q}}{\Omega} \ \chi_S} \end{array} \right),   \nonumber \\
 \bar{u}({q'}) &=& u({q'})^{\dagger} \ \gamma_0 = \sqrt{\frac{\Omega'}{2 \ m}} \  \left( \begin{array}{c} \chi_{S'}^{\dagger}  \hspace{.5cm} \displaystyle\frac{{\mbox{\boldmath$\sigma$}}
\cdot\bf{q'}}{\Omega'} \ \chi_{S'}^{\dagger} \end{array} \right) \ \gamma_0 =  \sqrt{\frac{\Omega'}{2 \ m}} \  \left( \begin{array}{c} \chi_{S'}^{\dagger}  -  \displaystyle\frac{{\mbox{\boldmath$\sigma$}}
\cdot\bf{q'}}{\Omega'} \ \chi_{S'}^{\dagger} \end{array} \right),
\end{eqnarray}
where the energies $\Omega$ and $\Omega'$ are 
\be \label{om}
\Omega = m + E_q\hspace{2cm}{\rm and}\hspace{2cm}\Omega' = m + \sqrt{({\bf{q}-\bf{k})}^2 + m^2}, 
\ee
\be \label{Eq}
E_q =  \sqrt{q^2 + m^2}, 
\ee
with $m$ the nucleon mass.
In the following we use the Bjorken and Drell definitions of the Pauli $\sigma$ and $\gamma$ matrices\cite{Bjorken64}. 
  Hence,  from Eq.~(\ref{AV0}) for the vector term we can write 
  \be \label{AV1}
   {A}_V(q',q,\epsilon^*(\lambda))    = N \   \Big [ \big ( 1 +  \frac{{\mbox{\boldmath$\sigma$}}
\cdot\bf{q'}}{\Omega'} \  \frac{{\mbox{\boldmath$\sigma$}} \cdot\bf{q} }{\Omega}
 \big ) \ \epsilon^*_0(\lambda) -  \big (\frac{{\mbox{\boldmath$\sigma$}}
\cdot{\mbox{\boldmath$\epsilon^*$}}
(\lambda)\ {\mbox{\boldmath$\sigma$}} \cdot\bf{q}}{\Omega} + \frac{{\mbox{\boldmath$\sigma$}}\cdot\bf{q'}\ {\mbox{\boldmath$\sigma$}} \cdot{\mbox{\boldmath$\epsilon^*$}} (\lambda)}{\Omega'} \big )  \Big ], 
\ee 
with the  normalization factor $N$ 
\be \label{zeta}
N = \frac{\sqrt{\Omega \ \Omega'}}{2m}= \frac{\Omega\ \zeta}{2m}
\hspace{1cm}{\rm with}\hspace{1cm} \zeta = \sqrt{\frac{\Omega'}{\Omega}}. \ee
 \noindent Upon using the standard relation for any two vectors $\bf{v}$ and $\bf{w}$
\be \label{sigvw}
{({\mbox{\boldmath$\sigma$}}\cdot\bf{v}) \  ({\mbox{\boldmath$\sigma$}}
\cdot\bf{w})} = {\bf{v}\cdot \bf{w}} + i \ {{\mbox{\boldmath$\sigma$}} \cdot(\bf{v}\wedge \bf{w})},
\ee

\noindent the spin operator $A_V(\bf{k},\bf{q}, \epsilon^*(\lambda)) $ becomes
 \begin{eqnarray}
\label{AV2}
 A_V({\bf{q},\bf{k}}, \epsilon^*(\lambda))  &=& {\frac{\bf{k}\cdot{\mbox{\boldmath$\epsilon^*$}}(\lambda)}{2 m k_0}} \ 
 \Big \{ \Omega \zeta + {\frac{\bf{q}\cdot(\bf{q}-\bf{k})}{\Omega \zeta}}  - i \ {\frac{{\mbox{\boldmath$\sigma$}}\cdot (\bf{k} \wedge\bf{q})}{\Omega \zeta}} \Big \} \nonumber \\
  &-& \frac{1}{2 m} \ \Big \{\big (\zeta+ \frac{1}{\zeta} \big )\ {\bf{q}  \cdot{\mbox{\boldmath$\epsilon$}}^*(\lambda)} - \frac{1}{\zeta} \ {\bf{k}\cdot{\mbox{\boldmath$\epsilon$}}^*(\lambda)}  - i \  \big (\zeta - \frac{1}{\zeta} \big )   {{\mbox{\boldmath$\sigma$}}\cdot (\bf{q}\wedge{\mbox{\boldmath$\epsilon$}}^*(\lambda))} - i  \frac{1}{\zeta}  \ {{\mbox{\boldmath$\sigma$}} \cdot (\bf{k}\wedge{\mbox{\boldmath$\epsilon$}}^*(\lambda))}  \Big \}.
  \end{eqnarray}
For the emission from the antinucleon with momentum $-\bf{q}$, we simply have to change in Eqs.~(\ref{AV2}) and~(\ref{AVgamma}) $\bf{q}$  into $-\bf{q}$ 
\be \label{AVNbar}
\overline{A}_V({\bf{q},\bf{k}}, {\mbox{\boldmath$\epsilon^*$}}(\lambda))= {A}_V(-\bf{q},\bf{k}, {\mbox{\boldmath$\epsilon^*$}}(\lambda)).
\ee
The substitution $\bf{q}\to-\bf{q}$ induces $\zeta \to \overline{\zeta}$  where
\be \label{zetabar}
\overline{\zeta} = \sqrt{\frac{\overline{\Omega}}{\Omega}}
\hspace{1cm}{\rm with}\hspace{1cm} \overline{\Omega} = m + \sqrt{({\bf{q}+\bf{k})}^2+m^2}. \ee

\noindent The tensor piece is more elaborate
\be \label{tensor}
{\mathcal{A}_T(k,q, \epsilon^*(\lambda))} = \bar{u} ({q'}) \ \sigma_{\mu\nu}\ k^{\mu} \epsilon^{\nu*}(\lambda)\ u({q}) =   \chi_{S'}^{\dagger} \  A_T(\bf{k},\bf{q}, \epsilon^*(\lambda)) \ \chi_S, 
\ee
 with
 \be \label{tensor1}
 {A_T({\bf{q},\bf{k}}, \epsilon^*(\lambda))} = \frac{\Omega\zeta}{2m} \ \left (1- \displaystyle\frac{{\mbox{\boldmath$\sigma$}}
\cdot\bf{q'}}{\Omega'} \right )  \ \sigma_{\mu\nu}\ k^{\mu} \epsilon^{\nu*}(\lambda) \ \left( \begin{array}{c} 1 \\
\displaystyle\frac{{\mbox{\boldmath$\sigma$}}\cdot\bf{q}}{\Omega}
\end{array} \right).
\ee
Since
 $$\sigma_{\mu\nu}\ k^{\mu} \epsilon^{\nu}(\lambda)(k) =
 - i \ k_0 \ \left( \begin{array}{c c} 0& {\mbox{\boldmath$\sigma$}}
\cdot{\mbox{\boldmath$\epsilon$}}^*(\lambda) \\  {\mbox{\boldmath$\sigma$}}
\cdot{\mbox{\boldmath$\epsilon$}}^*(\lambda)& 0 \end{array} \right)
 + i \ \epsilon^*_0(\lambda) \  \left ( \begin{array}{c c} 0& {\mbox{\boldmath$\sigma$}}\cdot\bf{k} \\  {\mbox{\boldmath$\sigma$}}\cdot\bf{k} & 0 \end{array} \right )
+  \left( \begin{array}{c c}  {\mbox{\boldmath$\sigma$}}
\cdot (\bf{k} \wedge{\mbox{\boldmath$\epsilon$}}^*(\lambda) ) & 0\\  0 &   {\mbox{\boldmath$\sigma$}}
\cdot (\bf{k}\wedge{\mbox{\boldmath$\epsilon$}}^*\lambda)) \end{array} \right). $$
we obtain
 \bqa \label{tensor2}
{A_T(\bf{k},\bf{q}, \epsilon^*(\lambda))}  &=& \frac{\Omega\zeta}{2m}  \ \left \{
- i \ k_0 \ \big [  {\frac{{\mbox{\boldmath$\sigma$}}
\cdot{\mbox{\boldmath$\epsilon$}}^* (\lambda)\ {\mbox{\boldmath$\sigma$}}
\cdot\bf{q} }{\Omega} } -{\frac{{\mbox{\boldmath$\sigma$}}\cdot\bf{q'}\ {\mbox{\boldmath$\sigma$}}
\cdot{\mbox{\boldmath$\epsilon$}}^*(\lambda)}{\Omega'} }\big ]
+ i \ \epsilon_0^*(\lambda) \ \big [{\frac{{\mbox{\boldmath$\sigma$}} \cdot\bf{k}
\ {\mbox{\boldmath$\sigma$}} \cdot\bf{q} }{\Omega}} -{\frac{{\mbox{\boldmath$\sigma$}} \cdot\bf{q'}\ {\mbox{\boldmath$\sigma$}} \cdot\bf{k}}{\Omega'}} \big ]   \right. \nonumber \\
&+& \left.  {{\mbox{\boldmath$\sigma$}} \cdot  [\bf{k} \wedge{\mbox{\boldmath$\epsilon$}}^*(\lambda) ]}
 - {\frac{{\mbox{\boldmath$\sigma$}}\cdot\bf{q'}}{\Omega'}} \
{{\mbox{\boldmath$\sigma$}}\cdot [\bf{k} \wedge{\mbox{\boldmath$\epsilon$}}^*(\lambda) ]} \
{\frac{{\mbox{\boldmath$\sigma$}} \cdot\bf{q}}{\Omega}}  \right \}.
\eqa

\noi With the repeated use of Eq.~(\ref{sigvw}) and of double vectorial product properties,  we are led to the following explicit expression
 \bqa \label{AT}
{A_T({\bf{q},\bf{k}}, \epsilon^*(\lambda))} &=& \frac{i}{2  m}\  \frac{\bf{k}\cdot{\mbox{\boldmath$\epsilon$}}^*(\lambda) }{k_0} \Big
[\big (\zeta - \frac{1}{\zeta} + \frac{k_0}{\Omega \ \zeta} \big )\ {\textbf{k} \cdot\textbf{q}}
 + \frac{1}{\zeta} \ (\textbf{k}^2 - k_0^2) + i \ \big (\zeta + \frac{1}{\zeta} +  \frac{k_0}{\Omega \ \zeta}  \big )  \ {{\mbox{\boldmath$\sigma$}}\cdot (\bf{k}\wedge\bf{q})}
\Big ]  \nonumber\\
&- & \frac{i\ k_0}{2  m} \ {\bf{q}\cdot{\mbox{\boldmath$\epsilon$}}^* (\lambda)}  \ \Big [\zeta
- \frac{1}{\zeta} + \frac{\bf{k}^2}{k_0 \ \Omega \ \zeta} + \frac{i}{k_0\ \Omega \ \zeta} \  {\mbox{\boldmath$\sigma$}}
\cdot{(\bf{k}\wedge\bf{q})}\Big ]  
+ \frac{\Omega \ \zeta}{2  m} \  {{{\mbox{\boldmath$\sigma$}}\cdot [\bf{k}\wedge{\mbox{\boldmath$\epsilon$}}^*(\lambda) ]} } \ \Big [ 1+ \frac{k_0}{\Omega \ \zeta^2}\Big ]  \nonumber\\
&- &  \frac{k_0}{2 \ m} \  {{{\mbox{\boldmath$\sigma$}}\cdot (\bf{q}\wedge{\mbox{\boldmath$\epsilon$}}^*
(\lambda) )}
 \ \Big [\zeta + \frac{1}{\zeta} -  \frac{1}{k_0 \ \Omega \ \zeta} \ (\bf{q}-\bf{k})\cdot\bf{k} \Big ]} 
-   \frac{1}{2 \ m}\  \frac{1}{\Omega \ \zeta} \ \bf{q} \cdot[\bf{k} \wedge{\mbox{\boldmath$\epsilon$}}^*(\lambda) ] \ \ {\mbox{\boldmath$\sigma$}} \cdot\bf{q},
\eqa
where $k_0 = \sqrt{{\bf{k}}^2+m_{\mathcal{B}}^2}$. The tensor amplitude for the emission from the antinucleon will be obtained from the replacements $\bf{q}\to -\bf{q}$ and hence  $\zeta \to \overline{\zeta}$ in Eq.~(\ref{AT}).\\  

 \subsection{The specific case of the photon} \label{photoncase}

For the photon, since  $\epsilon^*_0(\lambda)= 0 $ and thus ${\bf{k}
\cdot{\mbox{\boldmath$\epsilon$}}^*(\lambda)}=0$, the vector amplitude reduces to
 \be \label{AVgamma}
 {A_{V,\gamma}({\bf{q},\bf{k}}, {\mbox{\boldmath$\epsilon$}^*}(\lambda)) } = - \frac{1}{2 \ m} \ \Big \{\big (\zeta + \frac{1}{\zeta} \big )\ {\bf{q} \cdot{\mbox{\boldmath$\epsilon$}}^*(\lambda)} 
 - i \  \big (\zeta - \frac{1}{\zeta} \big )  \ {{\mbox{\boldmath$\sigma$}}
\cdot [\bf{q}\wedge{\mbox{\boldmath$\epsilon$}}^*(\lambda)]} - i \  \frac{1}{\zeta}  \ {{\mbox{\boldmath$\sigma$}}\cdot [\bf{k} \wedge{\mbox{\boldmath$\epsilon$}}^*(\lambda)]}  \Big \},
 \ee
  while the tensor amplitude becomes
\bqa \label{ARgamma}
A_{T,\gamma}({\bf{q},\bf{k}},{\mbox{\boldmath$\epsilon$}}^*(\lambda)) &=& -\frac{i\ k_0}{2  m} \ {\bf{q}\cdot{\mbox{\boldmath$\epsilon$}}^* (\lambda)}  \ \Big [\zeta
- \frac{1}{\zeta} + \frac{\bf{k}^2}{k_0 \ \Omega \ \zeta} + \frac{i}{k_0\ \Omega \ \zeta} \  {{\mbox{\boldmath$\sigma$}}
\cdot(\bf{k}\wedge\bf{q})}\Big ]  
+ \frac{\Omega \ \zeta}{2  m} \  {{\mbox{\boldmath$\sigma$}}\cdot [\bf{k}\wedge{\mbox{\boldmath$\epsilon$}}^*(\lambda)]}  \ \Big [ 1+ \frac{k_0}{\Omega \ \zeta^2}\Big ]  \nonumber\\
&- &  \frac{k_0}{2 \ m} \  {{{\mbox{\boldmath$\sigma$}}\cdot [\bf{q}\wedge{\mbox{\boldmath$\epsilon$}}^*(\lambda) ]} \ \Big [\zeta + \frac{1}{\zeta} -  \frac{1}{k_0 \ \Omega \ \zeta} \ (\bf{q}-\bf{k})\cdot\bf{k} \Big ]}  
-   \frac{1}{2 \ m}\  \frac{1}{\Omega \ \zeta} \ \bf{q} \cdot[\bf{k} \wedge{\mbox{\boldmath$\epsilon$}}^*(\lambda) ] \ \ \ {\mbox{\boldmath$\sigma$}} \cdot\bf{q}, 
\eqa
where, here, $k_0= {\vert \bf{k} \vert} = k$. The corresponding photon amplitude for the antinucleon emission,  $\overline{A}_{T,\gamma} $, will be obtained with the replacement $\bf{q}\to -\bf{q}$ which induces $\zeta \to \overline{\zeta}$.\\

The vertex coupling yields for the photon emission from the nucleon 
\bqa \label{VNg}
V_N^{\gamma}({\bf{q},\bf{k}})&=& e\ \left (A_{V,\gamma} + i \ \frac{\kappa}{2m} \ A_{T,\gamma} \right ) \nonumber \\
 &=& - \frac{e}{2m}\left \{ \left (\zeta + \frac{1}{\zeta} \right ) - \frac{\kappa\ k_0}{2m}\ \left [
\zeta - \frac{1}{\zeta} + \frac{k_0}{\Omega \ \zeta}+i\ \frac{{\mbox{\boldmath$\sigma$}}\cdot(\bf{k}\wedge\bf{q})}{k_0\ \Omega \ \zeta}   \right ] \right \} {\bf{q}\cdot{\mbox{\boldmath$\epsilon$}}^* (\lambda)}\nonumber \\
&+& \frac{ie}{2m}\left \{ \left ( \zeta- \frac{1}{\zeta} \right )- \frac{\kappa\ k_0}{2m}\ \left (\zeta + \frac{1}{\zeta} - 
\frac{{\bf{q}\cdot\bf{k}} - k_0^2}{k_0\ \Omega \ \zeta} \ \right ) \right \} {{\mbox{\boldmath$\sigma$}} \cdot (\bf{q}\wedge{\mbox{\boldmath$\epsilon$}}^*(\lambda))} \nonumber \\
&+& \frac{ie}{2m}\left \{  \frac{1}{\zeta} + \frac{\kappa}{2m}\ \left ( \Omega \ \zeta + \frac{k_0}{\zeta}
 \right ) \right \}\ {{\mbox{\boldmath$\sigma$}}\cdot [\bf{k}\wedge{\mbox{\boldmath$\epsilon$}}^*(\lambda) ]}   \nonumber \\
  &-& \frac{ie\kappa}{2m} \ \frac{\bf{q} \cdot[\bf{k} \wedge{\mbox{\boldmath$\epsilon$}}^*(\lambda) ]}{2m\Omega \ \zeta  } \  {\mbox{\boldmath$\sigma$}} \cdot\bf{q}.
\eqa
and correspondingly for the emission from the antinucleon
\be \label{VNbarg}
V_{\overline{N}}^{\gamma} ({\bf{q}},{\bf{k}})= - e\ \left [\overline{A}_{V,\gamma}({-\bf{q}},{\bf{k}}) + i \ \frac{\kappa}{2m} \ \overline{A}_{T,\gamma}({-\bf{q}},{\bf{k}}) \right ] 
\ee
obtained by making the substitutions, $e \to -e$,  ${\bf{q}} \to - {\bf{q}}$ and, hence, $\zeta \to \overline{\zeta} $. 

\noi Labelling the nucleon part by $1$ and the antinucleon part by $2$ we will get the potential for the photon emission and recombining these expressions 
\bqa \label{VNNbarg}
U^0_{\gamma}({\bf{q},\bf{k}})
&=& \frac{e}{2m} \ \left \{ f({\bf{q}},{\bf{k}}) + \bar{f}({\bf{q}},{\bf{k}}) + i\   \frac{\kappa}{2m\Omega} \  \ \left ( \frac{{\mbox{\boldmath$\sigma$}}_1}{\zeta} -  \frac{{\mbox{\boldmath$\sigma$}}_2}{\bar{\zeta}}\right ) \cdot (\bf{k}\wedge\bf{q}) \right \} \ {\bf{q}}\cdot{\mbox{\boldmath$\epsilon$}}^*(\lambda) \nonumber \\
&+&  \frac{ie}{2m}\ \left [ g({\bf{q}},{\bf{k}}) \ {\mbox{\boldmath$\sigma$}}_1 +  \bar{g}({\bf{q}},{\bf{k}}) \ {\mbox{\boldmath$\sigma$}}_2+ \frac{\kappa}{2m\Omega} \ {\bf{q}\cdot\bf{k}}\ \left  \{\frac{{\mbox{\boldmath$\sigma$}}_1}{\zeta} - \frac{{\mbox{\boldmath$\sigma$}}_2} {\bar{\zeta}} \right \}\right ] \cdot [
{\bf{q}\wedge{\mbox{\boldmath$\epsilon$}}^*(\lambda)}]  \nonumber \\
&+&  \frac{ie}{2m}\ \left [ h({\bf{q}},{\bf{k}}) \ {\mbox{\boldmath$\sigma$}}_1 -  \bar{h}({\bf{q}},{\bf{k}}) \ {\mbox{\boldmath$\sigma$}}_2
\right ] \cdot[{\bf{k}\wedge{\mbox{\boldmath$\epsilon$}}^*(\lambda)} ] \nonumber \\
&-&  \frac{ie\kappa}{2m}\ \frac{{\bf{q}\cdot({\bf{k}\wedge{\mbox{\boldmath$\epsilon$}}^*(\lambda)})} }{2m\Omega} 
 \left ( \frac{{\mbox{\boldmath$\sigma$}}_1}{\zeta} -  \frac{{\mbox{\boldmath$\sigma$}}_2}{\bar{\zeta}}\right ) \cdot{\bf{q}}\ , 
\eqa
\noi where we have introduced 
\bqa \label{fgh}
f({\bf{q}},{\bf{k}}) &=& - \left (\zeta + \frac{1}{\zeta} \right ) + \frac{k\kappa}{2m}\ \left (\zeta - \frac{1}{\zeta} + \frac{k}{\Omega \zeta} \right ), \nonumber \\
g({\bf{q}},{\bf{k}})&=& \left (\zeta - \frac{1}{\zeta} \right ) - \frac{k\kappa}{2m}\ \left (\zeta + \frac{1}{\zeta} + \frac{k}{\Omega \zeta} \right ), \nonumber \\
h({\bf{q}},{\bf{k}})&=&  \frac{1}{\zeta} +  \frac{k\kappa}{2m}\ \left (\frac{\Omega \zeta}{k}+ \frac{1}{\zeta} \right ) ,
\eqa
where we have used $k=k_0$. The bar functions are identical but with $\zeta \to \bar{\zeta}$.\\

\subsection{The case for the $\omega$ meson} \label{omegam}

For completeness we present below the exact expressions for the $\omega$ meson emission. Only a few terms are really exploited in the present work.  With now 
\be \label{k0om} 
k_0 = E_{\omega} = \sqrt{\mathbf{k}^2 + m_{\omega}^2},
\ee
 where $m_{\omega}$ denotes the mass of the $\omega$ meson, and, taking into account the change of sign of the $\omega$ coupling at  the $\overline{N}$ vertex because of G parity, from Eqs.~(38) and~(C8), the vector piece will read,

\bqa  \label{AVom}
A_{V,\omega} &=& g_{V\omega}\  {\frac{\bf{k}\cdot{\mbox{\boldmath$\epsilon^*$}}(\lambda)}{2 m E_{\omega}} }\ 
\left [ \Omega (\zeta - \overline{\zeta})   + \frac{\bf{q}^2}{\Omega}\left (\frac{1}{\zeta} - \frac{1}{\overline{\zeta}} \right ) -   {\frac{\bf{k}\cdot\bf{q}}{\Omega}} \left (\frac{1}{\zeta} + \frac{1}{\overline{\zeta}} \right )
 - i \ \frac{1}{\Omega} \left (  {\frac{{\mbox{\boldmath$\sigma_1$}}}{\zeta}} +  \frac{{\mbox{\boldmath$\sigma_2$}}}{\overline{\zeta}} \right )\cdot (\bf{k} \wedge \bf{q}) \right ]  
\nonumber \\
 &  -& \frac{g_{V\omega}}{2m}\ \left [ \left ( \zeta+ \frac{1}{\zeta} + \overline{\zeta} + \frac{1}{ \overline{\zeta}}\right )\ {\bf{q}  \cdot{\mbox{\boldmath$\epsilon$}}^*(\lambda)} - \left (  \frac{1}{\zeta}  - \frac{1}{ \overline{\zeta}}  \right ) \ {\bf{k}  \cdot{\mbox{\boldmath$\epsilon$}}^*(\lambda)} \right. \nonumber \\
 &-&\left.  i \left (\frac{{\mbox{\boldmath$\sigma_1$}}}{\zeta} - \frac{{\mbox{\boldmath$\sigma_2$}}}{\overline{\zeta}}\right ) \cdot [{\bf{k}\wedge{\mbox{\boldmath$\epsilon$}}^*(\lambda)} ]
- i \left \{ \left (\zeta- \frac{1}{\zeta}\right ){\mbox{\boldmath$\sigma_1$}} + \left ( \overline{\zeta} -\frac{1}{\overline{\zeta}} \right )\ {\mbox{\boldmath$\sigma_2$}}  \right \}\cdot \left \{ {\bf{q}\wedge{\mbox{\boldmath$\epsilon$}}^*(\lambda)} \right \} \right ], 
\eqa
which to order $k^2/4m^2$ reduces to 
\bqa \label{AVomapp}
A_{V,\omega} &\approx&   \frac{g_{V\omega}}{2m}\ \left [ 2 \left ( -1 + \frac{q^2}{2\Omega E_q}-\frac{\Omega}{2 E_q} \right ) \frac{\bf{k}\cdot\bf{q}}{\Omega}- \frac{i}{\Omega} ({\mbox{\boldmath$\sigma_1$}}+ {\mbox{\boldmath$\sigma_2$}})\cdot(\bf{k} \wedge \bf{q}) \right ] \ {\frac{\bf{k}\cdot{\mbox{\boldmath$\epsilon^*$}}(\lambda)}{E_{\omega}} } \nonumber \\
&-&    \frac{g_{V\omega}}{2m}\ \left \{4 \ {\bf{q}  \cdot{\mbox{\boldmath$\epsilon$}}^*(\lambda)} 
-i ({\mbox{\boldmath$\sigma_1$}}- {\mbox{\boldmath$\sigma_2$}})\cdot{[\bf{k}\wedge{\mbox{\boldmath$\epsilon$}}^*(\lambda)]} + i  \frac{{\bf{k}\cdot\bf{q}}}{\Omega E_q}\ ({\mbox{\boldmath$\sigma_1$}}- {\mbox{\boldmath$\sigma_2$}})\cdot { [\bf{q}\wedge{\mbox{\boldmath$\epsilon$}}^*(\lambda)]} \right \},  \eqa
where $E_q$ has been defined in Eq.~(\ref{Eq}) and $\Omega$ in Eq.~(\ref{om}). 
Then we have to add the tensor piece, knowing that the coupling constant for this part is rather ill known but most likely small 
\bqa \label{ATom}
E_{\omega}
A_{T,\omega}&=& i\  \frac{g_{T\omega}}{2  m}\ \left \{
i\   \left ( \left [\zeta + \overline{\zeta} - \left ( 1 - \frac{E_{\omega}}{\Omega} \right ) \ \left ( \frac{1}{\zeta} +\frac{1}{ \overline{\zeta}} \right )\right ] {\textbf{k} \cdot\textbf{q}}+ \left ( \frac{1}{\zeta} -\frac{1}{ \overline{\zeta}} \right )\ ({\bf{k}^2}-E_{\omega}^2)\right. \right.  
\nonumber \\
& +&  \left. \left. i\ (\zeta  {\mbox{\boldmath$\sigma_1$}}+ \overline{\zeta}   {\mbox{\boldmath$\sigma_2$}}) 
\cdot ({\bf{k}\wedge\bf{q}}) + i\ \left ( 1 + \frac{E_{\omega}}{\Omega} \right ) \left( \frac{{\mbox{\boldmath$\sigma_1$}}}{\zeta} + \frac{{\mbox{\boldmath$\sigma_2$}}}{\overline{\zeta}}
\right )\cdot ({\bf{k}\wedge\bf{q}})   \right ) \ \frac{\bf{k}\cdot{\mbox{\boldmath$\epsilon$}}^*(\lambda) }{2mE_{\omega}}\right. \nonumber \\
&-& \left. i\ \frac{E_{\omega}}{2m} \ \left (\left [\zeta + \overline{\zeta} - \left ( 1- \frac{k^2}{E_{\omega}\Omega}\right ) \left(\frac{1}{\zeta}+ \frac{1}{\overline{\zeta}} \right ) \right ] +\frac{i}{E_{\omega}\Omega} \left ( \frac{{\mbox{\boldmath$\sigma_1$}}}{\zeta} - \frac{{\mbox{\boldmath$\sigma_2$}}}{\overline{\zeta}}
\right )\cdot ({\bf{k}\wedge\bf{q}}) \right ) {\textbf{q} \cdot{\mbox{\boldmath$\epsilon$}}^*(\lambda) } \right.\nonumber \\
& +& \left. \frac{\Omega}{2m} \left [ \left(\zeta + \frac{E_{\omega}}{\Omega\zeta} \right){\mbox{\boldmath$\sigma_1$}} - \left(\overline{\zeta} +\frac{E_{\omega}}{\Omega\overline{\zeta}} \right){\mbox{\boldmath$\sigma_2$}}
\right ]\cdot [\bf{k}\wedge{\mbox{\boldmath$\epsilon$}}^*(\lambda) ]\right.\nonumber \\
&-&\left. \frac{E_{\omega}}{2m}\ \left [ \left(
\zeta + \frac{1}{\zeta} + \frac{E_{\omega}}{\Omega\zeta} - \frac{ {\textbf{k} \cdot\textbf{q}}}{E_{\omega}\Omega\zeta} \right ){\mbox{\boldmath$\sigma_1$}}+ \left ( \overline{\zeta} + \frac{1}{\overline{\zeta}}+ \frac{k^2}{E_{\omega}\Omega\overline{\zeta}} + \frac{ {\textbf{k} \cdot\textbf{q}}}{E_{\omega}\Omega\overline{\zeta}}
\right ){\mbox{\boldmath$\sigma_2$}} \right ]\cdot [\bf{q}\wedge{\mbox{\boldmath$\epsilon$}}^*(\lambda) ]
\right. \nonumber \\
&-& \left. \frac{\mathbf{q}\cdot[\mathbf{k}\wedge {\mbox{\boldmath$\epsilon$}}^*(\lambda)]}{2m \Omega} \ \left (
 \frac{{\mbox{\boldmath$\sigma_1$}}}{\zeta} - \frac{{\mbox{\boldmath$\sigma_2$}}}{\overline{\zeta}}\right )
 \cdot\bf{q} \right \}. 
\eqa 

We have then
\be \label{Vom}
V_{\omega} = A_{V,\omega} +  A_{T,\omega}.
\ee

\section{Amplitudes for boson emission } \label{exact}

The boson with 3-momentum $\bf{k}$ is emitted either from the nucleon (labelled $1$) or from the antinucleon (labelled $2$) lines. In the rest frame of the $J/\psi$, the nucleon momentum prior to the photon emission is denoted $\bf{q}$ and  $-\bf{q}$ is that of the antinucleon. The mass of the nucleon and antinucleon is denoted by $m$.
The various variables that occur in this appendix, $\Omega$, $\Omega'$, $\overline{\Omega}$, $\zeta$ and $\bar{\zeta}$, are given in Eqs.~(\ref{om}), (\ref{zeta}) and (\ref{zetabar}).

\subsection{Photon emission} \label{gammaemission}
\subsubsection{Vector coupling} \label{vector}

When the photon is emitted from the nucleon line the contribution of the vector term to the amplitude is given by Eq.~(\ref{AVgamma}) multiplied by the charge $e$ and where the helicity $\lambda$ takes the values $\pm1$. To obtain the amplitude corresponding to the photon emission from the antinucleon line one has simply to substitute $e\to -e$ and $\mathbf{q} \to - \mathbf{q} $. Thus the full amplitude arising from the vector coupling reads 
\be \label{AVtot}
\overline{A}_V ({\mathbf{q}, \mathbf{k}},{\mbox{ \boldmath $\epsilon^*$}}(\lambda))= \overline{A}_{V,a}^{\gamma}+  
\overline{A}_{V,b}^{\gamma} + \overline{A}_{V,c}^{\gamma}\ , 
\ee
with 
\be.  \label{AV1gamma} 
\overline{A}_{V,a}^{\gamma}  ({\mathbf{q}, \mathbf{k}},{\mbox {\boldmath $\epsilon^*$}}(\lambda)) =
-\frac{e}{2m}\left [\left ( \zeta +\frac{1}{\zeta}\right )+\left(\bar{\zeta} +\frac{1}{\bar {\zeta}}\right ) \right ] {\bf{q}}
\cdot {\mbox {\boldmath $\epsilon^*$}}(\lambda), 
\ee
\be \label{AV2gamma} 
\overline{A}_{V,b}^{\gamma}  ({\mathbf{q}, \mathbf{k}},{\mbox {\boldmath $\epsilon^*$}}(\lambda)) =
\frac{ie}{2m}\left [\left (\zeta -\frac{1}{\zeta}\right ){\mbox{\boldmath $\sigma$}}_1 \cdot {\bf q} 
\wedge{\mbox {\boldmath$\epsilon^*$}}(\lambda) +\left (\bar {\zeta} -\frac{1}{\bar{\zeta}}\right )
{\mbox {\boldmath $\sigma$}}_2 \cdot {\bf q}\wedge{\mbox {\boldmath $\epsilon^*$}}(\lambda)  \right ], 
\ee
\be \label{AV3gamma} 
\overline{A}_{V,c}^{\gamma}  ({\mathbf{q}, \mathbf{k}},{\mbox {\boldmath $\epsilon^*$}}(\lambda))
= \frac{ie}{2m} \ \left (\frac{{{\mbox {\boldmath $\sigma$}}_1}}{\zeta}-\frac{{{\mbox
{\boldmath $\sigma$}}_2}}{\tilde{\zeta}}\right ) \cdot {\bf k}
\wedge{\mbox {\boldmath $\epsilon^*$}}(\lambda) .
\ee
In these equations $\lambda= \pm 1$ and since, for the photon, $\epsilon^*_0(\lambda) =0 $ and $\mathbf{k}$
is chosen to lie along the $z$-axis, the orthogonality relation $\mathbf{k}\cdot {\mbox {\boldmath $\epsilon^*$}}(\lambda) =0$ implies
\be
{\mbox {\boldmath $\epsilon^*$}}(\lambda=\pm 1) = \frac{1}{\sqrt{2}} \left (-\lambda,-i,0 \right ) .
\ee
The $ \overline{A}_{V,a}^{\gamma}$ amplitude contributes to the $^3S_1 \to ^3P_1$
electric coupling amplitude~(\ref{VEphoton}). The $ \overline{A}_{V,b}^{\gamma}$
amplitude leads to a final $^3P_0$ state not included, so far, in
our work. Note that, it vanishes in the approximation $\zeta \sim
\tilde{\zeta}$. The $ \overline{A}_{V,c}^{\gamma}$ amplitude contributes to the
$^3S_1 \to ^1S_0$ transition. 

\subsubsection{Tensor coupling} \label{sstensor}

With the same notations as above, the exact amplitude for the tensor coupling when the photon is emitted from the nucleon line reads
\bqa
\label{A1gammaN}
A_{1}^{\gamma} ({\mathbf{q}, \mathbf{k}},{\mbox {\boldmath $\epsilon^*$}}(\lambda))&=&\frac{e\kappa_N}{4m^2}\left \{k\ {\bf q}\cdot {\mbox {\boldmath $\epsilon^*$}}(\lambda) \left (\zeta -\frac{1}{\zeta}+\frac{k}{ \Omega \zeta}+\frac{i}{k \Omega \zeta}{\mbox {\boldmath $\sigma$}}_1 \cdot {\bf k}\wedge {\bf q}\right )+i {\mbox {\boldmath $\sigma$}}_1 \cdot {\bf k}\wedge{\mbox {\boldmath $\epsilon^*$}}(\lambda) \left (\Omega \zeta+\frac{k}{\zeta} \right ) \right. \nonumber \\
&-& \left. ik {\mbox {\boldmath $\sigma$}}_1 \cdot {\bf
q}\wedge{\mbox {\boldmath $\epsilon^*$}}(\lambda) \left [\zeta
+\frac{1}{\zeta}-\frac{1}{k\Omega \zeta} ({\bf {q-k}})\cdot {\bf
k}\right ] - \frac{i}{\Omega \zeta} {\bf q} \cdot \left ({\bf k}\wedge
{\mbox {\boldmath $\epsilon^*$}}(\lambda) \right )\ {\mbox {\boldmath
$\sigma$}}_1 \cdot {\bf q} \right \},
\eqa
where $\kappa_N$ is either $\kappa_p$ or  $\kappa_n$. In the following we do not keep the terms 
proportional to ${\mbox{\boldmath $\sigma$}}_1 \cdot {\bf q}\wedge
{\mbox {\boldmath$\epsilon^*$}}(\lambda) $ and depending on $q^2$ as they will contribute to final
$^3P_0$ and $D$ waves, respectively. These $P$ and $D$ wave are
absent in our model. Thus, the amplitude is reduced to
\be 
A_{1}^{\gamma} ({\mathbf{q}, \mathbf{k}},{\mbox {\boldmath $\epsilon^*$}}(\lambda)) \approx 
\frac{e\kappa_N}{4m^2}\left \{k\ {\bf q}\cdot {\mbox {\boldmath $\epsilon^*$}}(\lambda) \left (\zeta -\frac{1}{\zeta}+\frac{k}{ \Omega \zeta}\right )+i {\mbox {\boldmath $\sigma$}}_1 \cdot 
{\bf k}\wedge{\mbox {\boldmath $\epsilon^*$}}(\lambda) \left (\Omega \zeta+\frac{k}{\zeta} \right ) \right \},  
\ee
to which we have to add the part associated to the emission from the antinucleon 
$A_{2}^{\gamma} ({\mathbf{q}, \mathbf{k}},{\mbox {\boldmath $\epsilon^*$}}(\lambda))$. Thus the total tensor amplitude in this approximation $A_{T}^{\gamma} ({\mathbf{q}, \mathbf{k}},{\mbox {\boldmath $\epsilon^*$}}(\lambda))$ can be split into two contributions (with $\lambda= \pm 1$)

\be \label{ATgammatot}
 A_{T}^{\gamma} ({\mathbf{q}, \mathbf{k}},{\mbox {\boldmath $\epsilon^*$}}(\lambda)) =
 A_{1}^{\gamma} ({\mathbf{q}, \mathbf{k}},{\mbox {\boldmath $\epsilon^*$}}(\lambda)) + 
 A_{2}^{\gamma} ({\mathbf{q}, \mathbf{k}},{\mbox {\boldmath $\epsilon^*$}}(\lambda)) =
 A_{T,a}^{\gamma} ({\mathbf{q}, \mathbf{k}},{\mbox {\boldmath $\epsilon^*$}}(\lambda)) +  A_{T,b}^{\gamma} ({\mathbf{q}, \mathbf{k}},{\mbox {\boldmath $\epsilon^*$}}(\lambda)), 
\ee
with 
\be \label{AT1gammatot} 
A_{T,a}^{\gamma} ({\mathbf{q}, \mathbf{k}},{\mbox {\boldmath $\epsilon^*$}}(\lambda)) =
 \frac{e\kappa_N}{4m^2}\ k \ \left [\zeta +\bar{\zeta}+\left(\frac{1}{\zeta}+ \frac{1}{\bar{\zeta}} \right )
\left (\frac{k}{\Omega} -1\right ) \right ] {\bf q}\cdot {\mbox{\boldmath $\epsilon^*$}}(\lambda),
\ee
and 
\be \label{AT2gammatot} 
A_{T,b}^{\gamma} ({\mathbf{q}, \mathbf{k}},{\mbox {\boldmath $\epsilon^*$}}(\lambda)) =
\frac{e\kappa_N}{4m^2}\left [i {\mbox{\boldmath $\sigma$}}_1 \cdot {\bf k}\wedge{\mbox {\boldmath
$\epsilon^*$}}(\lambda) \left (\Omega \zeta+\frac{k}{\zeta} \right )-i {\mbox {\boldmath $\sigma$}}_2 \cdot {\bf k}\wedge{\mbox {\boldmath$\epsilon^*$}}(\lambda) \left (\Omega\bar{\zeta}+\frac{k}{\bar{\zeta}} \right )\right].
\ee

The amplitude (\ref{AT1gammatot})  will add up to the $\overline{A}_{V,a}^{\gamma}  ({\mathbf{q}, \mathbf{k}},{\mbox {\boldmath $\epsilon^*$}}(\lambda))$ term (\ref{AV1gamma}) to
give the $^3S_1 \to ^3P_1$ transition amplitude while the (\ref{AT2gammatot}) together with the (\ref{AV3gamma}) amplitude will contribute to the  $^3S_1 \to ^1S_0$ transition.

\subsubsection{Magnetic and electric transitions}  \label{3S11S0}
\begin{itemize}
\item 
{\bf  $^3S_1 \to ^1S_0$ transitions}\\

The $^3S_1 \to ^1S_0$ magnetic coupling amplitude will be given by the sum of the (\ref{AV3gamma})
 and (\ref{AT2gammatot}) amplitudes. For any vector $\mathbf{a}$ the spin matrix elements read 
  \be \label{sigma1.a} 
 \left \langle ^1S_0 \right \vert {\mbox {\boldmath$\sigma_1$}} \cdot {\bf a} \left \vert ^3S_1 \right \rangle = 
- \left \langle ^1S_0 \right \vert {\mbox {\boldmath $\sigma_2$}} \cdot {\bf a}
\left \vert ^3S_1 \right \rangle =  \frac{1}{\sqrt{3} }\left ( -i \sqrt{2}a_y + a_z \right ), 
 \ee
 since the spin contents of the $^1S_0 $ and $^3S_1$ states are, with $\vert + \rangle$ the $1/2$  and $\vert - \rangle$ the $-1/2$
spin states, respectively,
 \be \label{spins}
 \vert ^1S_0 \rangle = \frac{1}{\sqrt{2}}\ \left [ \vert +-\rangle - \vert -+ \rangle  \right ],
\hspace{1cm}
  \vert ^3S_1 \rangle = \frac{1}{\sqrt{3}}\ \left [ \vert ++ \rangle + \frac{1}{\sqrt{2}}\left \{ \vert +-\rangle + \vert -+ \rangle \right \} +  \vert -- \rangle \right ]. \ee

\noi  With ${\bf k}$ along the $z$ axis, as already defined, we have (see Ref.~\cite{pil67} p. 62)
\be
\label{epsi} {\mbox {\boldmath$\epsilon^*$}}(\lambda= \pm 1) = \left(-\frac{\lambda}{\sqrt{2}}, -\frac{i}{\sqrt{2}}, 0\right ),
\hspace{1cm}
{\bf k} \wedge {\mbox {\boldmath $\epsilon^*$}}(\lambda= \pm 1) =\left ( \frac{ik}{\sqrt{2}},
-\frac{\lambda k}{\sqrt{2}}, 0 \right ), 
\ee
and therefore to 
\bqa \label{VMphoton}
V_M^{\gamma}({\mathbf{q}, \mathbf{k}},{\mbox {\boldmath $\epsilon^*$}}(\lambda)) &=& \langle ^1S_0 \vert \overline{A}_{V,c}^{\gamma} ({\mathbf{q}, \mathbf{k}},{\mbox {\boldmath $\epsilon^*$}}(\lambda)) +
 A_{T,b}^{\gamma} ({\mathbf{q}, \mathbf{k}},{\mbox {\boldmath $\epsilon^*$}}(\lambda))  \vert ^3S_1 \rangle  \nonumber \\
 &=& 
 - \frac{e\lambda k}{2m \sqrt{3}}
 \ \left \{ \left (1+ \frac{\kappa_N \ k}{2m} \right ) \left [\frac{1}{\zeta} + \frac{1}{\bar{\zeta}} \right ] 
+  \frac{\kappa_N \ \Omega}{2m}\ \left (\zeta+ \bar{\zeta} \right ) \right \}.
 \eqa
Upon summing over $\lambda$ the modulus squared we get 
\be
\sum_{\lambda= \pm 1}\vert
V_M^{\gamma}({\mathbf{q}, \mathbf{k}},{\mbox {\boldmath $\epsilon^*$}}(\lambda)  
\vert^2 = \frac{e^2}{4m^2}\ \frac{2k^2}{3} \
 \left \{ \left (1+ \frac{\kappa_N \ k}{2m} \right ) \left [\frac{1}{\zeta} + \frac{1}{\bar{\zeta}} \right ] 
+  \frac{\kappa_N \ \Omega}{2m}\ \left (\zeta+ \bar{\zeta} \right ) \right \}^2.
\ee
Note that in the small $k/2m$ limit, this reduces to 
\be 
\sum_{\lambda \pm 1}\vert
V_M^{\gamma}({\mathbf{q}, \mathbf{k}},{\mbox {\boldmath $\epsilon^*$}}(\lambda)  \vert^2 \approx \frac{8 e^2}{3}\ \frac{k^2}{4m^2}\ 
\left [1 + \frac{\kappa_N \Omega}{2m}\right ]^2,
\ee
a small contribution indeed. 
Using only the vector piece of the amplitude one would have obtained 
\be
\label{sumsigdotkcroepsi} \frac{e^2}{4m^2}\ \sum_{\lambda=\pm 1} \left \vert \left \langle ^1S_0 \right \vert 
\left (\frac{\mbox{\boldmath$\sigma_1$}}{\zeta}-\frac{\mbox{\boldmath$\sigma_2$}}{\bar{\zeta}}\right ) \cdot  {\bf k} \wedge 
{\mbox{\boldmath $\epsilon^*$}} (\lambda)\left \vert ^3S_1 \right \rangle\right \vert^2= \frac{2 e^2}{3}\ 
 \frac {k^2}{4m^2}\ \left [\frac{1}{\zeta} + \frac{1}{\bar{\zeta}} \right ] ^2.
\ee

Considering the final state scattering contributions with intermediate $p \bar p$
and $n \bar n$ states~(\ref{ampfsi}) one obtains the following magnetic amplitude

\bqa \label{A1VMgam} 
V_M^{\gamma}({\mathbf{q}, \mathbf{k}},{\mbox {\boldmath $\epsilon^*$}}(\lambda))
&=&-\frac{\lambda e k}{4 m \sqrt{3}} \left \{ \left (\frac{1}{\zeta}
+\frac{1}{\bar{\zeta}}\right )\left (T_0+T_1\right ) \right.\nonumber \\
&+& \left. \frac{1}{2m} \left[ \Omega \left ( \zeta +\bar{\zeta} \right ) + k  \left (\frac{1}{\zeta}
+\frac{1}{\bar{\zeta}}\right )\right ] \left [ T_0\left(\kappa_p +\kappa_n \right ) + T_1\left(\kappa_p 
-\kappa_n \right )\right]\right \}, 
\eqa
where $T_{0,1}$ are the $N\bar N$ scattering amplitudes in the corresponding $I=0,1$ isospin states.\\ 

\item {\bf $^3S_1 \to ^3P_1$ transitions} \label{3S13P1} \\

The $^3S_1 \to ^3P_1$ electric coupling amplitude is given by the sum of the amplitudes (\ref{AV1gamma}) and (\ref{AT1gammatot})

\be \label{VEphoton}
{\overline{A}_{V,a}^{\gamma}  ({\mathbf{q}, \mathbf{k}},{\mbox {\boldmath $\epsilon^*$}}(\lambda)) +
A_{T,a}^{\gamma} ({\mathbf{q}, \mathbf{k}},{\mbox {\boldmath $\epsilon^*$}}(\lambda))} 
= -\frac{e}{2 m} \ {{\bf q} \cdot {\mbox{ \boldmath $\epsilon^*$}}(\lambda)}  \left \{ \zeta
+\frac{1}{\zeta}+\bar{\zeta} +
  \frac{1}{\bar{\zeta}}-\frac{k\kappa_p}{2m}\left [\zeta+\bar{\zeta}
  +\left ( \frac{1}{\zeta}+\frac{1}{\bar{\zeta}}\right) \left (\frac{k}{\Omega}-1\right)   \right]\right \}.
\ee
Summing the squared ${{\bf q} \cdot {\mbox{ \boldmath $\epsilon^*$}}(\lambda)}$ term over $\lambda$ gives 
\be
\sum_{\lambda =\pm 1} \vert {{\bf q} \cdot {\mbox{ \boldmath $\epsilon^*$}}}(\lambda) \vert^2
= \frac{1}{2}\Sigma_{\lambda=\pm 1} \vert (-\lambda
q_x+iq_y) \vert^2 =q^2\ \left [1-\cos^2 \theta_q \right],
\ee
where
 \be \cos\theta_q= \frac{{\bf k}\cdot {\bf q}}{k\ q}.
 \ee 
 Hence 
\bqa
\sum_{\lambda =\pm 1} \vert {\overline{A}_{V,a}^{\gamma}  ({\mathbf{q}, \mathbf{k}},{\mbox {\boldmath $\epsilon^*$}}(\lambda))} &+& {A_{T,a}^{\gamma} ({\mathbf{q}, \mathbf{k}},{\mbox {\boldmath $\epsilon^*$}}(\lambda))} \vert^2 \nonumber \\
&=& \frac{e^2 q^2}{4m^2} \left [1-\cos^2 \theta_q \right] \left \{ \zeta+\frac{1}{\zeta}+\bar{\zeta} +  \frac{1}{\bar{\zeta}}-\frac{k\kappa_p}{2m}\left [\zeta+\bar{\zeta}+\left (\frac{1}{\zeta}+\frac{1}{\bar{\zeta}}\right) \left (\frac{k}{\Omega}-1\right)   \right]\right \}^2,
\eqa
which in the small $k/2m$ limit reduces to
$$ \frac{4 e^2 q^2}{m^2}\ \left [1-\cos^2 \theta_q \right]. $$
\end{itemize}

  \subsection{$\omega$ emission} \label{omegaemission}
 
For the $\omega$ meson, the polarization vector reads (see Ref.~\cite{pil67} p. 62) 
\be \label{epsiomeg}
\epsilon^*(\lambda=\pm 1) = \frac{1}{\sqrt{2}}\left (0, -\lambda,-i, 0 \right ), \hspace{2cm}
\epsilon^*(\lambda =0) = \frac{1}{m_\omega}\left (k, 0, 0, E_\omega \right ),
\ee
 As before the momentum of the emitted $\omega$ meson is assumed to lie along  the $z$-axis, so that 
one has 
\be \label{kdotepsi}
{\bf k} \cdot {\mbox {\boldmath $\epsilon^*$}}(\lambda=\pm 1) = 0, 
\ \ \ {\bf k} \cdot {\mbox {\boldmath $\epsilon^*$}}(\lambda= 0) = \frac{kE_\omega}{m_\omega},
\ \ \ {\bf k} \wedge {\mbox {\boldmath $\epsilon^*$}}(\lambda=\pm 1) = \frac{k}{\sqrt{2}} \ (i,-\lambda,0), 
\ {\rm and}\ \ \ \ \ 
{\bf k} \wedge {\mbox {\boldmath $\epsilon^*$}}(\lambda= 0) = 0.
\ee
The Lorentz condition, $k^\mu \cdot \epsilon^*(\lambda)_\mu = E_{\omega} \ \epsilon_0^*(\lambda) - {\bf k}
\cdot {\mbox {\boldmath $\epsilon$}}^*(\lambda) =0 $, implies
\be
\epsilon_0^*(\lambda) = \frac{{\bf k}\cdot {\mbox {\boldmath $\epsilon$}}^*(\lambda)}{E_\omega}.
\ee
The  $\omega$ tensor coupling being small we shall neglect its
contribution~\cite{rij06}. If the $\omega$ is emitted from the nucleon line, the exact amplitude for the 
vector coupling is given by Eq.~(\ref{AV2}) multiplied by the coupling constant $g_{V\omega}$. To obtain the full amplitude one has to add the amplitude corresponding to the $\omega$ emission from the antinucleon given by Eq.~(\ref{AVNbar}). It is given by Eq.~(\ref{AVom}). In addition to Eq.~(\ref{sigma1.a}), we have, for any vector $\bf{a}$, the equality 
\be
\langle ^3P_1 \vert \ {\mbox{\boldmath$\sigma_1$}}\cdot \mathbf{a} \vert ^3S_1\rangle =
\langle ^3P_1 \vert \ {\mbox{\boldmath$\sigma_2$}}\cdot \mathbf{a} \vert ^3S_1\rangle.
\ee 
The dominant electric contribution for $\lambda =\pm 1$ will then read 
\be
\label{VEomegapm1} V_E^\omega({\lambda=\pm 1}) =
-\frac{g_{V\omega}}{2 m} \ {\bf q} \cdot {\mbox{ \boldmath$\epsilon^*$}}(\lambda \pm 1)  \left 
( \zeta +\frac{1}{\zeta}+\bar{\zeta} + \frac{1}{\bar{\zeta}} \right ), 
\ee
while for $\lambda = 0$, since $ \epsilon_0^*(\lambda=0) = {k}/{m_\omega}$, one has 
\be
V_E^\omega({\lambda=0}) =
- \frac{g_{V\omega}}{2 m} \ \left \{ {\bf q} \cdot {\mbox{ \boldmath$\epsilon^*$}}(\lambda = 0)  \left 
( \zeta +\frac{1}{\zeta}+\bar{\zeta} + \frac{1}{\bar{\zeta}} \right )
+ \frac{k}{m_\omega}\ \left [\frac{1}{\zeta} + \frac{1}{\bar{\zeta}}\right ]\
\frac{\mathbf{k}\cdot\mathbf{q}}{\Omega} \right \}, 
\ee
and, thus,
\be
V_E^\omega({\lambda=0}) = - g_{V\omega} \ \frac{q \ \cos\theta_q}{2 m m_\omega} \ \left \{ 
E_\omega  \left ( \zeta +\frac{1}{\zeta}+\bar{\zeta} + \frac{1}{\bar{\zeta}} \right )
+ \frac{k^2}{\Omega} \  \left [\frac{1}{\zeta} + \frac{1}{\bar{\zeta}}\right ] \right \}.
\ee
This contributes to  the  $^3S_1 \to ^3P_1$ transition.
Summing over  the helicity $\lambda$ the squared amplitudes we obtain
\bqa \label{SVEom2}
\sum_{\lambda= -1}^1 \vert V_E^\omega({\lambda}) \vert^2 = g^2_{V\omega} \ \frac{q^2}{4m^2}\
 &&\! \! \! \left \{ (1-\cos^2\theta_q)  \ \left [ \zeta +\frac{1}{\zeta}+\bar{\zeta} + \frac{1}{\bar{\zeta}}  \right]^2  \right .\nonumber \\
&& \left . \ \ \ \ +\  \cos^2\theta_q  \ \left [ \frac{E_\omega}{m_\omega}  \left ( \zeta +\frac{1}{\zeta}+\bar{\zeta} + \frac{1}{\bar{\zeta}} \right )+ \frac{k^2}{m_\omega\Omega} \  \left (\frac{1}{\zeta} + \frac{1}{\bar{\zeta}} \right ) \right ]^2  \right \}. 
\eqa
In the small $k/2m$ limit, this reduces to 
$$\sum_{\lambda= -1}^1 \vert V_E^\omega({\lambda}) \vert^2 \approx g^2_{V\omega} \ \frac{4 q^2}{m^2}\
\left \{(1-  \cos^2\theta_q ) + \left (\frac{E_{\omega}}{m_{\omega}} \right )^2 \  \cos^2\theta_q \ \left [ 1 +\frac{k^2}{E_{\omega} \Omega}
\right ] \right \},
$$
giving rise again to a very small correction to the dominant piece $4 g^2_{V\omega} q^2 /m^2$.\\

For the magnetic contribution, using Eqs.~(\ref{kdotepsi}) and (\ref{AVom}) we retain the dominant piece and get 
\be
V_M^\omega({\lambda}= \pm 1) = i \ \frac{g_{V\omega}}{2 m} \ \left \{
\frac{\mbox{\boldmath$\sigma_1$}}{\zeta}-\frac{\mbox{\boldmath$\sigma_2$}}{\bar{\zeta}} \right \} \ 
\cdot [{\bf k} \wedge {\mbox {\boldmath $\epsilon^*$}}(\lambda = \pm 1)].
\ee
It is null for $\lambda =0$. Then, let us thus look at the $^3S_1 \to ^1S_0$ magnetic transition using Eqs.~(\ref{spins})
\bqa
\langle ^1S_0\vert V_M^\omega({\lambda}= \pm 1) \vert ^3S_1 \rangle &=&   i \ \frac{g_{V\omega}}{2 m} \ 
\langle ^1S_0\vert \left \{
\frac{\mbox{\boldmath$\sigma_1$}}{\zeta}-\frac{\mbox{\boldmath$\sigma_2$}}{\bar{\zeta}} \right \} \ 
\cdot [{\bf k} \wedge {\mbox {\boldmath $\epsilon^*$}}(\lambda = \pm 1)] \vert ^3S_1 \rangle 
\nonumber \\
&=& -\frac{\lambda g_{V\omega}
k}{2 m \sqrt{3}}  \left (\frac{1}{\zeta}+\frac{1}{\bar{\zeta}}\right ).
\eqa 
Summing over the helicities the amplitude squared, one gets 
\be \label{SVMom2}
\sum_{\lambda=\pm 1} \vert \langle ^1S_0\vert V_M^\omega({\lambda}= \pm 1) \vert ^3S_1 \rangle
\vert^2 = \frac{g^2_{V\omega}}{4 m^2}\ \frac{2 k^2}{3} \ \left (\frac{1}{\zeta}+\frac{1}{\bar{\zeta}}\right )^2,
\ee
and thus about 
$$\frac{8 g^2_{V\omega}}{3} \ \frac{k^2}{4m^2}$$
in the small $k/2m$ limit, i.e., a very small contribution. 
The $J/\psi$ and the $\omega$ being isospin 0 states only the $I=0$
component of the $p\bar p \to p \bar p$  and $n \bar n  \to p \bar
p$ rescattering terms contributes. In the Paris potential the
modulus of  these two components being equal, there will be a
cancellation either in the convention given by the Eqs.~(22) and
(23) of Ref.~\cite{lac09} or in that of Eqs.~(49) and (50) [there,
one will have to change the sign of the Paris  $n \bar n  \to p \bar p$ amplitude].

\section{ Numerical calculation of the loop integrals} \label{numloop}

In this appendix we outline 
the numerical calculation of the loop integrals in Eqs.~(\ref{ampfsi2}) or (\ref{s3}) with propagators (\ref{GNNbarV}) and (\ref{Gtilde}). The structure of these equations goes like
\be \label{Iloop}
I(\mathbf{q},\mathbf{k}) = 
\int~ \frac{d\mathbf{q}'}{(2\pi)^3} \ T_{ N \overline{N}}(\mathbf{q}-\mathbf{k}/2,\mathbf{q}'-\mathbf{k}/2, E_{ N\overline{N}})\ 
G^+_{0, N\overline{N}\mathcal{B}}( \mathbf{q}',\mathbf{k }) \ \widetilde{G}_{p\overline{p}}(q')\ U^0(\mathbf{q}',\mathbf{k}), 
 \ee
where we will use the half off-shell values of the $N\overline{N}$ scattering matrix evaluated from the Paris potential. \\

\noi With $\mathbf{k}$ along the $z$-axis, $\mathbf{q}$ in the $x$-$z$-plane, $\mathbf{q}=(q\ \sin\theta_q,0, q\ \cos\theta_q)$, we  have 
\be
\mathbf{q}\cdot\mathbf{q}'= qq'\ \left (\sin\theta_q  \sin\theta' \ \cos\varphi' + \cos\theta_q  \cos\theta' \right )
\hspace{1cm}{\rm and} \hspace{1cm} 
\mathbf{k}\cdot\mathbf{q}'= kq'\   \cos\theta'.
\ee 
Let us assume that the half off-shell dependence of the scattering matrix depends on the momentum transfer
\be
T_{ N \overline{N}}(\mathbf{q}-\mathbf{k}/2,\mathbf{q}'-\mathbf{k}/2, E_{ N\overline{N}}) = 
T_{ N \overline{N}}(\mbox{\boldmath$\chi'$} , E_{ N\overline{N}}), 
\ee
with
\be
\mbox{\boldmath$\chi'$} = {(\mathbf{q}-\frac{\mathbf{k}}{2})} -{(\mathbf{q}' -\frac{\mathbf{k}}{2})}= \mathbf{q}-\mathbf{q}'
\hspace{0.7cm} {\rm so\ \ that} \hspace{0.7cm} \chi' = \sqrt{q^2+q'^2 - 2 qq'\ \left (\sin\theta_q  \sin\theta' \ \cos\varphi' + \cos\theta_q  \cos\theta' \right )}.
\ee
Then, 
\be
G^+_{0, N\overline{N}\mathcal{B}}( \mathbf{q}',\mathbf{k })= \frac{1}
{M_{J/\psi}+i\epsilon -E_{\mathcal{B}}(k) -\sqrt{\mathbf{q}'^2+m^2}- E}, 
\ee
where $E_{\mathcal{B}}(k)= \sqrt{k^2+m_{\mathcal{B}}^2}$, the energy of the emitted boson while $E$ is the off-shell energy of the nucleon from which the boson $\mathcal{B}$ has been emitted:
\be 
E = \sqrt{{(\mathbf{q}' - \mathbf{k})}^2 + m^2 }.
\ee
 Using $E$ as a variable, rather than $\cos\theta'$, we may reexpress Eq.~(\ref{Iloop}) as
\bqa \label{Iloop2}
I(\mathbf{q},\mathbf{k})& =& \frac{1}{(2\pi)^3\ k \sqrt{\mathcal{V}_0}}\ \int_0^\infty q' dq'\ \frac{\mathcal{F}(q')}{M_{J/\psi}+ i \epsilon- 2 \sqrt{\mathbf{q}'^2+m^2}} \int_{E_-}^{E_+} \frac{EdE}{M_{J/\psi}+i\epsilon - E_{\mathcal{B}}(k) -\sqrt{\mathbf{q}'^2+m^2} - E} \nonumber \\ 
 &&\hspace{8cm}\times \int_0^{2\pi} d\varphi' \ T_{ N \overline{N}}(\mbox{\boldmath$\chi'$}, E_{ N\overline{N}})\ U^0(\mathbf{q}',\mathbf{k}),
\eqa
where we have used the relation $E\ dE = - k\ q'\ d \cos\theta' $ and where $\mathcal{F}(q')$ denotes the source function as given by Eq.~(\ref{dir3}). The explicit dependence on $\mathbf{q}$ in $I(\mathbf{q},\mathbf{k})$ comes from the $\mbox{\boldmath$\chi'$}$ dependence in the  half off-shell scattering matrix. The limits of integration for the $E$-integration are 
\be E_{\pm} = \sqrt{m^2+(k\pm q')^2 }. \ee
The invariant mass squared of the $N\overline{N}$ pair, $s=M_{N\overline{N}}^2$ is 
\be
s= \left ( \sqrt{\mathbf{q}'^2+m^2} + \sqrt{{(\mathbf{q}' - \mathbf{k})}^2 + m^2 }\right )^2 -\mathbf{k}^2, 
\ee
since the total momentum of this pair is 
$- \mathbf{k}$  and the relative energy~\footnote{In the spirit of the Paris potential and its parametrization, the non-relativistic approximation to this expression would read 
$$E_{ N\overline{N}}\approx \frac{T_{lab}}{2} \ \left \{ 1-\left ( \frac{T_{lab}}{8m}+\frac{k^2}{8m^2} \right ) \right \}
\hspace{1cm}{\rm with} \hspace {1cm} T_{lab} = \frac{s-4m^2}{2m}.$$}

\be
E_{ N\overline{N}} = \sqrt{s+ \mathbf{k}^2} - \sqrt{4m^2+ \mathbf{k}^2}.\ee
Note that in the non-relativistic limit this expression goes to 
\be 
 E_{ N\overline{N}} \approx \frac{s-4m^2}{4m}.
 \ee
 At threshold $ E_{ N\overline{N}}=0$, so that $s = 4 m^2$ and the emitted boson reaches its maximum momentum value, i.e., $k=979.9$ MeV/c for the photon and $ k=742.5$ MeV/c for the omega. At the other end of the spectrum for the emitted boson $k=0$
 and $s= (M_{J/\psi}- M_{\mathcal{B}})^2$ and we point out that there is no singularity in the integral due to this value of $k=0$.\\
 
The integral over  $\varphi'$ is performed numerically without difficulty and displays no singularity. In each of the other two integrations, there is the presence of a pole. In the $q'$-integral the pole in $q'$ lies at $q_p=\sqrt{{M}_{J/\psi}^2-4m^2}/2=1231.82$~MeV)  as one can write
\be \label{Ed} 
\frac{1} {M_{J/\psi}- 2 E(q')+ i \varepsilon}= -\frac{M_{J/\psi}+2 E(q')}{4(q'-q_p-i \varepsilon)(q'+q_p)}.
\ee
For practical calculation of the integral~(\ref{Iloop2}) it is sufficient to integrate the $q'$ variable up to the maximum value ${q}_{Max}=12$~fm$^{-1}$ (2367.94~MeV). \\
  
In Eq.~(\ref{Iloop2}) there will be a pole in $E$  at

\be \label{Ep} 
E_p={M}_{J/\Psi}-E_{\mathcal{B}}(k)-E(q'),
\ee
if $E_-\leq E_p \leq E_+$, i.e., if

\be \label{PosEp}
 \sqrt{({q'}-k)^2+m^2} \leq {M}_{J/\Psi}-\sqrt{k^2+{M_{\mathcal{B}}}^2}-\sqrt{{q'}^2+m^2} \le \sqrt{({q'}+k)^2+m^2}.
\ee
Study of the inequalities~(\ref{PosEp}) allows to write the integral~(\ref{Iloop2}) as

\be \label{A1eqI1pI2pI3} 
I{(\mathbf{q},\mathbf{k})}= I_1+I_2+I_3,
\end{equation}
with
\be
\label{I1}
 I_1= \int^{{q}_1}_{0} dq' ... ({\rm {no~pole~in~}} q')  \int^{E_+}_{E_-}~dE ...
({\rm {no~pole~in~}}E),
\ee
\be
\label{I2}
 I_2= \int^{{q}_2}_{{q}_1} dq' ... ({\rm {no~pole~in~}} q')  \int^{E_+}_{E_-}~dE ...
({\rm {pole~in~}} E),
\ee
and
\be
\label{I3}
 I_3= \int^{{q}_{Max}}_{{q}_2} dq' ... ({\rm {pole~in~}} q')  \int^{E_+}_{E_-}~dE ...
({\rm {no~pole~in~}} E).
\ee
In the above integrals, $I_1$, $I_2$ and $I_3$ the ... are to be identified with the corresponding functions given in Eq.~(\ref{Iloop2}). Defining 
\be \label{qpm} 
q^\pm=\frac {-ks \pm \sqrt{s^2 (k^2 +s-4 m^2)-4m^2 k^2 s}} {-2s} =
\frac{k}{2} \mp \frac{1}{2} \ \sqrt{(s+k^2)\frac{(s-4m^2)}{s}},
\ee
and
\be \label{Ed1} 
E_{ N\bar{N}}^0=\frac {{M}_{J/\psi}^2-{m_{\mathcal{B}}}^2} {4({M}_{J/\psi}-m)}-m= \frac{(M_{J/\psi}-2m)^2-{m_{\mathcal{B}}}^2}{4({M}_{J/\psi}-m)},
\ee
one finds, for $E_{ N\bar{N}}\leq E_{ N\bar{N}}^0$, $q_1=q^+$, $q_2=q^-$ and for $E_{ N\bar{N}} \geq E_{ N\bar{N}}^0$, $q_1=-q^+$, $q_2=q^-$. For $E_{
N\bar{N}} = E_{ N\bar{N}}^0$, $q_1=0$. The boundary $E_{ N\bar{N}}^0$ is equal to 172.48~MeV for the photon case and to 101.54~MeV for the omega one. Note that at the $p\bar p$ threshold there is no pole in the integrals over $dE$, and
\be \label{A1ENbN0}
 I(0,\mathbf{k})= \int^{q_{Max}}_{0} dq' ... ({\rm {pole~in~}} q')  \int^{E_+}_{E_-}~dE ... ({\rm {no~pole~in~}}E).
\ee
In the numerical program for the loop calculation,  
the principal value integrals in $E$ and in $q'$ are calculated using the FORTRAN subroutine dqawce.f download from the quadpack-netlib website~( http://www.netlib.org/quadpack/).\\


\end{document}